\documentclass[12pt]{article}
\usepackage[top=30truemm,bottom=30truemm,left=25truemm,right=25truemm]{geometry}
\usepackage{comment}
\usepackage[]{hyperref}
\usepackage{amsmath}
\usepackage{amsthm}
  \theoremstyle{definition}

  \newtheorem{example}{Example}
  
\usepackage{amssymb}
\usepackage[all]{xy}
\usepackage[pdftex]{graphicx}
\usepackage{here}
\usepackage{mathrsfs}
\usepackage{braket} 
\usepackage{mwe}

\newcommand{\imineq}[2]{\vcenter{\hbox{\includegraphics[height=#2ex]{#1}}}}

\newcommand{\tr}{\mathrm{tr}}
\newcommand{\Tr}{\mathrm{Tr}}

\newcommand{\de}{\partial}
\newcommand{\be}{\begin{equation}}
\newcommand{\ba}{\begin{eqnarray}}
\newcommand{\ea}{\end{eqnarray}}
\newcommand{\ee}{\end{equation}}

\newcommand{\s}{\sqrt}
\newcommand{\vp}{\varphi}

\newcommand{\ti}{\tilde}

\newcommand{\ddd}{\cdot\cdot\cdot}
\newcommand{\no}{\nonumber \\}
\newcommand{\la}{\langle}
\newcommand{\lb}{\rangle}
\newcommand{\bea}{\begin{eqnarray}}
\newcommand{\eea}{\end{eqnarray}}
\newcommand{\bes}{\begin{equation*}}
\newcommand{\beas}{\begin{eqnarray*}}
\newcommand{\eeas}{\end{eqnarray*}}
\newcommand{\bas}{\begin{array*}}
\newcommand{\eas}{\end{array*}}
\newcommand{\ees}{\end{equation*}}
\newcommand{\nn}{\nonumber}

\newcommand{\ep}{\epsilon}

\def\CO{{\mathcal{O}}}
\def\CT{{\mathcal{T}}}
\def\CM{{\mathcal{M}}}
\def\CH{{\mathcal{H}}}
\def\a{{\alpha}}
\def\vv{{\varphi}}

\makeatletter

\@addtoreset{equation}{section}
\makeatother

\begin{document}

\begin{titlepage}
\thispagestyle{empty}

\begin{flushright}
YITP-20-71
\\
IPMU20-0060
\\
\end{flushright}

\bigskip

\begin{center}
\noindent{{ \huge {Holographic Pseudo Entropy}}}\\
\vspace{1.5cm}

Yoshifumi Nakata$^{a,b}$,
Tadashi Takayanagi$^{a,c,d}$,
Yusuke Taki$^{a}$,\\
Kotaro Tamaoka$^{a}$
and
Zixia Wei$^{a}$
\vspace{1cm}\\

{\it $^a$Center for Gravitational Physics,\\
Yukawa Institute for Theoretical Physics,
Kyoto University, \\
Kitashirakawa Oiwakecho, Sakyo-ku, Kyoto 606-8502, Japan}\\

{\it $^b$JST, PRESTO, \\
4-1-8 Honcho, Kawaguchi, Saitama 332-0012, Japan}

{\it $^c$Inamori Research Institute for Science,\\
620 Suiginya-cho, Shimogyo-ku,
Kyoto 600-8411 Japan}\\

{\it $^{d}$Kavli Institute for the Physics and Mathematics
 of the Universe (WPI),\\
University of Tokyo, Kashiwa, Chiba 277-8582, Japan}

\end{center}

\begin{abstract}
We introduce a quantity, called pseudo entropy, as a generalization of entanglement entropy via post-selection. In the AdS/CFT correspondence, this quantity is dual to areas of minimal area surfaces in time-dependent Euclidean spaces which are asymptotically AdS. We study its basic properties and classifications in qubit systems. In specific examples, we provide a quantum information theoretic meaning of this new quantity as an averaged number of Bell pairs when the post-selection is performed. We also present properties of the pseudo entropy for random states. We then calculate the pseudo entropy in the presence of local operator excitations for both the two dimensional free massless scalar CFT and two dimensional holographic CFTs. We find a general property in CFTs that the pseudo entropy is highly reduced when the local operators get closer to the boundary of the subsystem. We also compute the holographic pseudo entropy for a Janus solution, dual to an exactly marginal perturbation of a two dimensional CFT and find its agreement with a perturbative calculation in the dual CFT. We show the linearity property holds for holographic states, where the holographic pseudo entropy coincides with a weak value of the area operator. Finally, we propose a mixed state generalization of pseudo entropy and give its gravity dual.
\end{abstract}

\end{titlepage}

\tableofcontents
\newpage
\section{Introduction}

In recent developments of theoretical physics, entanglement entropy has played crucial roles, unifying theoretical frameworks in quantum information theory, condensed matter physics, and high energy physics. Entanglement entropy measures the amount of quantum entanglement, which is an important resource in quantum information theory \cite{HHHH,NC}. It also becomes a useful quantum order parameter in condensed matter physics \cite{Vidal:2002rm,Kitaev:2005dm}.
In high energy physics, this quantity provides a universal characterization of degrees of freedom in quantum field theories \cite{BKLS,Sr,HLW,CC04,Casini:2004bw}. 
Moreover, the geometric formula of entanglement entropy \cite{RT,HRT,RTreview}, based on
 the anti de-Sitter space/ conformal field theory (AdS/CFT) correspondence \cite{Maldacena:1997re,GKPW}
or holography in a more general context \cite{tHooft:1993dmi,Susskind:1994vu}, motivates us expect that spacetimes in gravity emerge from quantum entanglement \cite{Swingle:2009bg,VanRaamsdonk:2010pw}.

The AdS/CFT relates path integrals in conformal field theories (CFTs) to gravitational partition functions in anti-de Sitter (AdS) spaces in an equivalent way \cite{Maldacena:1997re,GKPW}. This allows us to understand quantum states in quantum many-body systems, which are normally algebraically complicated, in the light of a much simpler geometrical language. One typical example of this advantage of AdS/CFT is the computation of entanglement entropy. In AdS/CFT, the entanglement entropy in CFTs can be computed as the area of extremal surface \cite{RT,HRT}, whose derivations based on the bulk-boundary correspondence were given in \cite{Lewkowycz:2013nqa,Dong:2016hjy}. The entanglement entropy is defined by dividing the Hilbert space as
$\mathcal{H}=\mathcal{H}_A\otimes \mathcal{H}_B$ and by calculating the von Neumann entropy
\ba
S(\rho_A)=-\mbox{Tr}[\rho_A\log\rho_A],
\ea
where $\rho_A=\mbox{Tr}_B[|\Psi\lb\la\Psi|]$ is the reduced density matrix for the 
total quantum state $|\Psi\lb\in \mathcal{H}$.
At the AdS boundary, the extremal surface, whose area gives the above entanglement entropy, is supposed to end on the boundary of $A$. 

In a static asymptotically AdS spacetime, the extremal surface sits on the canonical time slice and therefore coincides with the minimal area surface on that slice \cite{RT}. This static setup can be Wick rotated into a Euclidean AdS/CFT setup in a straightforward way. On the other hand, in a generic asymptotically AdS spacetime which is time-dependent, there is no common canonical time slice on which extremal surfaces at a fixed boundary time extend \cite{HRT}. The Wick rotation which brings this Lorentzian time-dependent metric into a Euclidean one is subtle as the metric can become complex valued in general under this procedure.

\begin{figure}[H]
\centering
\includegraphics[width=12cm]{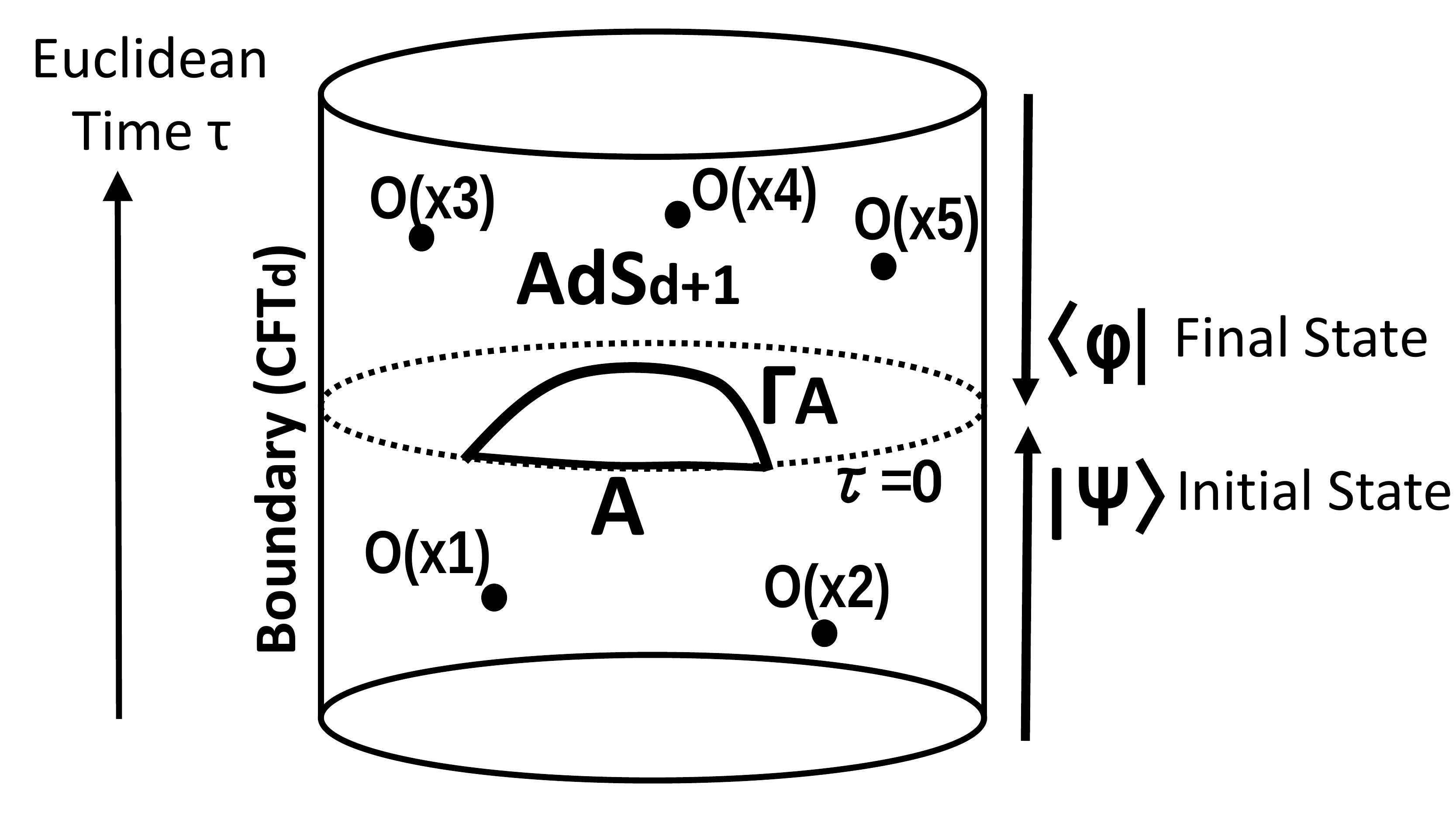}
\caption{The calculation of holographic pseudo entropy. At the time specified as the dotted circle, the bra state and ket state are different. Accordingly, the asymptotically AdS Euclidean geometry is time-dependent. The dots are excitations by inserting external sources or operators to CFTs. }\label{fig:EAdSPE}
\end{figure}

This raises a simple question: what is the meaning of the area of minimal surfaces in Euclidean asymptotically AdS  spaces which are time-dependent? Via the AdS/CFT, such a geometry is typically dual to a path-integral in the dual CFT with time-dependent sources as sketched in figure \ref{fig:EAdSPE}. We write the coordinate of the Euclidean time as $\tau$.
By focusing on a particular Euclidean time $\tau=0$, the upper path-integral (i.e. $0\leq \tau<\infty$) defines a quantum state $|\vv\lb$, while the lower one (i.e. $-\infty<\tau\leq 0)$ defines another quantum state $|\psi\lb$. In particular, the gravity partition function on the whole Euclidean space, which is asymptotically AdS, coincides with the inner product 
$\la \vv|\psi\lb$.  

In this paper we argue that the area of a minimal area surface in a (generically time-dependent) Euclidean asymptotically AdS space calculates the following CFT quantity analogous to the von Neumann entropy, which we call pseudo entropy:
\ba
S(\mathcal{T}^{\psi|\vv}_A)=-\mbox{Tr}[\mathcal{T}^{\psi|\vv}_A\log \mathcal{T}^{\psi|\vv}_A],
\label{entrops}
\ea 
where $\mathcal{T}^{\psi|\vv}_A=\mbox{Tr}_B[ \mathcal{T}^{\psi|\vv}]$ is defined from the transition matrix:
\begin{equation}
  \mathcal{T}^{\psi|\vv} \equiv \frac{|{\psi}\rangle\langle{\vv}|}{\braket{\vv|\psi}}.
  \label{tmintro}
\end{equation}
We call  $\mathcal{T}^{\psi|\vv}_A$ a reduced transition matrix.
Notice that these matrices $\mathcal{T}^{\psi|\vv}_A$ and $\mathcal{T}^{\psi|\vv}$ are not Hermitian in general and therefore we cannot regard them as quantum (mixed) states in an ordinary sense. Indeed, in general the entropic quantity $S(\mathcal{T}^{\psi|\vv}_A)$ is complex valued. Nevertheless, such generalized density matrices naturally arise in the context of the post-selection where the initial state $|\psi\lb$ is post-selected into the state $|\vv\lb$. In the post-selection setup, a quantity analogue to the expectation value of an observable operator ${\cal O}$ can be defined by
\ba
\la {\cal O} \lb=\frac{\la \vv|{\cal O}|\psi\lb}{\la \vv|\psi\lb}=\mbox{Tr}[{\cal O} \mathcal{T}^{\psi|\vv} ].
\ea
This is known as a weak value \cite{AAV} and has been studied since it can be used in certain experiments (refer e.g. \cite{WV} for a review). 
Note that this quantity also takes complex values in general.
Refer to \cite{SSW} for studies of conditional entropy of post-selected states.

The main purpose of this paper is to introduce the novel quantity (\ref{entrops}), which we call pseudo entropy, in quantum many-body systems and field theories mainly motivated by the above holographic consideration.\footnote{We should distinguish the pseudo entropy, defined through the post-selection process as we explained in the above,  
from the (standard) entanglement entropy after  projection measurements. The latter was studied in 
\cite{Rajabpour:2015uqa,Numasawa:2016emc} for two dimensional CFTs.}
Since this quantity provides a new fundamental relationship between geometries in gravity and quantum information in CFT, we expect that this helps us to understand the basic principle of the AdS/CFT correspondence and eventually quantum gravity. Also, its involving the information of two independent quantum states makes it largely different from known quantum informational quantities which are related to geometries in gravity. 
At the same time, this quantity can serve as a new class of order parameters in quantum many-body systems. 

This paper is organized as follows. In section \ref{sec:PEbasics}, we will present the basic definitions and properties of pseudo entropy. We will also give a general replica method which will be used to calculate the pseudo entropy in quantum field theories. Moreover, we will explain the holographic calculation of pseudo entropy in AdS/CFT. In section \ref{sec:qubitpr}, we examine properties of pseudo entropy in qubit systems. Firstly we classify the reduced transition matrices and also point out a monotonicity property in two qubit systems. Next we will provide an interpretation of pseudo entropy for a class of transition matrices in terms of averaged number of Bell pairs that could have been distilled from the intermediate state when a final state is post-selected. We also show a couple of typical properties of pseudo entropy. In section \ref{sec:rfreecft}, we will present computations of pseudo R\'enyi entropy in a two dimensional free scalar CFT by choosing the quantum states to be locally excited states. In section \ref{secHPE}, we study holographic pseudo entropy. After we explain its general properties, we give explicit results in a Janus AdS/CFT, which corresponds to the pseudo entropy for CFT states with two different external fields. We also present results of holographic pseudo entropy for locally excited states. We will also show the linearity is satisfied in our holographic formula. In section \ref{sec:MixedGen}, we give a mixed state generalization of pseudo entropy, based on reflected entropy. In section \ref{sec:Conclusions}, we summarize our conclusions and discuss future problems. In appendix \ref{sec:REandvNE}, we give a detailed discussion of the expressions of pseudo (R\'enyi) entropy. In appendix \ref{tfdpe}, we present a calculation of pseudo entropy for thermo field double states.  In appendix \ref{thppe}, we calculate pseudo entropy for two coupled harmonic oscillators. In appendix~\ref{App:Haar}, we provide how to compute the average of the pseudo entropy for random states.
In appendix \ref{pertep}, we analyze pseudo entropy for quantum states with perturbative external fields. In appendix \ref{opepea}, we study the pseudo entropy and fidelity for local operator excited states. In appendix \ref{app:linearity}, we leave a proof for the linearity of pseudo entropy associated with linear combination of heavy states.

\section{Basics of Pseudo Entropy}\label{sec:PEbasics}

Here we introduce a quantity, called pseudo entropy, which is a generalization of entropy from an ordinary quantum state to a process of a post-selection. After we give its definition and basic properties, we explain general methods to calculate this quantity in quantum field theories and the AdS/CFT.

\subsection{Definition of Pseudo Entropy}
Consider two pure quantum states $|\psi\lb$ and $|\vv\lb$ which satisfy $\braket{\vv|\psi}\neq0$. We introduce the transition matrix between them $\mathcal{T}^{\psi|\vv}$ as follows:
\begin{equation}
  \mathcal{T}^{\psi|\vv} \equiv \frac{|{\psi}\rangle\langle{\vv}|}{\braket{\vv|\psi}},
  \label{eq:TMdef}
\end{equation}
which is normalized such that its trace is one.

Note that this satisfies
\ba
\left(\mathcal{T}^{\psi|\vv}\right)^n=\mathcal{T}^{\psi|\vv}, \label{prodp}
\ea
and thus we find
\ba
\mbox{Tr}[\left(\mathcal{T}^{\psi|\vv}\right)^n]=1, \label{prodpp}
\ea
for any $n\in\mathbb{N}^+$. Under the exchange of  $|\psi\lb$ and $|\vv\lb$, we find
\ba
 \mathcal{T}^{\psi|\vv}= \left[\mathcal{T}^{\vv|\psi}\right]^\dagger.\label{prodppp}
\ea

We divide the total system into two subsystems $A$ and $B$ such that the total Hilbert space $\mathcal{H}$ is given by a tensor product:
\ba
\mathcal{H}=\mathcal{H}_A\otimes \mathcal{H}_B.
\ea
Accordingly we introduce the reduced transition matrix 

\begin{equation}
  \mathcal{T}^{\psi|\vv}_A \equiv \Tr_B \left[\CT^{\psi|\vv}\right] = \mbox{Tr}_B\left[\frac{|\psi\rangle\langle{\vv}|}{\braket{\vv|\psi}}\right].
\end{equation}

Now we introduce ``$n$-th R\'enyi entropy" of the transition matrix $\mathcal{T}^{\psi|\vv}_A$ in the same way as we define the $n$-th R\'enyi entropy of a quantum state $\rho$:
\ba
S^{(n)}(\mathcal{T}^{\psi|\vv}_A) \equiv \frac{1}{1-n}\log\mbox{Tr}\left[(\mathcal{T}^{\psi|\vv}_A)^n\right]~~(n\in \mathbb{N}^+,n\geq2),
\ea
where we can simply choose the branch of the log function: $-\pi<{\rm Im} \left[\log(z)\right]\leq\pi$. We call this quantity $S^{(n)}(\CT^{\psi|\vv}_A)$ the pseudo $n$-th R\'{e}nyi entropy. In the following of this paper, we use $\log(z)$ to denote the principal value of the logarithmic function. Notice that, since  $\mathcal{T}^{\psi|\vv}_A$ is not a quantum state (i.e. Hermitian and non-negative), this pseudo R\'{e}nyi entropy takes a complex value in general. Note also that when $B$ is empty, i.e. $\mathcal{T}^{\psi|\vv}_A=\mathcal{T}^{\psi|\vv}$, it is obvious that its entropy is vanishing 
\ba
S^{(n)}(\mathcal{T}^{\psi|\vv})=0.
\ea
This can be easily seen from (\ref{prodpp}).

For $n\in\mathbb{N}^+$, $n\geq2$, $S^{(n)}(\CT^{\psi|\vv}_A)$ admits an alternative expression
\begin{align}
    S^{(n)}(\CT^{\psi|\vv}_A) = \frac{1}{1-n} \log \left[\sum_j\lambda_j(\CT^{\psi|\vv}_A)^n\right], \label{eq:PREeigen}
\end{align}
where $\lambda_j(\CT^{\psi|\vv}_A)$ are the eigenvalues of $\CT^{\psi|\vv}_A$.\footnote{
Note that since  $\CT^{\psi|\vv}_A$ is not Hermitian in general, we cannot always diagonalize it either by the unitary matrices or even by regular matrices. Nevertheless, its eigenvalues are always well-defined and (\ref{eq:PREeigen}) follows directly from a Jordan decomposition of $\CT^{\psi|\vv}_A$. Refer to appendix \ref{sec:REandvNE} for a review of Jordan normal form and an argument about pseudo R\'{e}nyi entropy using it.}

This expression can be extended to $n \in \mathbb{R}^+\backslash\{1\}$ where $a^n$ is defined as 
\begin{align}
    a^n \equiv e^{n\log(a)}~~(a \in \mathbb{C}, n\in\mathbb{R}^+).
\end{align}

This allows us to take the $n\rightarrow1$ limit and define a von Neumann version of pseudo entropy as 
\begin{align}
    S(\CT^{\psi|\vv}_A) \equiv \lim_{n\rightarrow1} S^{(n)}(\CT^{\psi|\vv}_A) = -\sum_{j}\lambda_j(\CT^{\psi|\vv}_A)\log\left[\lambda_j(\CT^{\psi|\vv}_A)\right].
\end{align}
We call this the pseudo entropy of $\CT^{\psi|\vv}_A$, or the entanglement pseudo entropy of $A$.

Finally, we would like to note that pseudo (R\'{e}nyi) entropy can be expressed in matrix form as follows:
\begin{align}
    &S^{(n)}(\CT^{\psi|\vv}_A) = \frac{1}{1-n}\log\Tr \left(\CT^{\psi|\vv}_A\right)^n,~~(n\in\mathbb{R}^+\backslash\{1\})\\
    &S(\CT^{\psi|\vv}_A) = -\Tr\left(\CT^{\psi|\vv}_A \log \CT^{\psi|\vv}_A \right),
\end{align}
if we give proper treatments to some subtle points. See appendix \ref{sec:REandvNE} for a discussion on it. 

A few comments are in order. In the main parts of this paper, especially in the context of quantum field theories and the AdS/CFT correspondence, we focus on the case where  $\mbox{Tr}\left[(\mathcal{T}^{\psi|\vv}_A)^n\right]$ is positive. Therefore, we do not need to worry about the choice of branch of the log function. In more generic examples,  $\mbox{Tr}\left[(\mathcal{T}^{\psi|\vv}_A)^n\right]$ takes complex values and the choice of the branch is important. Though in the above we just chose a simple one 
$-\pi<{\rm Im} \left[\log(z)\right]\leq\pi$,\footnote{We would like to thank very much a referee for pointing out that this particular choice of the branch may be relevant when one wants to apply Carlson's theorem to prove the uniqueness of extension of holographic pseudo R\'enyi entropy to non-integer replica number $n$, under the assumption that all the eigenvalues of the corresponding reduced transition matrix are not negative numbers. We would like to leave the precise discussion and justifications for the assumptions above as a future problem.}
we would like to keep open the possibility that there might be a better choice for applications in physics, leaving as a future problem.

\subsection{Basic Properties of Pseudo Entropy}

First we have to note that the pseudo entropy $S^{(n)}(\CT^{\psi|\vv}_A)$ is complex valued in general
because the eigenvalues of the transition matrix $\mathcal{T}^{\psi|\vv}_A$ can be complex. 
Only in special choices of states  $|\psi\lb$ and $|\vv\lb$, we find real and non-negative eigenvalues of 
$\mathcal{T}^{\psi|\vv}_A$, which we will be especially interested in this paper.

In general, we can easily prove the following basic properties for $n\in \mathbb{R}^+$,
\ba
&&(i)\ \mbox{If}\ |\psi\lb\ \mbox{has no entanglement, then }\ S^{(n)}(\mathcal{T}^{\psi|\vv}_A)=0. \label{propa} \\
&&(ii)\ \mbox{If $\Tr(\CT^{\psi|\vv}_A)^n$ and all eigenvalues of $\CT^{\psi|\vv}_A$ are in $\mathbb{C}\backslash\mathbb{R}^-$, }\nonumber\\
&&~~~~~\mbox{then }S^{(n)}(\mathcal{T}^{\psi|\vv}_A)=S^{(n)}(\mathcal{T}^{\vv|\psi}_A)^*.\label{propb} \\
&&(iii)\ S^{(n)}(\mathcal{T}^{\psi|\vv}_A)=S^{(n)}(\mathcal{T}^{\psi|\vv}_B).  \label{propc}
\ea 
The property $(i)$ follows because when $|\psi\lb=|\psi'\lb_A|\psi''\lb_B$, we can show
$\left(\mathcal{T}^{\psi|\vv}_A\right)^n=\mathcal{T}^{\psi|\vv}_A$.
The property $(ii)$\footnote{The preconditions of $(ii)$ are imposed since $\log(z) = \left(\log(z^*)\right)^*$ does not hold for $z\in \mathbb{R}^-$ in our convention. Note that these preconditions are sufficient but not necessary. The precondition ``eigenvalues of $\CT^{\psi|\vv}_A$ are in $\mathbb{C}\backslash\mathbb{R}^-$" can be removed if we restrict to $n\in \mathbb{N}^+\backslash\{1\}$. In section \ref{sec:randomstates}, we will use pseudo Tsallis entropy $T_q (\CT^{\psi|\vv}_A) =(1- \tr (\CT^{\psi|\vv}_A)^q)/(q-1)$ for analysis. It satisfies $T_q(\CT^{\psi|\vv}_A) = T_q(\CT^{\vv|\psi}_A)^*$ for $q\in\mathbb{N}^+\backslash\{1\}$.} can be found from the relation (\ref{prodppp}). 
The property $(iii)$ can also be proved via an explicit calculation.

\subsection{Classifications of Transition Matrices}
In this paper, we mainly focus on the transition matrices which give positive $S^{(n)}(\CT^{\psi|\vv}_A)$ for $n>0$. To understand how special this class is, we would like to classify the transition matrices. 

We start with a Hilbert space $\CH$ factorized into two parts as $\CH = \CH_A \otimes \CH_B$. Let us denote the set of transition matrices on $\CH$ as $\mathscr{T}(\CH)$. Consider several special classes of transition matrices given as follows.
\begin{itemize}
    \item $\mathscr{A}$: The set of transition matrices which give real-valued $\Tr\left(\CT^{\psi|\vv}_A\right)^n$ for $n>0$.
    \item $\mathscr{B}$: The set of transition matrices which give nonnegative real $S^{(n)}(\CT^{\psi|\vv}_A)$ for $n>0$.
    \item $\mathscr{C}$: The set of transition matrices which give $\CT^{\psi|\vv}_A$ whose eigenvalues are real and nonnegative. 
    \item $\mathscr{D}$: The set of transition matrices which give positive semi-definite Hermitian $\CT^{\psi|\vv}_A$.
    \item $\mathscr{E}$: The set of transition matrices which give positive semi-definite Hermitian $\CT^{\psi|\vv}_A$ and positive semi-definite Hermitian $\CT^{\psi|\vv}_B$.
\end{itemize}
It is not difficult to figure out that $\mathscr{E}\subseteq\mathscr{D}\subseteq\mathscr{C}\subseteq\mathscr{B}\subseteq\mathscr{A}\subseteq\mathscr{T}(\CH)$.  

For example, in a trivial setup where ${\rm dim}\CH = {\rm dim}\CH_A = {\rm dim}\CH_B = 1$ and $\CT^{\psi|\vv}=1$, these classes degenerate, i.e. $\mathscr{E}=\mathscr{D}=\mathscr{C}=\mathscr{B}=\mathscr{A}=\mathscr{T}(\CH)$.

In particular, for 2-qubit systems, we find $\mathscr{E}\subset\mathscr{D}\subset\mathscr{C}=\mathscr{B}\subset\mathscr{A}\subset\mathscr{T}(\CH)$, as we will see in section \ref{cltqsz}. We will explore various properties of pseudo entropy in qubit systems in section \ref{sec:qubitpr}.

With these in mind, the only nontrivial part remained in the general inclusion relation $\mathscr{E}\subseteq\mathscr{D}\subseteq\mathscr{C}\subseteq\mathscr{B}\subseteq\mathscr{A}\subseteq\mathscr{T}(\CH)$ is whether there is a system in which there exists a $\CT^{\psi|\vv}$ belonging to $\mathscr{B}$ but not to $\mathscr{C}$. The answer is yes. The following two states in a 2-qutrit system give such an example
\begin{align}
    &\ket{\psi} = \left(U\otimes I\right)\ket{\vv}, \\
    &\ket{\vv} = \frac{2}{\sqrt{5}}\ket{0}\otimes\ket{0} + \frac{1}{\sqrt{10}}\ket{1}\otimes\ket{1} + \frac{1}{\sqrt{10}}\ket{2}\otimes\ket{2},
\end{align}
where
\begin{align}
    U = |0\rangle\langle0| + \cos\frac{\pi}{4} |1\rangle\langle1| - \sin\frac{\pi}{4} |1\rangle\langle2| + \sin\frac{\pi}{4} |2\rangle\langle1| + \cos\frac{\pi}{4}|2\rangle\langle2|.
\end{align}
In this case 
\begin{align}
    \CT^{\psi|\vv}_A &= \frac{5\sqrt{2}}{4\sqrt{2}+1}\left(\frac{4}{5}|0\rangle\langle0| + \frac{1}{10\sqrt{2}}|1\rangle\langle1| - \frac{1}{10\sqrt{2}}|1\rangle\langle2| + \frac{1}{10\sqrt{2}}|2\rangle\langle1| + \frac{1}{10\sqrt{2}}|2\rangle\langle2| \right),\nonumber\\
\end{align}
and the three eigenvalues of $\CT^{\psi|\vv}_A$ are
\begin{align}
    \frac{8}{8+\sqrt{2}}\equiv p,~~ \frac{1+i}{\sqrt{2}(8+\sqrt{2})}\equiv q e^{i\omega},~~ \frac{1-i}{\sqrt{2}(8+\sqrt{2})}\equiv q e^{-i\omega}.
\end{align}
Clearly this example is in class $\mathscr{A}$ but not in class $\mathscr{C}$. We can see that, for $n<1$,
\begin{align}
    \Tr\left(\CT^{\psi|\vv}_A\right)^n=p^n+(qe^{i\omega})^n+(qe^{-i\omega})^n 
    = p^n + 2q^n \cos (n\omega) 
    > p + 2 q \cos \omega = 1, 
\end{align}
and hence $S^{(n)}(\CT^{\psi|\vv}_A) > 0$. For $n>1$,
\begin{align}
    \Tr\left(\CT^{\psi|\vv}_A\right)^n= p^n+(qe^{i\omega})^n+(qe^{-i\omega})^n > p^n-2q^n >0. 
\end{align}
Moreover, for $1<n\leq2$,
\begin{align}
    \Tr\left(\CT^{\psi|\vv}_A\right)^n= p^n+(qe^{i\omega})^n+(qe^{-i\omega})^n = p^n + 2q^n \cos (n\omega) < p+2q\cos\omega=1,
\end{align}
and for $2<n$,
\begin{align}
    \Tr\left(\CT^{\psi|\vv}_A\right)^n= p^n+(qe^{i\omega})^n+(qe^{-i\omega})^n \leq p^n + 2q^n < p^2 + 2q^2 = \frac{33}{33+\sqrt{2}} < 1.
\end{align}
Therefore, $0<\Tr\left(\CT^{\psi|\vv}_A\right)^n<1$ for $n>1$ and hence $S^{(n)}(\CT^{\psi|\vv}_A) > 0$. On the other hand, we can explicitly confirm that $S(\CT^{\psi|\vv}_A)>0$. As a result, this example is in class $\mathscr{B}$ but not in class $\mathscr{C}$.

\subsection{Pseudo Entropy in QFT} \label{sec:replica}
(Entanglement) pseudo entropy can also be computed in quantum field theories using a replica trick in path integral formalism. This is just a straightforward generalization of the replica trick for computing conventional entanglement entropy given in \cite{HLW,CC04}. 
The upshot is that we can simply generalize the replica trick such that the initial and final state are different, keeping the replicating procedure the same, to calculate the pseudo entropy.

For simplicity, let us firstly consider a QFT which lives on an infinite line $\mathbb{R}$ parameterized by $x$ and under a Euclidean time evolution (or imaginary time evolution). The following inner product can be evaluated using a path integral over a manifold with proper boundary conditions imposed. 
\begin{align}
    \braket{\beta|e^{-T H[\hat{\phi}(x)]}|\alpha} = \int \prod_{0<\tau<T}\prod_{x} d\phi(\tau,x)~ e^{-S[\phi(\tau,x)]} = \imineq{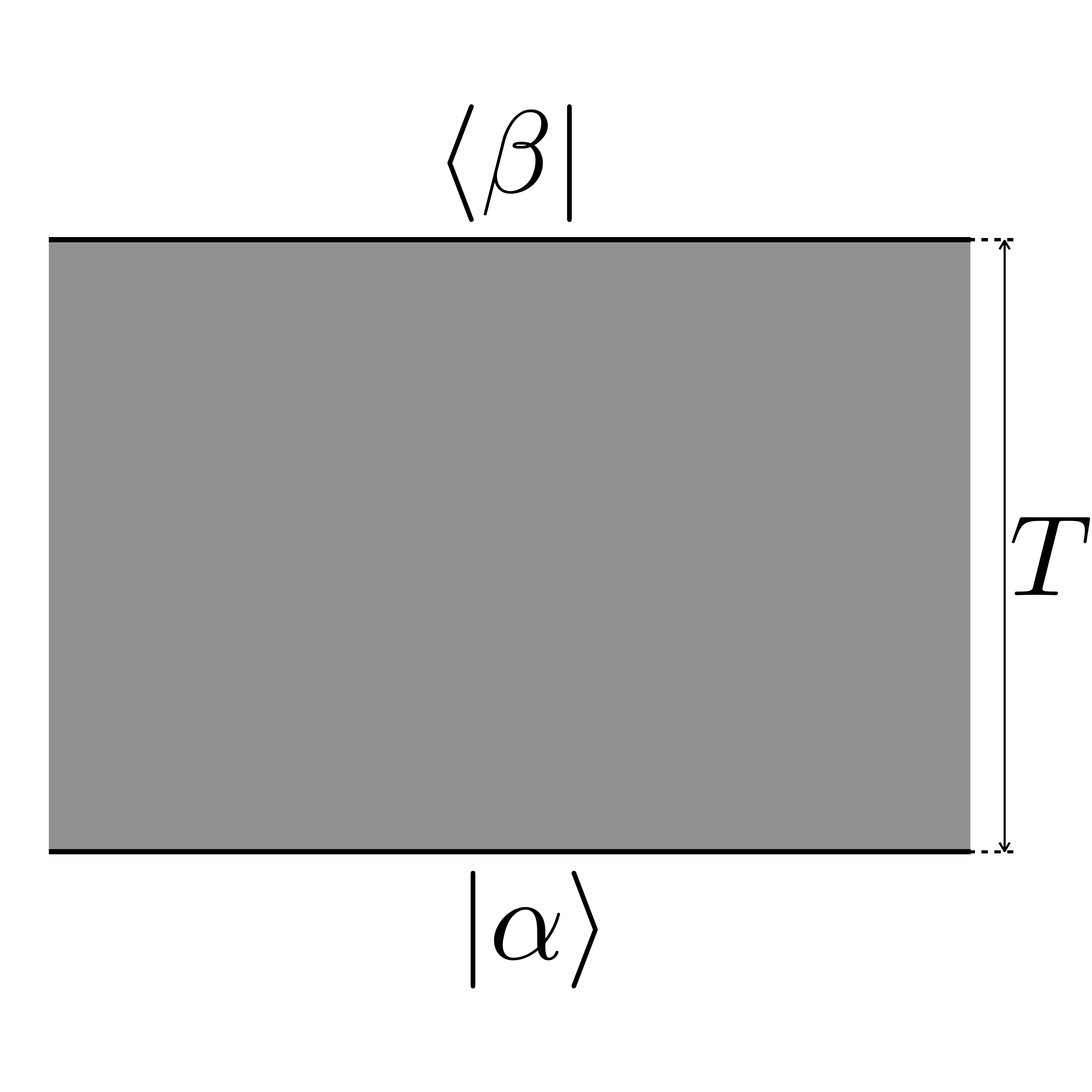}{20}
\end{align}
Here, $\ket{\alpha}$ and $\ket{\beta}$ are some pure states in the QFT, $\hat{\phi}(x)$ is the field operator, ${\phi}(\tau,x)$ is the field configuration, $H[\hat{\phi}(x)]$ is the Hamiltonian and $S[{\phi}(\tau,x)]$ is the (Euclidean) action. Therefore, we can regard a path integral over a manifold with a free boundary as a pure state. Here we use dashed lines to denote the free boundary. For example, 
\begin{align}
    e^{-T H[\hat{\phi}(x)]}\ket{\alpha} = \imineq{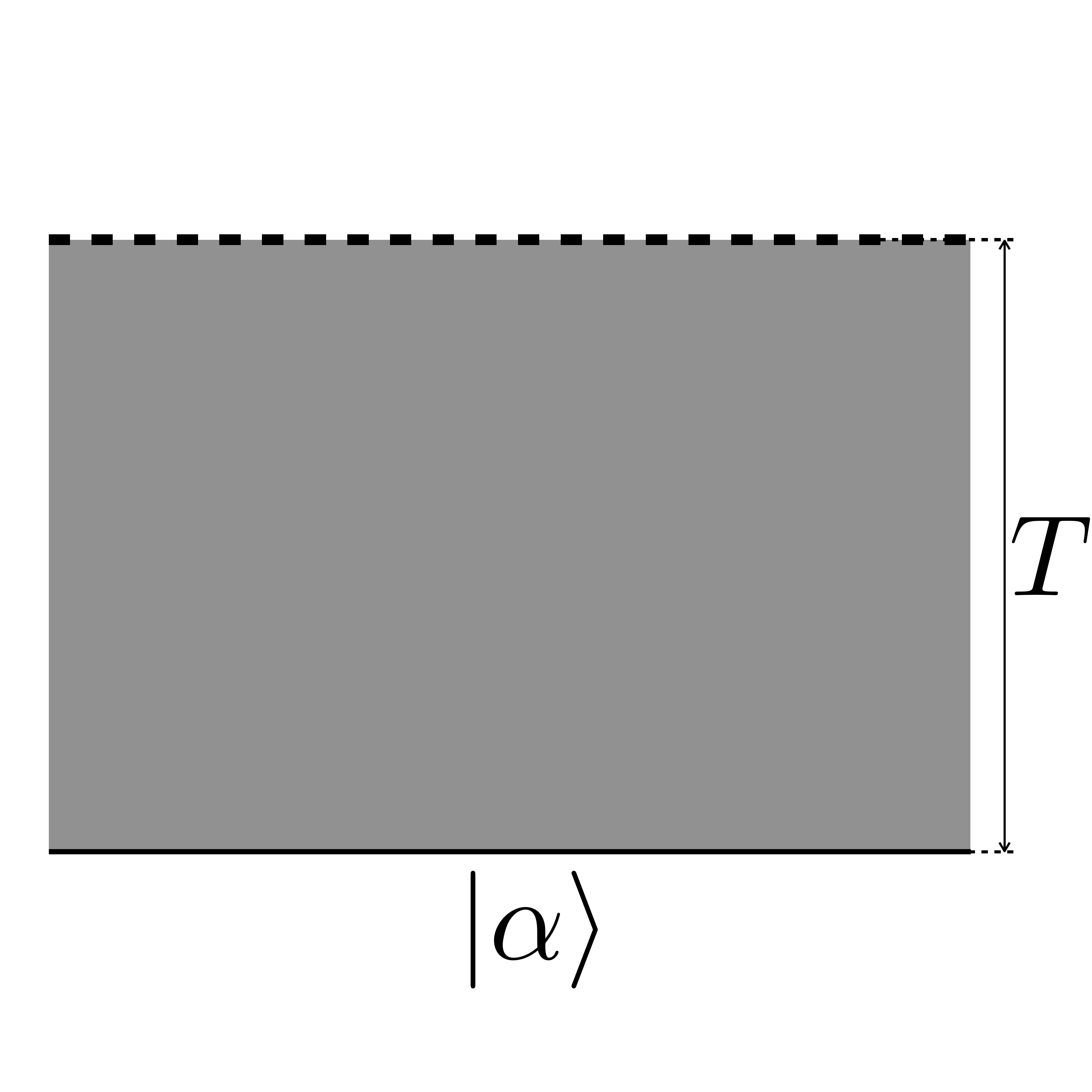}{20}
\end{align}

Now, let us consider the following two states given by path integral\footnote{
Notice that in the pictures below we just write specific  examples. Though we took  $\ket{\vv}$ to be 
a direct product state just for an illustration, we can consider other non-direct product states like  $\ket{\psi}$.} 

\begin{align}
    \ket{\psi} = \imineq{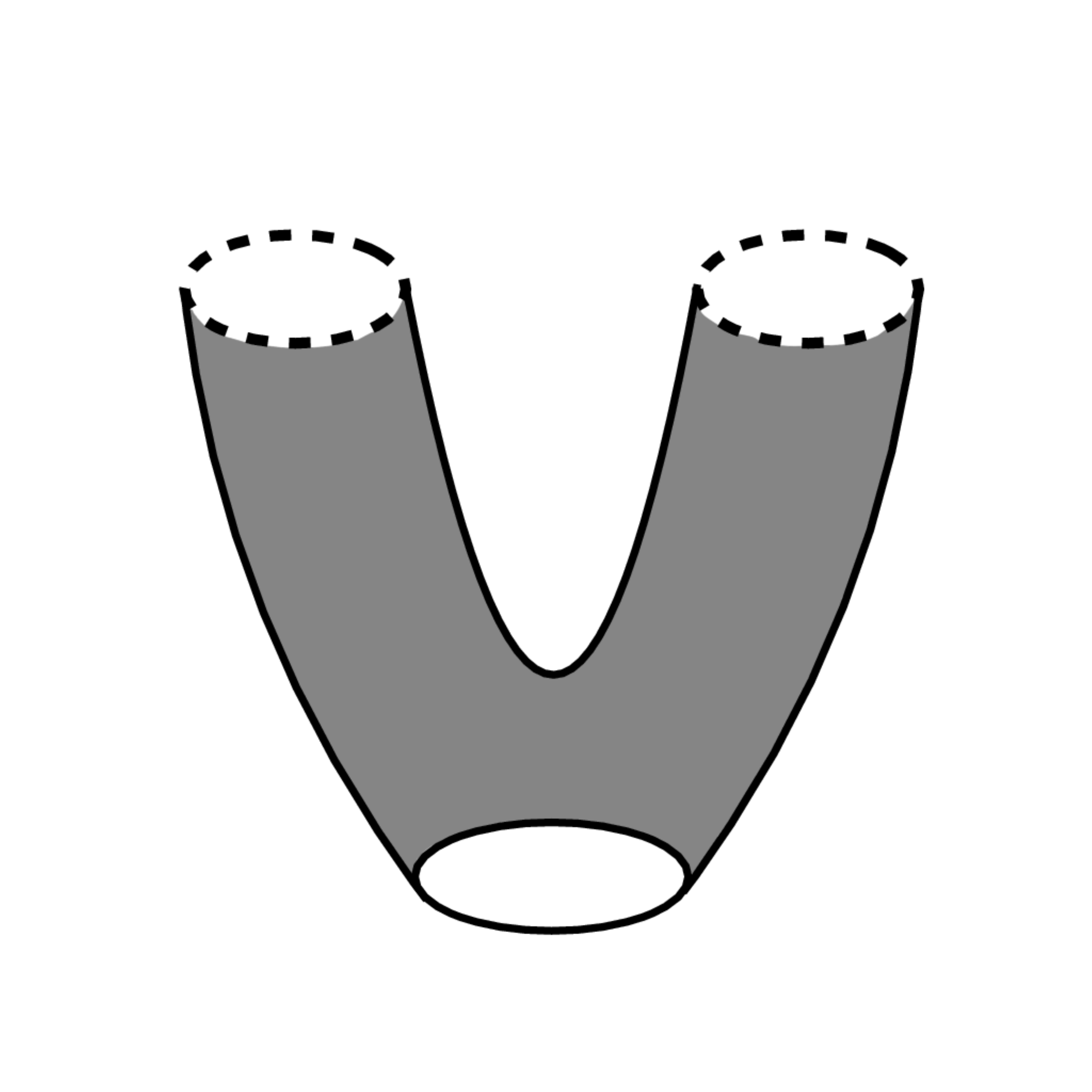}{20}\\
    \ket{\vv} = \imineq{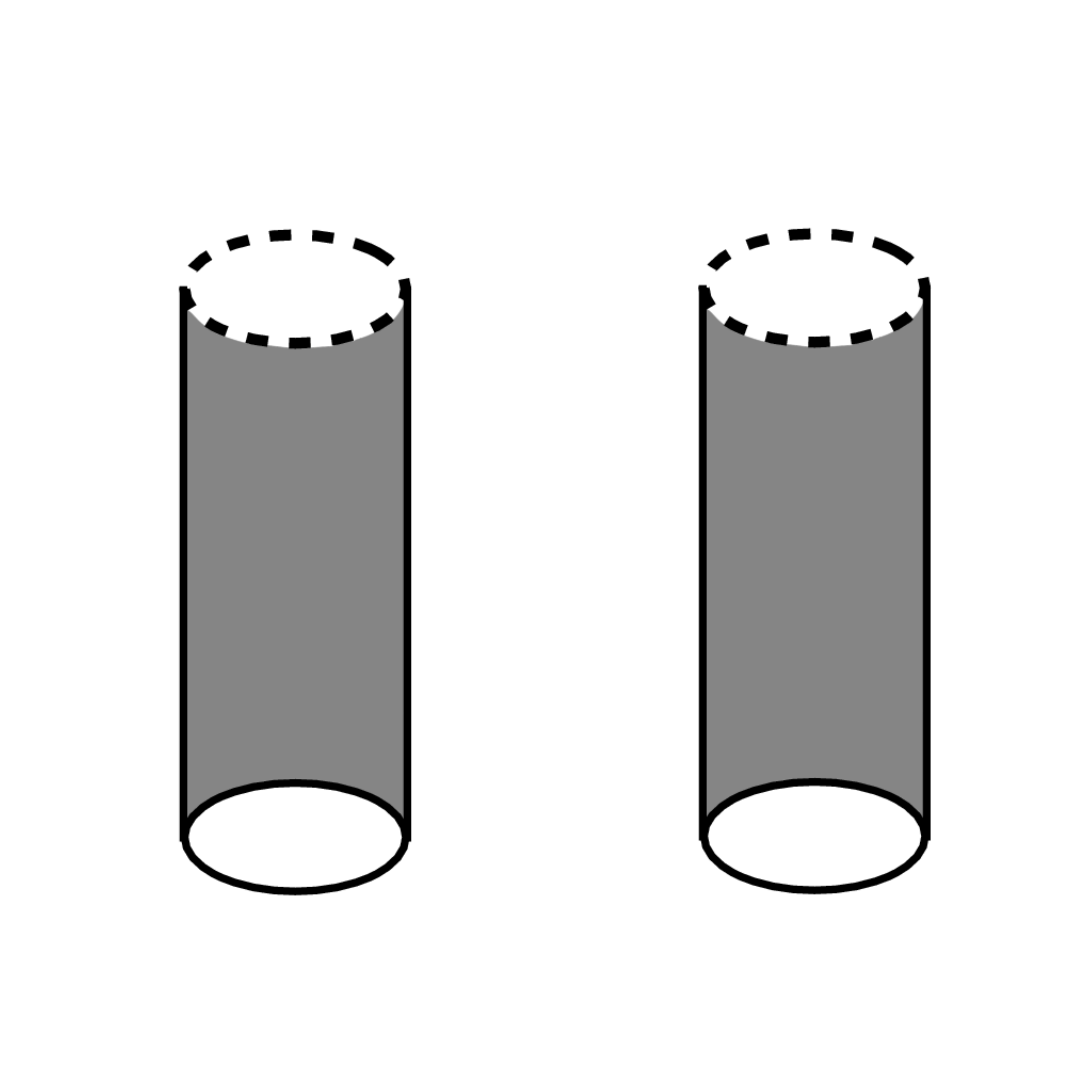}{20}
\end{align}
Then, 
\begin{align}
    |\psi\rangle\langle\vv| = \imineq{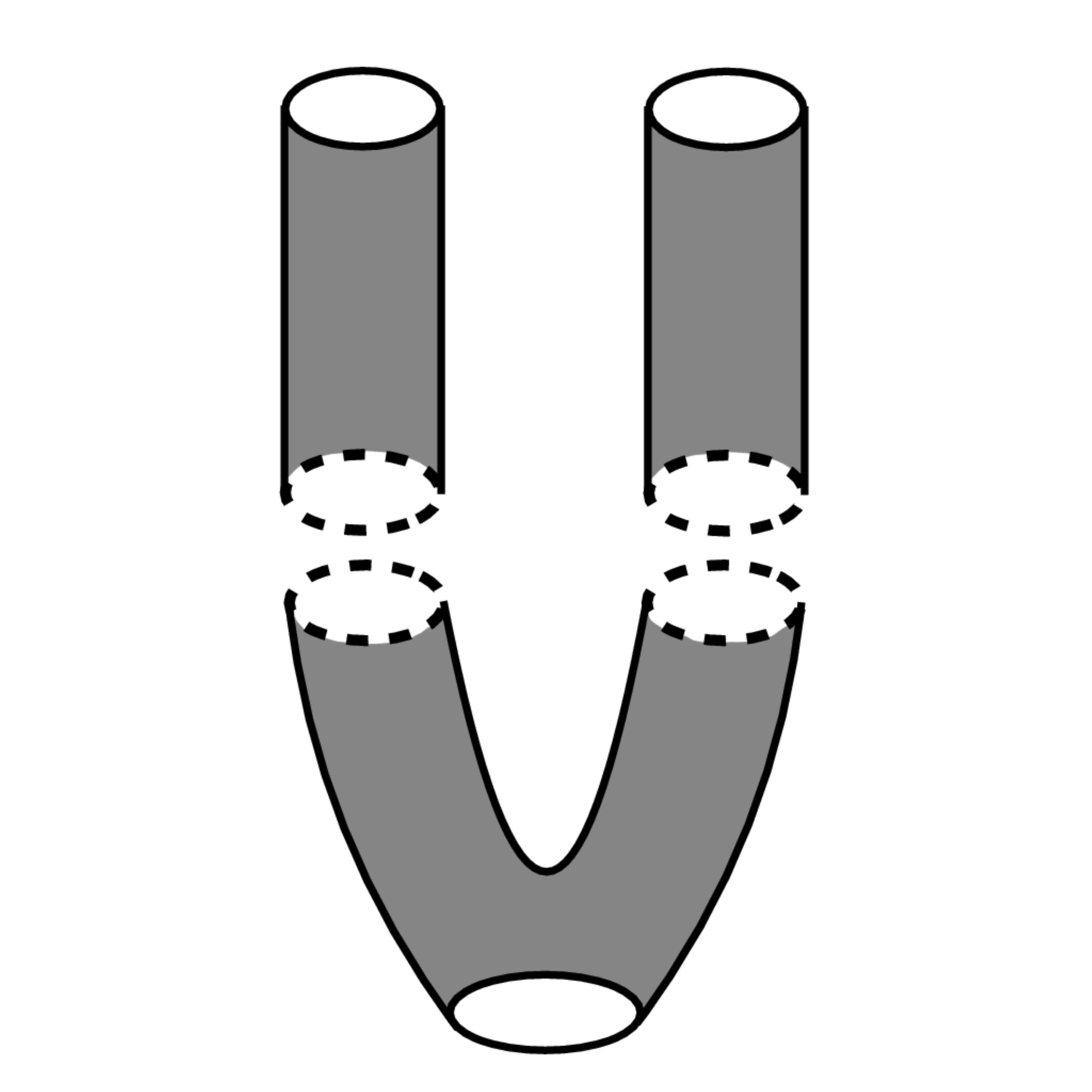}{20}
\end{align}
and 
\begin{align}
    \braket{\vv|\psi} = \imineq{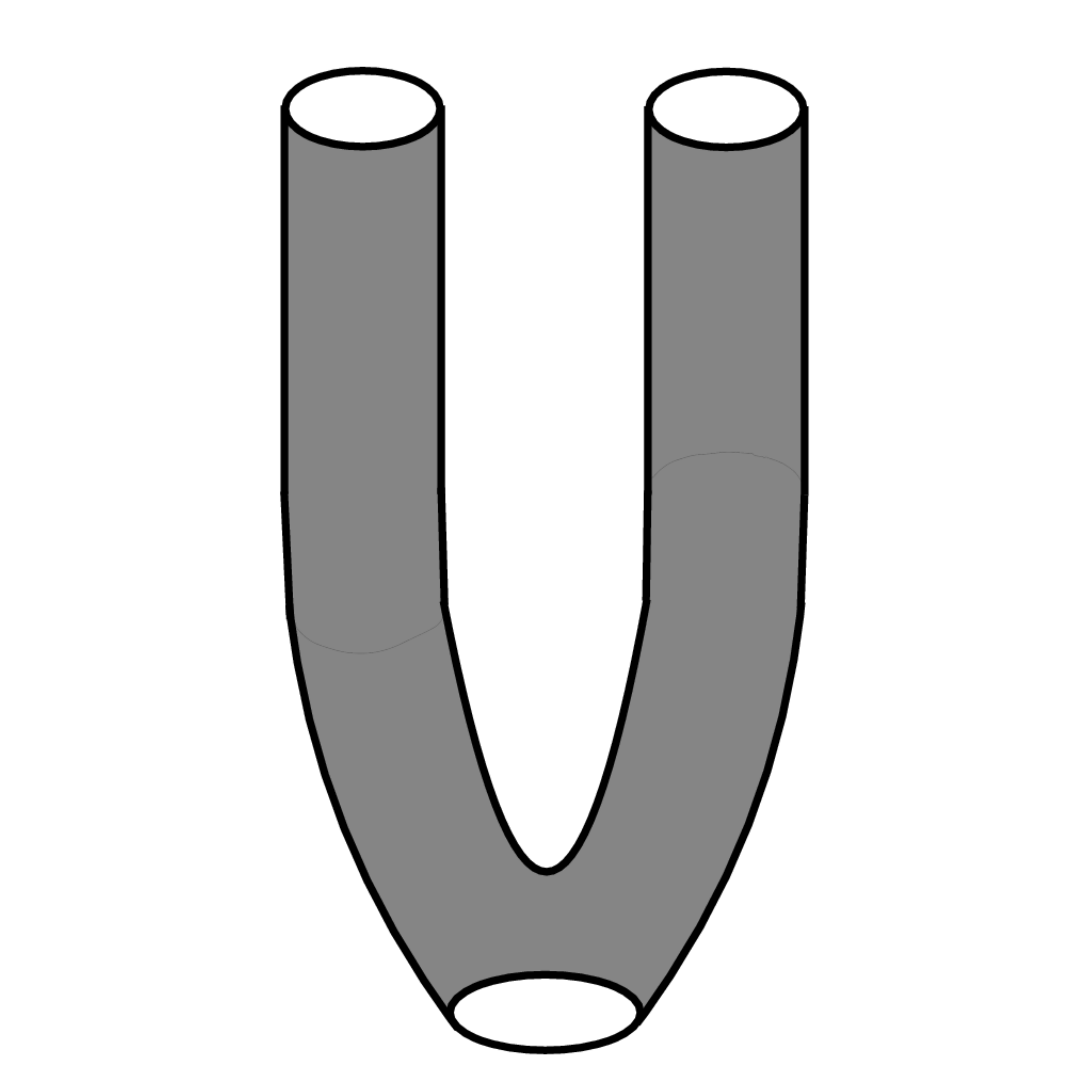}{20}
\end{align}
From these two path integrals we can get the transition matrix $\CT^{\psi|\vv}$. We can divide the time slice into two parts $A$ and $B$ and accordingly factorize the total Hilbert space as $\CH = \CH_A \otimes \CH_B$. Here, we take $A$ as the left-hand part and $B$ as the right-hand part. Then, we have, 
\begin{align}
    \Tr_B\left(|\psi\rangle\langle\vv|\right) = \imineq{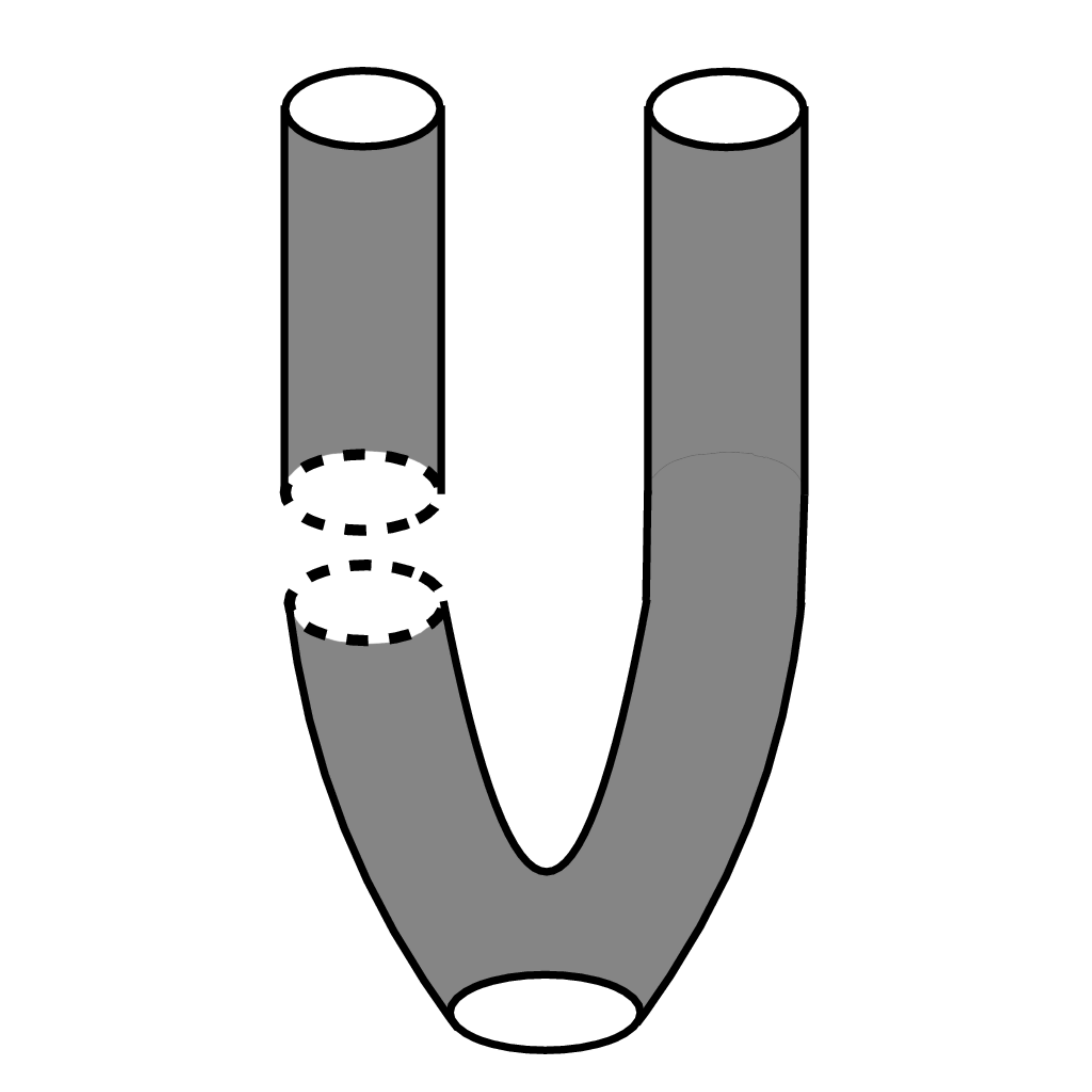}{20}
\end{align}
and the $n$-th replica turns out to be 
\begin{align}
    \Tr_A\left(\Tr_B\left(|\psi\rangle\langle\vv|\right)\right)^n = \imineq{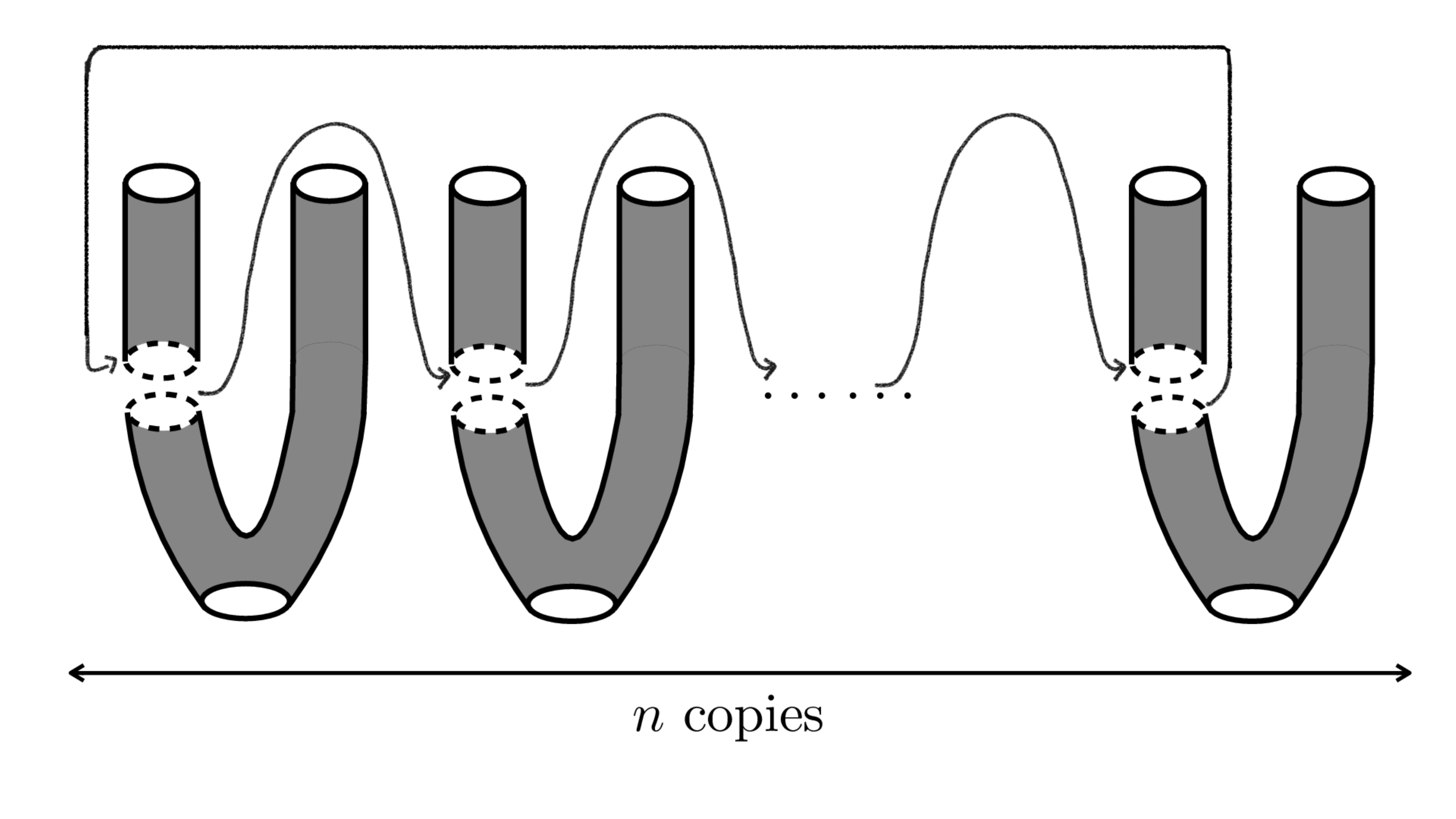}{30}
\end{align}
Let us denote the manifold corresponding to $\braket{\vv|\psi}$ as $\Sigma_1$ and the one corresponding to $\Tr_A\left(\Tr_B\left(|\psi\rangle\langle\vv|\right)\right)^n$ as $\Sigma_n$. Also let us use $Z_\CM$ to denote the path integral over the manifold $\CM$. Then, the $n$-th pseudo R\'{e}nyi entropy can be computed as 
\begin{align}
    S^{(n)}(\CT^{\psi|\vv}_A) = \frac{1}{1-n} \log\left(\frac{Z_{\Sigma_n}}{(Z_{\Sigma_1})^n}\right),
\label{qftsnf}
\end{align}
and the entanglement pseudo entropy can be obtained by taking the $n\rightarrow1$ limit.

It is useful to note that if we consider an Euclidean quantum field theory whose action is real valued and add a time-dependent source which is also real valued,  we expect the partition function is positive.
Therefore, in this case, $S^{(n)}(\CT^{\psi|\vv}_A)$ is real for any $n$. In this paper, especially via connection to the AdS/CFT, we have in mind this class of Euclidean field theory.

In section \ref{sec:rfreecft}, as one of the simplest examples of calculations of pseudo R\'enyi entropy in quantum field theories, we will present explicit calculations of pseudo entropy in a two dimensional conformal field theory (CFT), described by a massless free scalar field $\phi$. We will define $|\psi\lb$ and $|\vv\lb$ as excited states by acting a primary operator ${\cal O}(x,\tau)$ on the vacuum with different points $(x_1,\tau_1)$ and $(x_2,\tau_2)$ such that $|\psi\lb\propto \mathcal{O}(w_1,\bar{w}_1)\ket{0}$ and 
$|\vv\lb\propto \mathcal{O}(w_2,\bar{w}_2)\ket{0}$, where $w=x+i\tau$. The reduced transition matrix looks like
\begin{equation}
  \mathcal{T}^{\psi|\vv}_A=\mathcal{N}\cdot\mbox{Tr}_B[\mathcal{O}(w_1,\bar{w}_1)\ket{0}\bra{0}\mathcal{O}^{\dagger}(w_2,\bar{w}_2)],
\end{equation}
where $\mathcal{N}$ is a normalization factor to secure the unit norm.

\begin{figure}[H]
\centering
\includegraphics[width=10cm]{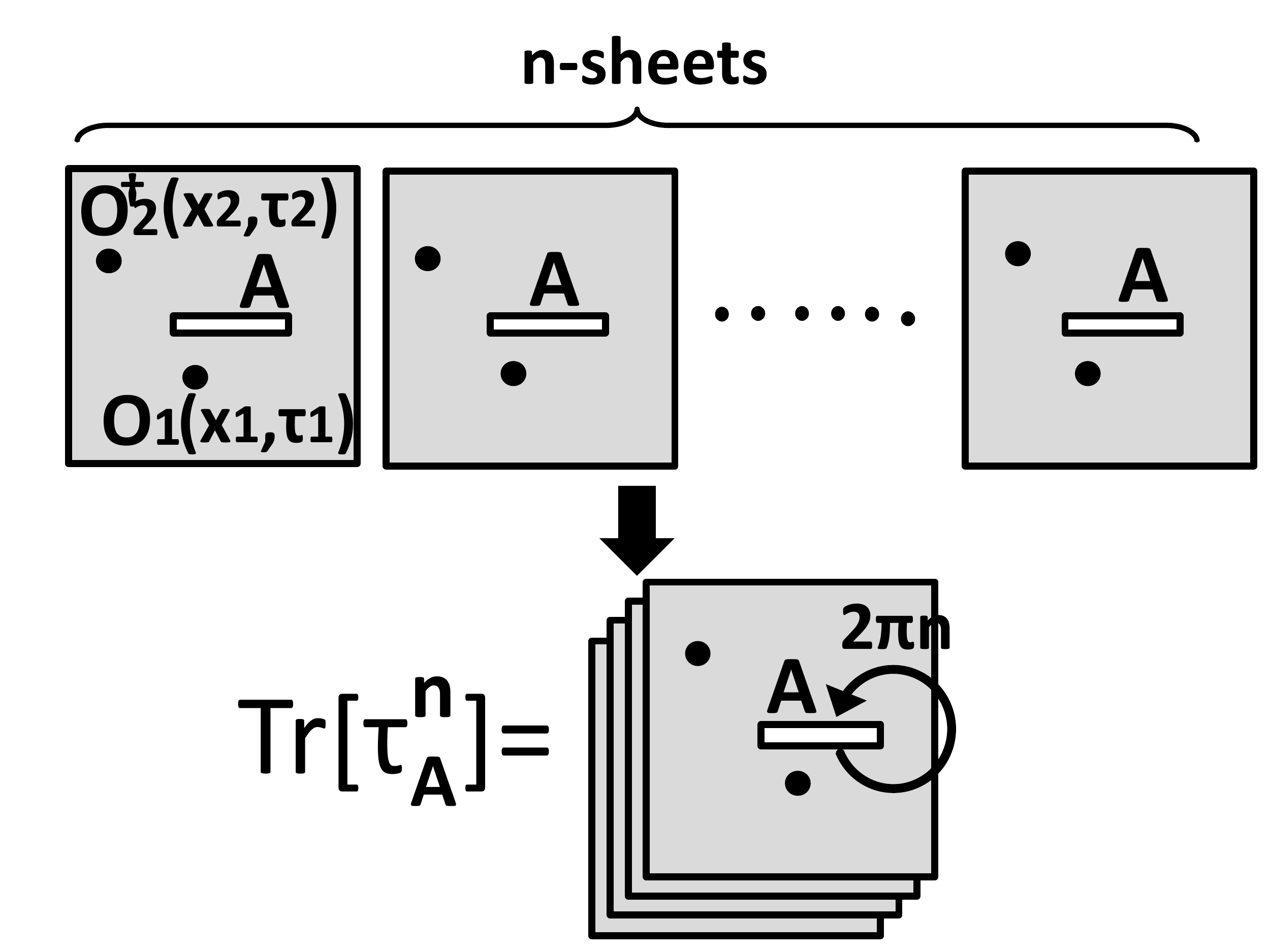}
\caption{The replica method calculation of pseudo entropy in two dimensional CFTs
for locally excited states.}\label{fig:rep2d}
\end{figure}

We are interested in the difference between the pseudo R\'{e}nyi entropy and the R\'{e}nyi entropy of the ground state. 
\begin{equation}
  \Delta S^{(n)}_A=S^{(n)}(\mathcal{T}^{\psi|\vp}_A)-S^{(n)}(\rho^{(0)}_A),  \label{difpe}
\end{equation}
where $\rho^{(0)}_A=\mbox{Tr}_B\ket{0}\bra{0}$, i.e. the reduced density matrix for the CFT vacuum.

The trace Tr$ (\mathcal{T}^{\psi|\vv}_A)^n$
 can be obtained from the path-integral by gluing $n$-sheets with two operators  $\mathcal{O}$ and $\mathcal{O}^{\dagger}$ inserted as shown in figure \ref{fig:rep2d}. We denote this replicated space as $\Sigma_n$. As in \cite{Nozaki,HNTW} for the ordinary R\'{e}nyi entropy via the replica method, we can calculate the difference (\ref{difpe}) from the $2n$ point function on $\Sigma_n$ as follows
\ba
\Delta S^{(n)}_A=\frac{1}{1-n}\left[\log\frac{\la \mathcal{O}^\dagger(w_1,\bar{w}_1)\mathcal{O}(w_2,\bar{w}_2)
\ddd \mathcal{O}(w_{2n},\bar{w}_{2n})\lb_{\Sigma_n}}{\left(\la \mathcal{O}^\dagger(w_1,\bar{w}_1)\mathcal{O}(w_2,\bar{w}_2)\lb_{\Sigma_1}\right)^n}\right].\label{localcpr}
\ea
We take $w_i=x_i+i\tau_i$ to be the location of the operator insertion in each sheet.
By applying the conformal map  (we chose the subsystem $A$ to be an interval $[x_l,x_r]$),
\ba
\frac{w-x_l}{w-x_r}=z^n,  \label{xxxcmap}
\ea
we can relate the above $2n$ point function on $\Sigma_n$ to that on  a standard complex plane.
We will evaluate  $\Delta S^{(n)}_A$ explicitly in section \ref{sec:rfreecft} for $n=2,3$. We will also present holographic pseudo entropy for such locally excited states in section \ref{sec:hpelocp}.

\subsection{Holographic Pseudo Entropy} 

As a special class of QFTs, holographic CFTs admit classical gravity duals via the AdS/CFT correspondence \cite{Maldacena:1997re}. 
This allows us to compute the pseudo entropy holographically. 
Consider a computation of the inner product $\la \vv|\psi\lb$ in the path-integral formalism of CFT.
This quantity is identical to the partition function obtained by gluing along a time slice two Euclidean path-integrals, each of which produces $|\psi\lb$ and $|\vv\lb$, respectively (depicted as the boundary of figure \ref{fig:EAdSPE}). If we insert operators $O_1(x_1), O_2(x_2), \ddd $ on this time slice in this total path-integral, then we have (after dividing the inner product)   
\ba
\frac{\la \vv|O_1(x_1)O_2(x_2)\ddd|\psi\lb}{\la \vv|\psi\lb}
=\mbox{Tr}[\CT^{\psi|\vv}\cdot O_1(x_1)O_2(x_2)\ddd].
\ea  

The AdS/CFT correspondence relates a $d+1$ dimensional gravity on an AdS$_{d+1}$ to a $d$ dimensional holographic CFT. The gravity dual of this inner product is the classical partition function of an asymptotically AdS Euclidean space,  which is obtained from the standard rule of the AdS/CFT. 
Note that since the bra and ket state on the time slice is different, there is neither (Euclidean) time-translational invariance nor the time reversal symmetry on the time slice. In this sense, we have the genuine Euclidean time-dependent geometry as the classical gravity dual. 

Now we take a subsystem $A$ on the time slice in the holographic CFT and consider its (von Neumann) pseudo entropy $S(\CT^{\psi|\vv}_A)$. We argue that this is simply given by generalizing the holographic entanglement entropy \cite{RT,HRT} to the gravity background with Euclidean time-dependence (refer to figure \ref{fig:EAdSPE}):
\ba
S(\CT^{\psi|\vv}_A)=\mbox{Min}_{\Gamma_A}\left[\frac{\mbox{A}(\Gamma_A)}{4G_N}\right],
\label{eq:HEPE}
\ea
where $\Gamma_A$ satisfies $\de \Gamma_A=\de A$ and is homologous to $A$. 

Since it is straightforward to derive this formula as the authors of  \cite{Lewkowycz:2013nqa}
did in the case of holographic entanglement entropy, here we give only a brief sketch of this.
In the replica calculation explained in the previous subsection, we can obtain the $n$-th pseudo R\'{e}nyi entropy by computing 
$\Tr_A\left(\Tr_B\left(|\psi\rangle\langle\vv|\right)\right)^n$ as a partition function on $n$-replicated geometry $\Sigma_n$. In this geometry, the deficit angle $2\pi (1-n)$ is present around the
 boundary of $A$ (i.e. $\de A$). In the AdS/CFT description, this partition function on $\Sigma_n$ is equal to that of classical gravity on an asymptotically AdS space whose boundary coincides with the replicated space $\Sigma_n$. Naively the solution to Einstein equation with this boundary geometry is given by extending the deficit angle on $\de A$ towards the bulk AdS. We call this extended surface with the deficit angle as $\Gamma_A$. However, the true solution to Einstein equation should be smooth and this deficit angle 
surface in the bulk, which is singular, is not an appropriate solution.  Nevertheless, this naive prescription gives a correct von Neumann entropy (but not correct R\'enyi entropy). This is because the difference between the true solution and the singular solution is $O(n-1)$ when $n$ is very closed to $1$ and thus the 
values of the gravity action evaluated on these two solutions differ by $O\left((n-1)^2\right)$,
which does not contribute in the von Neumann entropy limit $n\to 1$.
 To see this, note that at $n=1$ the two solutions coincide with the standard AdS solution which satisfies the Einstein equation. 

The gravity action takes the familiar form:
\ba
I_G=-\frac{1}{16\pi G_N}\int d^{d+1}x\s{g}\left(R-2\Lambda\right)+\ddd,
\ea
where the omission $\ddd$ includes the boundary term as well as matter field contributions which do not 
contribute to the entropy. In the presence of the deficit angle on $\Gamma_A$, the Ricci scalar behaves as 
$R=4\pi (1-n)\delta_{\Gamma_A}(x)$, where $\delta_{\Gamma_A}(x)$ is the delta-function which localizes on 
$\Gamma_A$. Therefore we can evaluate the $O(n-1)$ term of the gravity action 
\ba
I_G=\frac{n-1}{4G_N}A(\Gamma_A)+\ddd,
\ea
where the terms $\ddd$ are all proportional to $n$, which does not contribute to the entropy.
The gravity partition function is expressed as $Z_{G}=e^{-I_G}$ in terms of the on-shell gravity action $I_G$.
Finally the holographic pseudo entropy is computed as $\frac{A(\Gamma_A)}{4G_N}$ by taking $n\to 1$ limit 
of the formula (\ref{qftsnf}). Solving Einstein equation corresponds to minimizing the area with respect to the change of $\Gamma_A$. These arguments derive the holographic formula (\ref{eq:HEPE}).

Interestingly, this holographic formula tells us that the pseudo entropy computed for 
a classical gravity dual is always real and non-negative, assuming the bulk metric is real valued.
This is because the only source for the holographic entanglement entropy by the replica trick 
is the Einstein Hilbert term which gives rise to the area term as in  \cite{Lewkowycz:2013nqa},
as long as we consider Einstein gravity coupled to various matter fields. 
In the field theory side we can understand this as follows. Typically the Euclidean gravity with a real valued metric is dual to an Euclidean CFT with real valued external sources. In such a theory, the partition functions are positive and the same is true for Tr$ (\mathcal{T}^{\psi|\vv}_A)^n$. Also we expect the analyticity about the replica number $n$, which can be confirmed in explicit results from CFT calculations in section \ref{sec:hpelocp}. Therefore, we can compute the pseudo entropy without worrying about the choice of branch of the log function.

Notice also that the basic properties of (\ref{propa}),   (\ref{propb}), and  (\ref{propc}) are obvious in the holographic formula. Moreover, it is clear from the geometric property of asymptotically AdS backgrounds that the holographic pseudo entropy obeys the area law as in the standard entanglement entropy \cite{BKLS,Sr}. 
We will give more details of holographic pseudo entropy with several examples in section \ref{secHPE}. 

We would like to mention that if we Wick rotate a generic Lorentzian time-dependent solution of Einstein gravity into an Euclidean time-dependent solution, we encounter a complex valued metric. In this case we expect the minimal area becomes complex valued. Though, in the present paper we will not discuss such cases, leaving it as a future problem, we would like to note that this looks consistent with the fact that the pseudo entropy is generically complex valued. 

It is intriguing to compare this formula with the covariant holographic entanglement entropy \cite{HRT}, which calculates the entanglement entropy under time evolutions as the area of an extremal surface in a Lorentzian time-dependent asymptotically AdS spacetime. Our pseudo entropy formula in Euclidean asymptotically AdS spaces, may be regarded as a Wick rotation of the covariant holographic entanglement entropy from the gravity dual viewpoint. However, note that the former computes the von Neumann entropy for a standard quantum state, while the latter computes the von Neumann entropy for a transition matrix.

It is also useful to note that we can express the holographic pseudo entropy by using the area operator
$\hat{\cal{A}}$ introduced in  \cite{Faulkner:2013ana} as follows:
\ba
S(\CT^{\psi|\vv}_A)=\frac{\la \vv|\frac{\hat{\cal{A}}}{4G_N}|\psi\lb}{\la \vv|\psi\lb},
\ea
i.e. the weak value of the area operator. We assume that  $|\psi\lb$ and $|\vv\lb$  are in the same subspace of the 
low energy Hilbert space. 
Notice that if $|\psi\lb$ and $|\vv\lb$ are descendants of
 two different primaries, which are orthogonal to each other, then the $\la \vv|\hat{\cal{A}}|\psi\lb$ does vanish. 
Using this fact, we will confirm the expected linearity of the area operator 
in section \ref{subsec:linearity}.

\section{Pseudo Entropy in Qubit Systems}\label{sec:qubitpr}
In this section, we study pseudo entropy in qubit systems in detail.

\subsection{Classification in 2-Qubit Systems}\label{cltqsz}

An arbitrary pure state in a 2-qubit system can be written as (up to an overall normalization)
\begin{align}
    \ket{\psi} = \ket{00} + a \ket{11},~~(0\leq a \leq 1) \label{eq:psipar1}
\end{align}
by choosing the basis appropriately. Here, we use the notation $\ket{ij}\equiv|i\rangle_{A}\otimes|j\rangle_{B}$. In such a basis, another arbitrary bipartite state can be written as (up to an overall normalization)
\begin{align}
    &\ket{\vv} = \ket{00} + be^{-i\theta}\ket{11} + ce^{-i\xi}\ket{01} + de^{-i\eta}\ket{10},\label{eq:phipar1}\\
    &(b,c,d\geq0,~-\pi\leq\theta,\xi,\eta\leq\pi). \nonumber
\end{align}
The transition matrix between the two states is reduced to 
\begin{align}
    \CT^{\psi|\vv}_A &= \frac{1}{1+abe^{i\theta}}\left(|0\rangle\langle0| + ac e^{i\xi}|1\rangle\langle0|+ d e^{i\eta}|0\rangle\langle1|+ ab e^{i\theta}|1\rangle\langle1| \right) \nonumber\\
    &\equiv 
    \begin{pmatrix}
    {\frac{1}{1+abe^{i\theta}}} & {\frac{d e^{i\eta}}{1+abe^{i\theta}}}\\ 
    {\frac{ac e^{i\xi}}{1+abe^{i\theta}}} & {\frac{abe^{i\theta}}{1+abe^{i\theta}}}
    \end{pmatrix} \label{eq:TA}
\end{align}

It is easy to see 
\begin{align}
    \CT^{\psi|\vv}\in\mathscr{A}~\Longleftrightarrow~0\leq\det{(\CT^{\psi|\vv}_A)}.
    \label{eq:classA}
\end{align}

It is also not hard to find that the following three statements are equivalent:
\begin{align}
    &S^{(n)}(\CT^{\psi|\vv}_A){\rm ~takes ~nonnegative ~real ~values ~for} ~n>0{\rm , ~i.e.~}  \CT^{\psi|\vv}\in\mathscr{B}. \nonumber \\
    \Longleftrightarrow~& {\rm Both}~\lambda_1~{\rm and}~\lambda_2~ {\rm are~real~and~nonnegative,~i.e.~} \CT^{\psi|\vv}\in\mathscr{C}.
    \label{eq:nnreal} \\
    \Longleftrightarrow~& 0\leq\det(\CT^{\psi|\vv}_A)\leq1/4. \nonumber
\end{align}

Let us then figure out when $\CT^{\psi|\vv}$ gives positive semi-definite Hermitian $\CT^{\psi|\vv}_A$, i.e. $\CT^{\psi|\vv}\in\mathscr{D}$. For two states $\ket{\psi}$ and $\ket{\vv}$ where
\begin{align}
    \ket{\psi} = \ket{00} + a \ket{11},~~(0\leq a \leq 1),
\end{align}
$\CT^{\psi|\vv}_A$ is Hermitian if and only if 
\begin{align}
    &\exists \begin{cases}
                b\in \mathbb{R}, b\neq -1/a \\
                0\leq c \\
                -\pi\leq \xi\leq \pi
            \end{cases}
     \nonumber \\
    &{\rm s.t.} \nonumber\\
    \ket{\vv} = \ket{00} &+ b \ket{11} + ce^{-i\xi}\ket{01} + ace^{i\xi}\ket{10}.
    \label{eq:singleH1}
\end{align}

In this case, 
\begin{align}
    \CT^{\psi|\vv}_A {\rm~is~positive~semi\mathchar`-definite} \Longleftrightarrow~a=0~ {\rm or} ~b\geq ac^2.
    \label{eq:singleH2}
\end{align}

Note that $\CT^{\psi|\vv}_A$ and $\CT^{\psi|\vv}_B$ are not necessarily Hermitian at the same time. For two states $\ket{\psi}$ and $\ket{\vv}$ where
    \begin{align}
        \ket{\psi} = \ket{00} + a \ket{11},~~(0\leq a \leq 1),
    \end{align}
    both $\CT^{\psi|\vv}_A$ and $\CT^{\psi|\vv}_B$ are Hermitian if and only if it is either of the following two cases.\\
    Case I: $\exists b\in \mathbb{R}, b\neq -1/a$, s.t.
    \begin{align}
        \ket{\vv} = \ket{00}+b\ket{11}.
    \end{align}
    Case II: $a=1$ and 
   \begin{align}
        &\exists \begin{cases}
                    b\in \mathbb{R}, b\neq -1/a \\
                    0\leq c \\
                    -\pi\leq \xi\leq \pi
                \end{cases}
         \nonumber \\
        &{\rm s.t.} \nonumber\\
        \ket{\vv} = \ket{00} &+ b \ket{11} + ce^{-i\xi}\ket{01} + ce^{i\xi}\ket{10}.
    \end{align}

    In case I, 
    \begin{align}
        {\rm Both}~\CT^{\psi|\vv}_A~ {\rm and}~ \CT^{\psi|\vv}_B~ {\rm are~ positive~semi\mathchar`-definite}~\Longleftrightarrow~ a=0~ {\rm or}~b\geq 0.
    \end{align}
    In case II, 
    \begin{align}
        {\rm Both}~\CT^{\psi|\vv}_A~ {\rm and}~ \CT^{\psi|\vv}_B~ {\rm are~ positive~semi\mathchar`-definite}~\Longleftrightarrow~ b\geq c^2.
    \end{align}

\paragraph{Counting the Degrees of Freedom of 2-Qubit Transition Matrix}~\par
Choosing the basis like (\ref{eq:psipar1}) and (\ref{eq:phipar1}), it can be observed that there are 7 independent variables\footnote{We are counting the number of independent real variables which take continuous value.} to characterize the transition matrix $\CT^{\psi|\vv}$. However, as we can see in (\ref{eq:TA}), only 6 independent variables remain in $\CT^{\psi|\vv}_A$. Moreover, since the pseudo entropy can be computed from the two eigenvalues, there are only 2 independent variables that are relevant. 

Considering transition matrices which give nonnegative real $S^{(n)}(\CT^{\psi|\vv}_A)$ for $n>0$, as we can see from (\ref{eq:nnreal}), there are 6 independent variables for $\CT^{\psi|\vv}$ and 5 independent variables for $\CT^{\psi|\vv}_A$. 

If we focus on transition matrices which give positive semi-definite Hermitian $\CT^{\psi|\vv}_A$, according to (\ref{eq:singleH1}) and (\ref{eq:singleH2}), there are 4 independent variables for $\CT^{\psi|\vv}$ and 3 independent variables for $\CT^{\psi|\vv}_A$.

Moreover, when both $\CT^{\psi|\vv}_A$ and $\CT^{\psi|\vv}_B$ are positive semi-definite Hermitian, the transition matrices can be classified in two cases. In case I, there are 2 independent variables for $\CT^{\psi|\vv}$ and 1 independent variables for $\CT^{\psi|\vv}_A$. In case II, there are 3 independent variables for $\CT^{\psi|\vv}$ and 3 independent variables for $\CT^{\psi|\vv}_A$.

\begin{figure}[H]
    \centering
    \includegraphics[width=12cm]{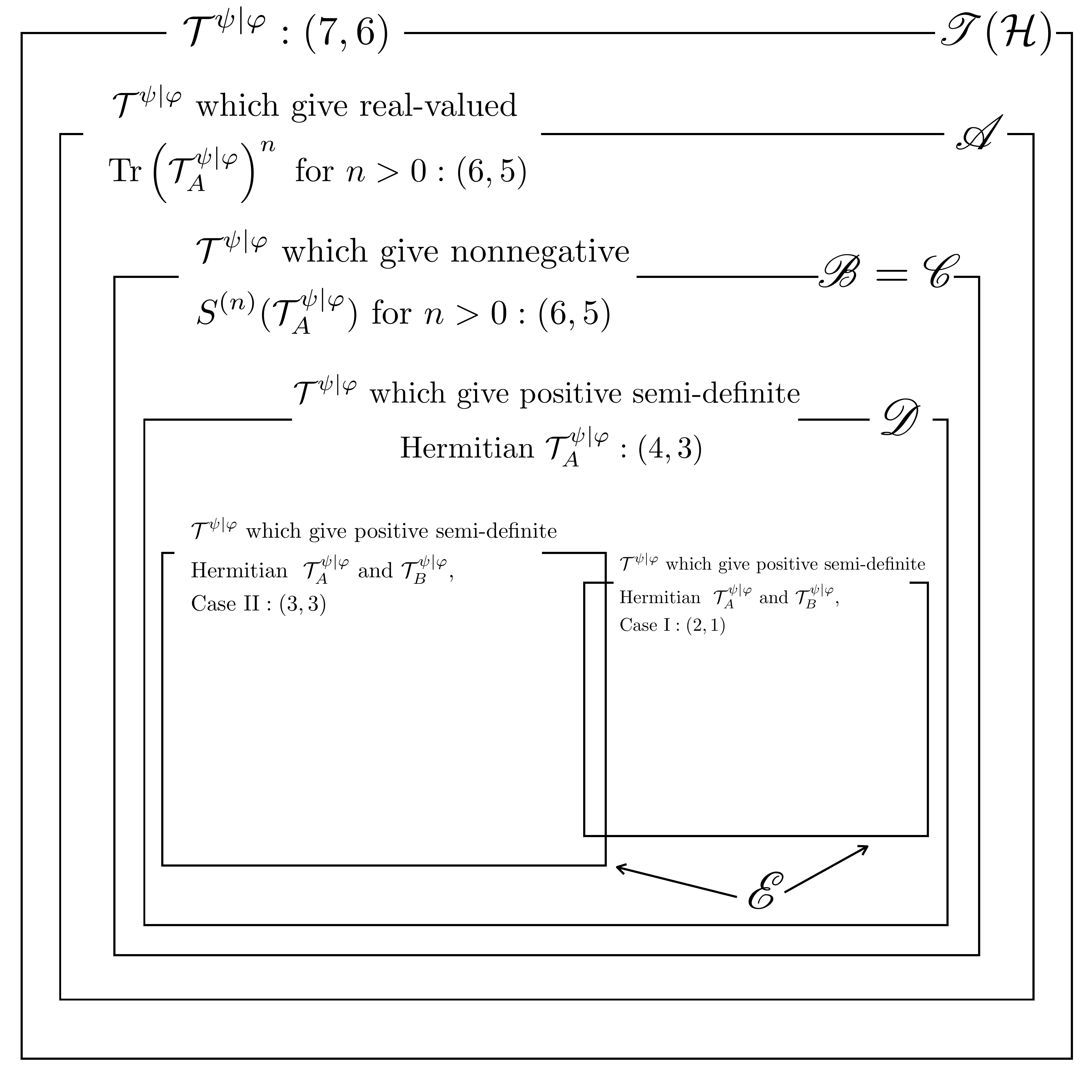}
    \caption{The Venn diagram of the classification of 2-qubit transition matrices. For each class of states, we use $(m,n)$ to represent that there are $m$ independent variables for $\CT^{\psi|\vv}$ and $n$ independent variables for $\CT^{\psi|\vv}_A$.}
    \label{fig:venn2qubit}
\end{figure}

Figure \ref{fig:venn2qubit} shows the Venn diagram of the classification of 2-qubit transition matrices we have discussed above. 

Let us present an exotic example as follows, which belongs to the class $\mathscr{A}$ but not to 
$\mathscr{B}$.
\begin{example}[]~\par
 For this we choose 
\ba
&& |\psi\lb=\frac{1}{\s{2}}(|00\lb+e^{i\theta}|11\lb),\no
&& |\vv\lb=\frac{1}{\s{2}}(|00\lb+|11\lb).\label{twoqttt1}
\ea
We obtain
\ba
\CT^{\psi|\vv}_A=
\frac{1}{1+e^{i\theta}}\left(|0\lb\la 0|+e^{i\theta}|1\lb\la 1|\right),
\ea
where the eigenvalues are complex valued and each of them is complex conjugate to the other. 
We can evaluate the pseudo R\'{e}nyi entropy
\ba
S^{(n)}(\CT^{\psi|\vv}_A)=\frac{1}{1-n}\log\left[\frac{\cos\frac{n\theta}{2}}{2^{n-1}\cos^n\left(\frac{\theta}{2}\right)}\right]. \label{twoqttt}
\ea
In particular for $n=2$, we find
\ba
S^{(2)}(\CT^{\psi|\vv}_A)=\log\left(\frac{1+\cos\theta}{\cos\theta}\right),\label{twoqtt}
\ea
which is larger than $\log 2$ for $0< |\theta|< \frac{\pi}{2}$ and is complex valued for 
$\frac{\pi}{2}< |\theta|< \pi$.

The pseudo von Neumann entropy reads
\ba
S(\CT^{\psi|\vv}_A)=\log\left[2\cos\frac{\theta}{2}\right]+\frac{\theta}{2}\cdot 
\tan\frac{\theta}{2}.
\ea
Also note that, in this case, the two states $\ket{\psi}$ and $\ket{\vv}$ satisfy $S(\rho^{\psi}_A) = S(\rho^{\vv}_A)$ where $\rho_A^\psi = \Tr_B\left(|\psi\rangle\langle\psi|\right)$. We have
\begin{align}
    &S(\CT^{\psi|\vv}_A)-S(\rho^{\psi}_A) = S(\CT^{\psi|\vv}_A)-S(\rho^{\vv}_A) \nonumber\\
    =& \log\left[\cos\frac{\theta}{2}\right]+\frac{\theta}{2}\cdot \tan\frac{\theta}{2} \geq 0.
\end{align}
\end{example} 
However, note that the original periodicity of $\theta$ disappears.

\subsection{A Monotonicity of Pseudo Entropy in 2-Qubit Systems}

Consider two states $\ket{\psi}$ and $\ket{\vv}$ in a 2-qubit system $\CH = \CH_A\otimes\CH_B$ which are related by a local basis transformation: 
\begin{align}
    \ket{\psi} = (U \otimes V
    )\ket{\vv}
\end{align}
where $U$ and $V$ are unitary transformations on $\CH_A$ and $\CH_B$ respectively. A significant feature of 2-qubit systems in this setup is the following monotonicity under the unitary transformations:
\begin{align}
        S^{(n)}(\CT^{\psi|\vv}_A)\geq S^{(n)}(\Tr_B(|\psi\rangle\langle\psi|)) = S^{(n)}(\Tr_B(|\vv\rangle\langle\vv|)){~~\rm for~~}n>0 , 
        \label{eq:mono2qbit}
\end{align}
if all the eigenvalues of $\CT^{\psi|\vv}_A$ are real and nonnegative. 

\subsubsection{Proof}

    We present a proof by explicitly writing down the reduced transition matrix. We can always choose the basis such that 
    \begin{align}
    &\ket{\psi} = c\ket{0'}\ket{0''} + s\ket{1'}\ket{1''}\label{eq:ETpsi} \\
    &\ket{\vv} = c\ket{0}\ket{0} + s\ket{1}\ket{1} \label{eq:ETphi}
    \end{align}
    where $0\leq s\leq c \leq 1$, $c^2+s^2=1$ and,
    \begin{align}
    &\ket{0'}=U\ket{0},~\ket{1'}=U\ket{1}, \\
    &\ket{0''}=V\ket{0},~\ket{1''}=V\ket{1}.
    \end{align} 
    Then, 
    \begin{align}
        &|\psi\rangle\langle\vv| \nonumber\\=& ~~c^2 \left(U |0\rangle\langle0|\right) \otimes \left(V |0\rangle\langle0|\right) + cs \left(U |0\rangle\langle1|\right) \otimes \left(V |0\rangle\langle1|\right)  \nonumber\\
        &+ cs \left(U |1\rangle\langle0|\right) \otimes \left(V |1\rangle\langle0|\right) + s^2 \left(U |1\rangle\langle1|\right) \otimes \left(V |1\rangle\langle1|\right).
    \end{align}
    Let us introduce the following notation:
    \begin{align}
        M = M_{00}|0\rangle\langle0| + M_{01}|0\rangle\langle1| + M_{10}|1\rangle\langle0| + M_{11}|1\rangle\langle1| =
        \begin{pmatrix}
            M_{00} & M_{01} \\
            M_{10} & M_{11} \\
        \end{pmatrix}.
    \end{align}
    Tracing out the subsystem $B$, we have 
    \begin{align}
        \Tr_B\left(|\psi\rangle\langle\vv|\right) = \begin{pmatrix}
            c^2V_{00}U_{00} + cs V_{01}U_{01} & s^2V_{11}U_{01} + cs V_{10}U_{00} \\
            c^2V_{00}U_{10} + cs V_{01}U_{11} & s^2V_{11}U_{11} + cs V_{10}U_{10} \\
        \end{pmatrix}.
    \end{align}
    Thus we get the reduced transition matrix
    \begin{align}
        \CT^{\psi|\vv}_A = \frac{1}{\Delta} \begin{pmatrix}
            c^2V_{00}U_{00} + cs V_{01}U_{01} & s^2V_{11}U_{01} + cs V_{10}U_{00} \\
            c^2V_{00}U_{10} + cs V_{01}U_{11} & s^2V_{11}U_{11} + cs V_{10}U_{10} \\
        \end{pmatrix},
    \end{align}
    where 
    \begin{align}
        \Delta = \Tr\left(\CT^{\psi|\vv}_A\right) = c^2V_{00}U_{00} + cs V_{01}U_{01} + s^2V_{11}U_{11} + cs V_{10}U_{10}.
    \end{align}
    It is sufficient to consider the situation in which $U,V\in{\rm SU(2)}$. This allows us to have
    \begin{align}
        U=\begin{pmatrix}
            p & -q^* \\
            q & p^* \\
        \end{pmatrix},~
        V=\begin{pmatrix}
            r & -t^* \\
            t & r^* \\
        \end{pmatrix}
    \end{align}
    where $|p|^2+|q|^2 = |r|^2+|t|^2 = 1$. We then get
    \begin{align}
        \Delta &= c^2 pr + csq^*t^* + s^2p^*r^* + csqt \nonumber\\
        &= {\rm Re}(pr) + 2cs {\rm Re}(qt) + i(c^2-s^2){\rm Im}(pr)
    \end{align}
    and 
    \begin{align}
        \det{\CT^{\psi|\vv}_A} = \frac{c^2s^2}{\Delta^2}.
    \end{align}
    The two eigenvalues of $\CT^{\psi|\vv}_A$ are the two solutions of 
    \begin{align}
        F(\lambda) \equiv \lambda^2 - \lambda + \frac{c^2s^2}{\Delta^2} = 0.
    \end{align}
    We denote them as $\lambda_- \leq \lambda_+$. Both $\lambda_-$ and $\lambda_+$ are real and nonnegative if and only if 
    \begin{align}
        0\leq\frac{c^2s^2}{\Delta^2}\leq\frac{1}{4}.
    \end{align}
    In this case $\Delta\in\mathbb{R}$ and hence 
    \begin{align}
        |\Delta| &= |{\rm Re}(pr) + 2cs {\rm Re}(qt)| \nonumber\\ 
        &\leq |{\rm Re}(pr)| + 2cs |{\rm Re}(qt)| \nonumber\\
        &\leq |{\rm Re}(pr)| + (c^2+s^2) |{\rm Re}(qt)| \nonumber\\
        &= |{\rm Re}(pr)| +  |{\rm Re}(qt)| \nonumber\\
        &\leq |p||r| + |q||t| \nonumber\\
        &= \cos(\alpha-\beta) \leq 1.
        \label{eq:TrT2q}
    \end{align}
    where in the last line we recognize $|p|=\cos\alpha, ~|q|=\sin\alpha, ~|r|=\cos\beta, ~|t|=\sin\beta$.
    Let us go back to our statement (\ref{eq:mono2qbit}). Under the condition that the eigenvalues of $\CT^{\psi|\vv}_A$ are real and positive,
    \begin{align}
        &S^{(n)}(\CT^{\psi|\vv}_A)\geq S^{(n)}(\Tr_B(|\psi\rangle\langle\psi|)) = S^{(n)}(\Tr_B(|\vv\rangle\langle\vv|)){~~\rm for~~}n>0 \nonumber \\
        \Longleftrightarrow~ & 0\leq s^2 \leq \lambda_- \leq \lambda_+ \leq c^2 \leq 1 ~~{\rm (Majorization\cite{Nielsen98})}\nonumber\\
        \Longleftrightarrow~ & F(c^2) = c^2s^2\left(\frac{1}{\Delta^2}-1\right) \geq 0,
    \end{align}
    and the last line follows directly from (\ref{eq:TrT2q}).

\subsubsection{Beyond Two Qubits}\label{entwasps}

We would like to note that the same statement is not true in general Hilbert space. Let us see a representative example in 2-qutrit systems. 

\begin{example}
    Consider the following two states. 
    \begin{align}
        &\ket{\psi} = \frac{1}{\sqrt{2}}\left(\ket{0}\ket{0}+\ket{1}\ket{1}\right), \\
        &\ket{\vv} = \frac{1}{\sqrt{2}}\left(\ket{0}\ket{0}+\ket{1}\ket{2}\right).
    \end{align}
    The two states are related by a local basis transformation and 
    \begin{align}
        S^{(n)}(\Tr_B(|\psi\rangle\langle\psi|)) = S^{(n)}(\Tr_B(|\vv\rangle\langle\vv|)) = \log2{~~\rm for~~}n>0.
    \end{align}
    On the other hand, 
    \begin{align}
        \CT^{\psi|\vv}_A=\begin{pmatrix}
            1 & 0 & 0 \\
            0 & 0 & 0 \\
            0 & 0 & 0
        \end{pmatrix}, 
        \CT^{\psi|\vv}_B=\begin{pmatrix}
            1 & 0 & 0 \\
            0 & 0 & 1 \\
            0 & 0 & 0
        \end{pmatrix}.
    \end{align}
    and hence 
    \begin{align}
        S^{(n)}(\CT^{\psi|\vv}_A) = 0 < \log2 =  S^{(n)}(\Tr_B(|\psi\rangle\langle\psi|)) = S^{(n)}(\Tr_B(|\vv\rangle\langle\vv|)){~~\rm for~~}n>0.
    \end{align}
    The third dimension plays a crucial role in the distinguishment from 2-qubit cases. 

    It is easy to understand this difference between 2-qubit systems and higher dimensional systems. Let us denote the eigenvalues of ${\rm Tr}_B\left(|\vv\rangle\langle \vv|\right)$ as $a>b>c$. Now in this case the local unitary transformation only changes the eigenspace corresponding to $b$ and $c$. Therefore, by using the result in the 2-qubit case, the eigenvalues change toward majorization\cite{Nielsen98}. On the other hand, if we take the normalization into consideration and look at the whole Hilbert space, we can see that the normalization changes the eigenvalues toward an opposite direction of majorization. As a result, we cannot justify whether $S^{(n)}(\CT^{\psi|\vv}_A)$ is larger than $S^{(n)}(\Tr_B(|\psi\rangle\langle\psi|)) = S^{(n)}(\Tr_B(|\vv\rangle\langle\vv|))$ in general.
\end{example}

We also would like to point out a closely related example in 4-qubit system:
~\par
\begin{example} 
Consider a 4-qubit system, where the qubits are denoted by $A_1,A_2,B_1$ and $B_2$, respectively.
We regard the first two qubits as $A$ i.e. $A=A_1A_2$ and similarly for the latter two qubits $B=B_1B_2$.
For the states
    \begin{align}
    &\ket{\psi} = (\ket{0000} + \ket{0110})/\sqrt{2}, \\
    &\ket{\vv} = (\ket{0000} + \ket{1001})/\sqrt{2},
\end{align}
it is straightforward to see $S(\CT^{\psi|\vv}_A)=0$ and $S(\Tr_B(|\psi\rangle\langle\psi|)) = S(\Tr_B(|\vv\rangle\langle\vv|)) = \log2$. In this example, $\ket{\vv}$ is obtained from 
$\ket{\psi}$ by replacing the Bell pair of $A_2B_1$ with that of $A_1B_2$. This implies the entanglement swapping (or transition)
reduces the values of pseudo entropy compared with the original amount of quantum entanglement for each of states.
\end{example}

Moreover,  as we discuss in appendix \ref{tfdpe} for thermofield double states and appendix \ref {thppe} for two coupled harmonic oscillators, we can find physical examples with larger degrees of freedom which tend to satisfy the opposite inequality, namely, the pseudo entropy gets smaller than the averaged original entanglement entropy.

\subsubsection{A Weaker Inequality}

Nevertheless, we would like to mention that in general Hilbert spaces, a weaker bound can be obtained. We consider 
\begin{align}
    \ket{\psi} = (U \otimes 1
    )\ket{\vv}
\end{align}
where the unitary transformation $U$ only acts on the Hilbert space $\mathcal{H}_A$. We assume 
$\mathcal{H}_A$ and $\mathcal{H}_B$ are arbitrary and 
$\ket{\vv}\in \mathcal{H}_A\otimes \mathcal{H}_B$ can be any quantum state.
In this setup, if we set $\rho_\vv=\mbox{Tr}_B|\vv\lb\la \vv|$, then the transition matrix looks like
\ba
\CT^{\psi|\vv}_A=\frac{U\cdot \rho_\vv}{\mbox{Tr}[U\cdot \rho_\vv]}.
\ea
By writing the eigenvalues of $\rho^\vv_A$ and $U\cdot \rho^\vv_A$ as
$\{\sigma_i\}$ and  $\{\lambda_i\}$ in decreasing order, respectively, the Weyl's inequality leads to the relation
\ba
\sum_{i=1}^k |\lambda_i|^p\leq \sum_{i=1}^k (\sigma_i)^p,
\ea
for any $k (\leq \mbox{dim} \mathcal{H}_A)$.

If we assume the eigenvalues of $U\cdot \rho^\vv_A$ are real and non-negative, then the above inequality gives 
the weaker bound:
\ba
S^{(n)}(\CT^{\psi|\vv}_A) \geq   S^{(n)}(\rho^\vv_A) +\frac{n}{n-1}\log\la \vv|\psi\lb.
\ea

\subsection{Pseudo Entropy as Number of Bell Pairs in Class \texorpdfstring{$\mathscr{E}$}{Lg}} \label{sec:Distill}

In this subsection we would like to present an interpretation of pseudo entropy as a number of Bell pairs included in intermediate states by focusing on the class $\mathscr{E}$, which is supposed to be least exotic.\footnote{
Even though there are case I and case II in class $\mathscr{E}$ (see Fig.\ref{fig:venn2qubit}), 
the case II can be reduced to case I via unitary transformation of 
the basis. Therefore we can focus on the case I.}.

For this, let us consider the following states:
~\par
\begin{example}
~\par
We choose the two quantum states as follows
\ba
&& |\psi_1\lb_{AB}=\cos \theta_1|00\lb_{AB}+\sin \theta_1|11\lb_{AB},\no
&& |\psi_2\lb_{AB}=\cos \theta_2|00\lb_{AB}+\sin \theta_2|11\lb_{AB}. \label{twoqubitf}
\ea
We assume $0\leq \theta_1,\theta_2\leq\frac{\pi}{2}$ or $\frac{\pi}{2}\leq \theta_1,\theta_2\leq \pi$
in order to be in class  $\mathscr{D}$ (i.e.
 $\CT^{\psi_1|\psi_2}_A$ is positive semi-definite Hermitian matrix).
The reduced density matrix in each state reads 
\ba
\rho^{(i)}_A=\cos^2\theta_i |0\lb\la 0|+ \sin^2\theta_i |1\lb\la 1|.  \  \ \ (i=1,2)
\ea
The entanglement entropy for each state is found as 
\ba
S(\rho^{(i)}_{A})=-\cos^2\theta_i \log \cos^2\theta_i 
-\sin^2\theta_i \log \sin^2\theta_i,\ \ \ (i=1,2).   \label{eewq}
\ea

The transition matrix is found as 
\ba
 \mathcal{T}^{\psi_1|\psi_2}_A=\frac{1}{\cos(\theta_1-\theta_2)}(\cos\theta_1\cos\theta_2|0\lb\la 0|
+\sin\theta_1\sin\theta_2|1\lb\la 1|).
\ea

The pseudo R\'{e}nyi entropy is computed as 
\ba
S^{(n)}( \mathcal{T}^{\psi_1|\psi_2}_A)=\frac{1}{1-n}
\log\left[\left(\frac{\cos\theta_1\cos\theta_2}{\cos(\theta_1-\theta_2)}\right)^n+\left(\frac{\sin\theta_1\sin\theta_2}{\cos(\theta_1-\theta_2)}\right)^n\right].
\ea
The entanglement pseudo entropy is found as
\ba
S( \mathcal{T}^{\psi_1|\psi_2}_A)=&&-\left(\frac{\cos\theta_1\cos\theta_2}{\cos(\theta_1-\theta_2)}\right)\cdot \log\left(\frac{\cos\theta_1\cos\theta_2}{\cos(\theta_1-\theta_2)}\right)\no
&&\ \ -\left(\frac{\sin\theta_1\sin\theta_2}{\cos(\theta_1-\theta_2)}\right)\cdot \log \left(\frac{\sin\theta_1\sin\theta_2}{\cos(\theta_1-\theta_2)}\right). \label{tqbwee}
\ea
This is plotted in the left picture of Fig.\ref{fig:twoqubitWE}.

When $\theta_1=\theta_2$, this is reduced to the ordinary entanglement entropy (\ref{eewq}).
Moreover, when $\theta_1+\theta_2=\frac{\pi}{2}(2m+1)$ (here $m$ is an integer), we always find
$S(\mathcal{T}^{\psi_1|\psi_2}_A)=\log 2$. This is intriguing because even though the entanglement entropies
for $|\psi_1\lb$ and $|\psi_2\lb$ are small, the pseudo entropy can be large.

It is useful to note that the following difference which measures such an enhancement:
\ba
S( \mathcal{T}^{\psi_1|\psi_2}_A)-\frac{1}{2}(S(\rho^{(1)}_A)+S(\rho^{(2)}_A))  \label{difave}
\ea
can be both positive and negative in general as in the right picture of Fig.\ref{fig:twoqubitWE}.

\begin{figure}[t]
  \begin{center}
    \begin{tabular}{cc}
      \begin{minipage}{0.5\hsize}
        \begin{center}
          \includegraphics[width=0.9\linewidth,clip]{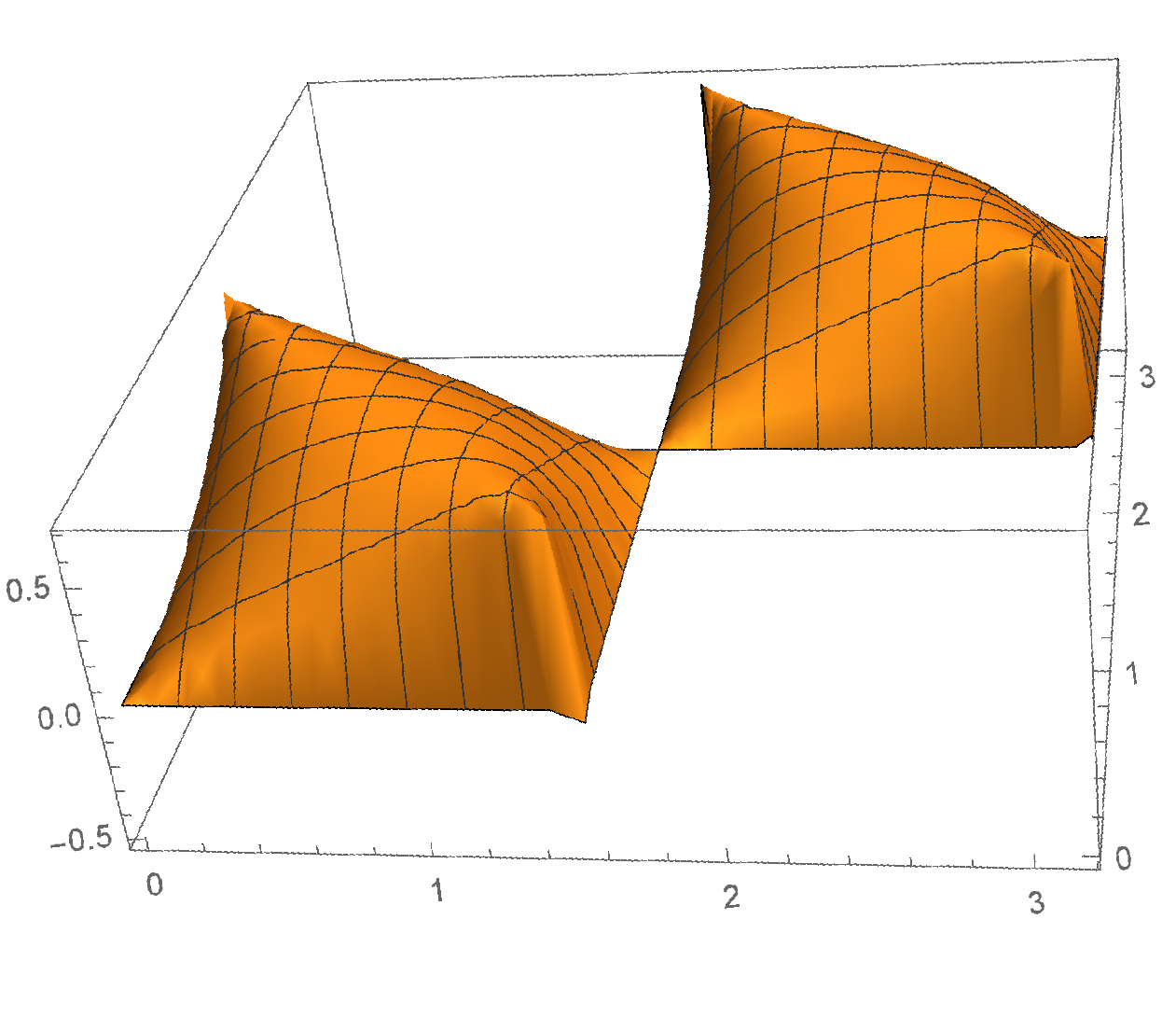}
        \end{center}
      \end{minipage}
      \begin{minipage}{0.5\hsize}
        \begin{center}
          \includegraphics[width=0.9\linewidth,clip]{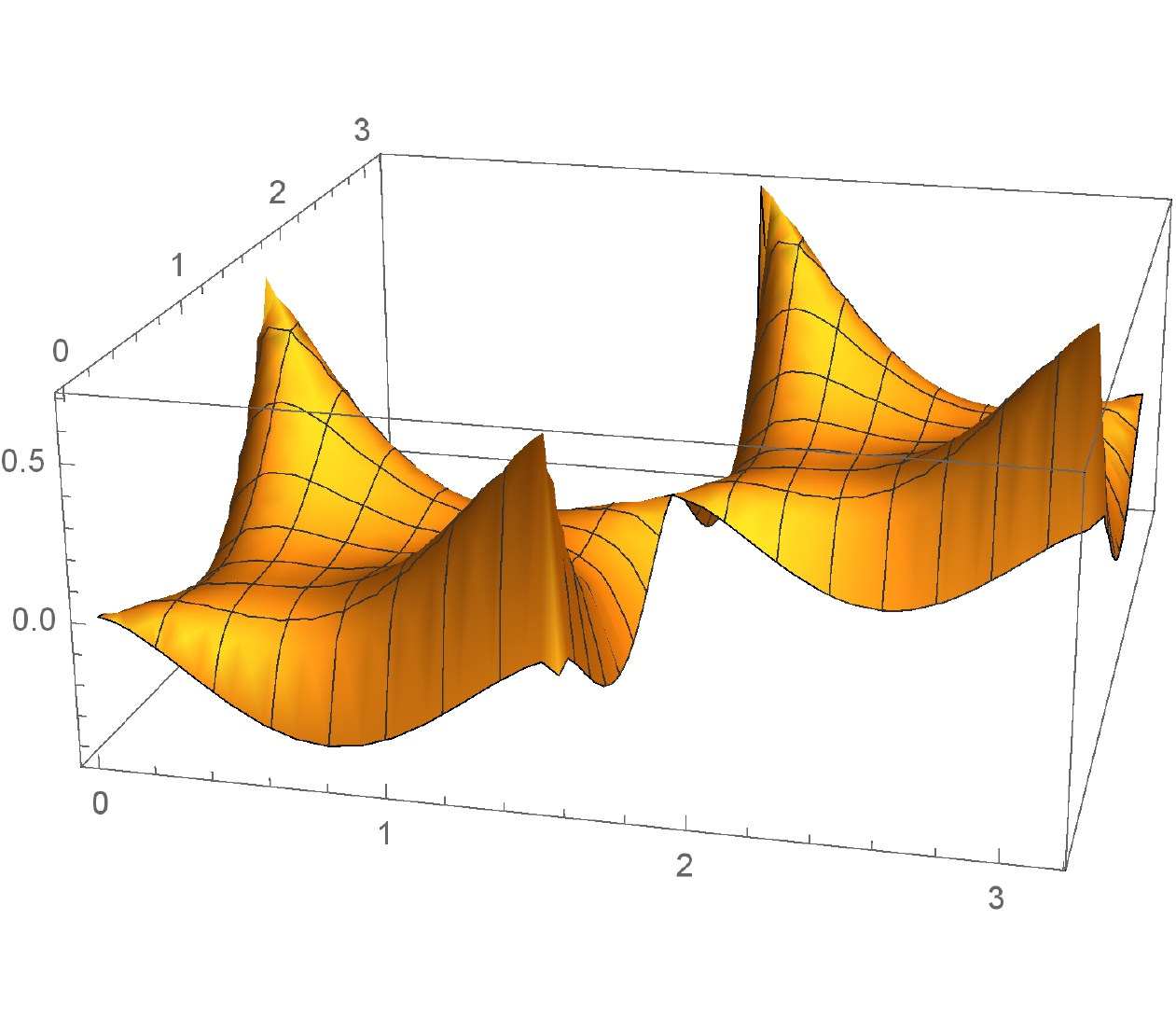}
        \end{center}
      \end{minipage}
    \end{tabular}
    \caption{We plot the pseudo entropy for the two qubit system as a function $\theta_1$ (horizontal axis) and $\theta_2$ (depth axis) in the left graph. The right one shows the pseudo entropy minus the averaged entanglement entropy i.e. (\ref{difave}). We took the range $0\leq \theta_{1,2}\leq\pi$. The region where no graph is shown gives complex valued pseudo entropy.}
    \label{fig:twoqubitWE}
  \end{center}
\end{figure}

\subsubsection{Interpretation via Quantum Entanglement}

Now let us consider an interpretation of pseudo entropy from the view point of quantum entanglement by extending the well-known argument \cite{Nielsen98,Bennett95} for the entanglement entropy.
Again we assume two quantum states are in the form (\ref{twoqubitf}). We regard $|\psi_1\lb$ as the initial state and $|\psi_2\lb$ as the final state of the post-selection.
We would like to estimate the averaged number of Bell pairs {that could have been potentially distilled from intermediate states before the state was actually post-selected.
More specifically, we rewrite $\la \psi_2| \psi_1\lb $ as $\sum_n
\la \psi_2|n\lb\la n|\psi_1\lb$ and consider the number of Bell pairs that can be distilled from the intermediate state $|n\lb$. Since the state would have appeared with probability $p_n=(\la \psi_2|n\lb\la n|\psi_1\lb)/\la \psi_2| \psi_1\lb$, we take the average over $p_n$.
This protocol is obviously not real since the post-selected state is $\ket{\psi_2}$, but not $\ket{n}$. We however think that this quantity is worth studying from a theoretical viewpoint since the quantity allows us to assess the amount of entanglement that is virtually involved in the post-selection process.}

Motivated by the LOCC interpretation of entanglement entropy we would like to take the asymptotic limit and consider the $M$ replicated states
\ba
|\psi_{i}\lb^{\otimes M}=(c_i|00\lb+s_i|11\lb)^{\otimes M}, 
\ea
where we defined $c_i=\cos\theta_i$ and $s_i=\sin\theta_i$. This is expanded in the form:
\ba
|\psi_{i}\lb^{\otimes M}=\sum_{k=0}^M c_i^{M-k} s_i^{k} \sum_{a=1}^{{}_M C_{k}}
|P^{(k)}_a \lb_A  |P^{(k)}_a \lb_B ,\ \ (i=1,2)
\ea
where ${}_M C_{k}=\frac{M!}{(M-k)!k!}$ is the combination factor and
we introduced orthonormal basis states $|P^{(k)}_a\lb$ $(k=0,1,\ddd,M, \ \ a=1,2,\ddd,{}_M C_k)$ for the $M$ qubits such that $|P^{(k)}_a\lb$ includes $M-k$ $|0\lb$-states and $k$ $|1\lb$-states as follows
\ba
&& |P^{(0)}_1\lb=|00\ddd 0\lb,\no
&& |P^{(1)}_1\lb=|10\ddd 0\lb,\ \ \  |P^{(1)}_2\lb=|01\ddd 0\lb,\ \ \ \ddd, \ \ \ \ 
 |P^{(1)}_M\lb=|00\ddd 1\lb, \no
\ea
For this basis we introduce the projection operators 
\ba
\Pi_k=\sum_{i=1}^{{}_M C_{k}}|P^{(k)}_i\lb_A  \la P^{(k)}_i|,
\ea
such that $\sum_{k=0}^{M} \Pi_k=1$. This projector $\Pi_k$ projects a given state into the maximally entangled state
\ba
|\Psi_k\lb=\frac{1}{\s{{}_M C_{k}}}\sum_{i=1}^{{}_M C_{k}}|P^{(k)}_i\lb_A |P^{(k)}_i\lb_B,
\ea
It is obvious that we can distill $\log_2 {}_M C_{k}$ Bell pairs from $|\Psi_k\lb$. 

The  probability $p_k$ of the appearance of the state $|\Psi_k\lb$ in the transition $\la \psi_2|\psi_1\lb$
is computed as  
\ba
p_k=\frac{\la \psi_2|^{\otimes M} \Pi_k |\psi_{1}\lb^{\otimes M} }
{\la \psi_2|^{\otimes M} |\psi_{1}\lb^{\otimes M}}
=\frac{(c_1c_2)^{M-k}(s_1s_2)^k}{(c_1c_2+s_1s_2)^M}\cdot {}_M C_{k}
\ea
Therefore we can estimate the number of Bell pairs which we could have distilled during the transition $\la \psi_2|\psi_1\lb$ as follows:
\ba
\bar{N}=\sum_{k=0}^M p_k  \log_2 {}_M C_{k}=
\sum_{k=0}^M  {}_M C_{k} \cdot\frac{(c_1c_2)^{M-k}(s_1s_2)^k}{(c_1c_2+s_1s_2)^M} \cdot \log_2 {}_M C_{k}, \label{bellnm}
\ea
By using the Stirling formula $n!\sim n^n e^{-n}\ (n\to \infty)$, we find that the summation over $k$ is localized at the point 
\ba
k_*=M\cdot \frac{\sin\theta_1\sin\theta_2}{\cos(\theta_1-\theta_2)}.
\ea
Thus we obtain
\ba
\lim_{M\to\infty}\frac{\bar{N}}{M}=&&-\left(\frac{\cos\theta_1\cos\theta_2}{\cos(\theta_1-\theta_2)}\right)\cdot \log\left(\frac{\cos\theta_1\cos\theta_2}{\cos(\theta_1-\theta_2)}\right)\no
&&\ \ -\left(\frac{\sin\theta_1\sin\theta_2}{\cos(\theta_1-\theta_2)}\right)\cdot \log \left(\frac{\sin\theta_1\sin\theta_2}{\cos(\theta_1-\theta_2)}\right). \label{tqbweeq}
\ea
This coincides with the entanglement pseudo entropy (\ref{tqbwee}).
\end{example}

In this way, the averaged number of Bell pairs which {could have been virtually} distilled during the post-selection process coincides with our pseudo entropy in  the class $\mathscr{D}$. Notice that formally the above argument can be analytically continued to arbitrary complex valued $\theta_1$ and $\theta_2$, though its physical interpretation is less clear.

\subsection{Violation of Sub-additivity}\label{sec:vssa}

Subadditivity and strong subadditivity are important properties of von Neumann entropy of quantum states. Considering a density matrix $\rho$ in a multipartite system whose Hilbert space is factorized as $\CH = \CH_A \otimes \CH_B \otimes \CH_C\otimes\cdots$, then subadditivity 
\begin{align}
    S(\rho_A) + S(\rho_B) - S(\rho_{AB}) \geq 0
\end{align}
and strong subadditivity\cite{LR73a,LR73b}
\begin{align}
    S(\rho_{AC})+S(\rho_{BC})-S(\rho_{ABC})-S(\rho_C) \geq 0
\end{align}
are always satisfied. 

It is obvious that neither subaddtivity nor strong subadditivity holds for pseudo entropy since it can be complex in general. However, one may wonder whether they hold if we restrict our transition matrices to a much more special class in which $\CT^{\psi|\vv}$ give $\CT^{\psi|\vv}_A$, $\CT^{\psi|\vv}_B$, ..., $\CT^{\psi|\vv}_{AB}$, ... whose eigenvalues are real and nonnegative. The answer to this question is no. Let us present a counter example. 

\begin{example}
    Consider the following two states on a 3-qubit system $\CH=\CH_A\otimes\CH_B\otimes\CH_C$:
    \begin{align}
        &\ket{\psi} = \left(I\otimes I\otimes U\right) \ket{\vv},\\
        &\ket{\vv} = \cos{\alpha}\ket{000} + \sin{\alpha}\ket{111},
    \end{align}
    where $I$ is the identity transformation and $U$ is a unitary transformation which satisfies 
    \begin{align}
        &U\ket{0} = \cos{\theta}\ket{0} + \sin{\theta}\ket{1}, \\
        &U\ket{1} = -\sin{\theta}\ket{0} + \cos{\theta}\ket{1}.
    \end{align}
    Then it is easy to figure out
    \begin{align}
        \ket{\psi} = \cos{\theta}\cos{\alpha}\ket{000} + \sin{\theta}\cos{\alpha}\ket{001} + \sin{\theta}\sin{\alpha}\ket{110} + \cos{\theta}\sin{\alpha}\ket{111}
    \end{align}
    and 
    \begin{align}
        \CT^{\psi|\vv}_A = \Tr_{BC}\left(\CT^{\psi|\vv}\right) =&~\cos^2{\alpha}|0\rangle\langle0| + \sin^2{\alpha}|1\rangle\langle1|, \\
        \CT^{\psi|\vv}_B = \Tr_{AC}\left(\CT^{\psi|\vv}\right) =&~\cos^2{\alpha}|0\rangle\langle0| + \sin^2{\alpha}|1\rangle\langle1|, \\
        \CT^{\psi|\vv}_{AB} = \Tr_{C}\left(\CT^{\psi|\vv}\right) =~&\cos^2{\alpha}|00\rangle\langle00| + \tan{\theta}\cos{\alpha}\sin{\alpha}|00\rangle\langle11| \nonumber\\
        -& \tan{\theta}\cos{\alpha}\sin{\alpha}|11\rangle\langle00| +\sin^2{\alpha}|11\rangle\langle11|. 
    \end{align}
    The eigenvalues of $\CT^{\psi|\vv}_A=\CT^{\psi|\vv}_B$ are
    \begin{align}
        \cos^2{\alpha},~\sin^2{\alpha}
        \label{eq:SAeigen1}
    \end{align}
    and the eigenvalues of $\CT^{\psi|\vv}_{AB}$ are 
    \begin{align}
        0,~0,~\frac{1}{2}\left(1+\sqrt{\frac{\cos{4\alpha}+\cos{2\theta}}{1+\cos{2\theta}}}\right),~\frac{1}{2}\left(1-\sqrt{\frac{\cos{4\alpha}+\cos{2\theta}}{1+\cos{2\theta}}}\right).
        \label{eq:SAeigen2}
    \end{align}
    In a similar manner, we can find that the nonzero eigenvalues of $\CT^{\psi|\vv}_{BC}$ and $\CT^{\psi|\vv}_{AC}$ are (\ref{eq:SAeigen1}) and the eigenvalues of $\CT^{\psi|\vv}_C$ are the same as the nonzero ones of (\ref{eq:SAeigen2}). Therefore, in this case, $\CT^{\psi|\vv}$ give $\CT^{\psi|\vv}_A$, $\CT^{\psi|\vv}_B$, ..., $\CT^{\psi|\vv}_{AB}$, ... whose eigenvalues are real and nonnegative if and only if $(\cos{4\alpha} + \cos{2\theta}) \geq 0$. Now we would like to check if $S(\CT^{\psi|\vv}_A) + S(\CT^{\psi|\vv}_B) - S(\CT^{\psi|\vv}_{AB})$ is nonnegative or not. Noticing the symmetry, it is sufficient to look at $\alpha\in[0,\pi/4]$ and $\theta\in[0,\pi/2]$. Figure \ref{fig:SAvio} shows this region. The colored part shows $(\cos{4\alpha} + \cos{2\theta}) \geq 0$ where $S(\CT^{\psi|\vv}_A) + S(\CT^{\psi|\vv}_B) - S(\CT^{\psi|\vv}_{AB}) < 0$ in the blue region and $S(\CT^{\psi|\vv}_A) + S(\CT^{\psi|\vv}_B) - S(\CT^{\psi|\vv}_{AB}) \geq 0$ in the yellow region. 
    \begin{figure}[H]
        \centering
        \includegraphics[width=8cm]{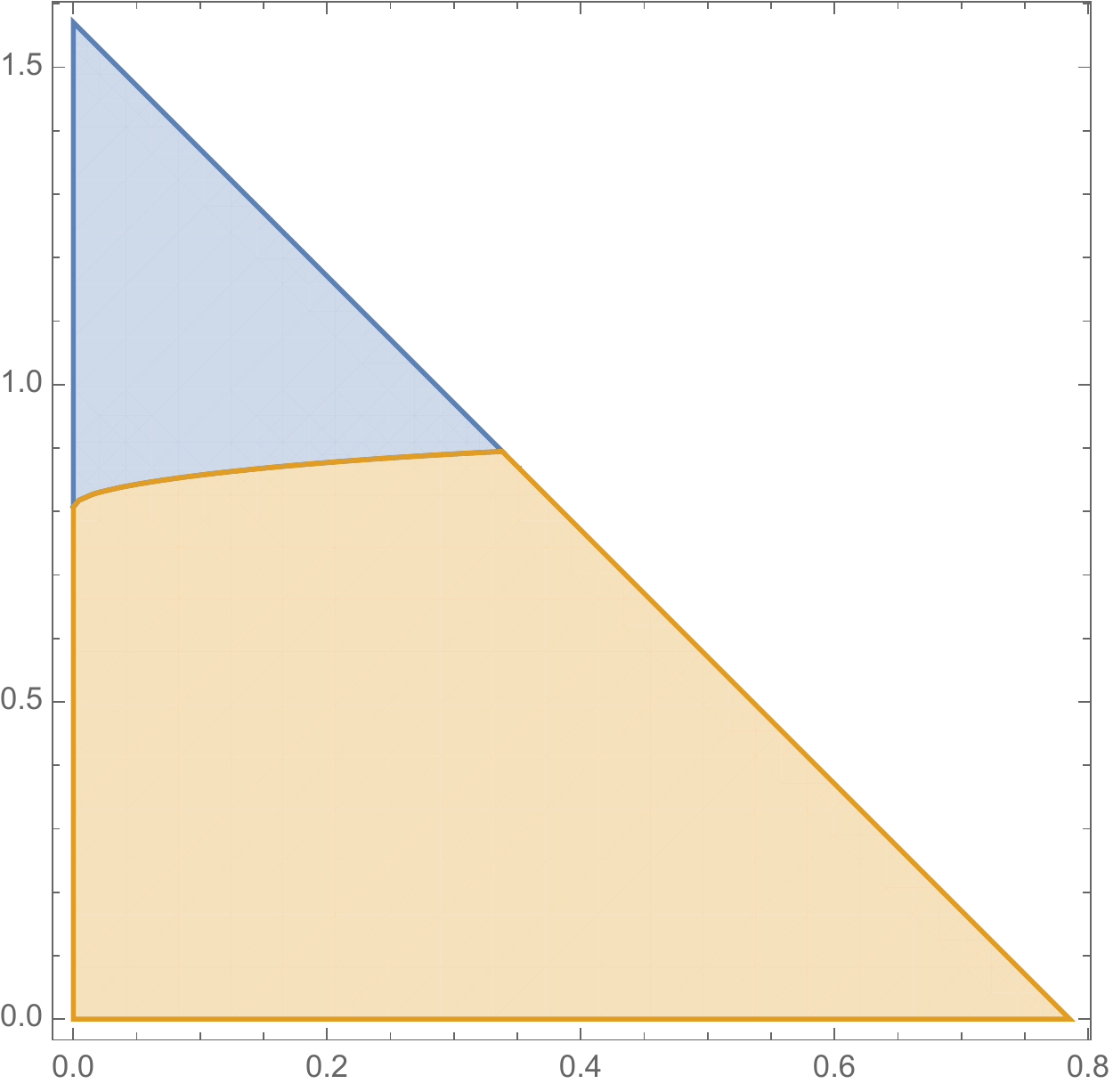}
        \caption{The region of $\alpha\in[0,\pi/4]$ and $\theta\in[0,\pi/2]$. The colored part shows the region in which all eigenvalues of all reduced transition matrices are real and nonnegative. $S(\CT^{\psi|\vv}_A) + S(\CT^{\psi|\vv}_B) - S(\CT^{\psi|\vv}_{AB}) < 0$ in the blue region and $S(\CT^{\psi|\vv}_A) + S(\CT^{\psi|\vv}_B) - S(\CT^{\psi|\vv}_{AB}) \geq 0$ in the yellow region.}\label{fig:SAvio}
    \end{figure}
    Due to the existence of the blue region, subadditivity is violated in this example. Then it is obvious that strong subadditivity is also violated. 
\end{example}

\subsection{Pseudo Entropy for Random States}\label{sec:randomstates}

We here provide generic properties of pseudo entropy by investigating the pseudo entropy for \emph{Haar random} states. 
A Haar random state is defined as a pure state drawn from a Hilbert space uniformly at random according to the unitarily invariant probability measure ${\sf H}$. Such a measure is unique and is called the Haar measure. A Haar random state is often denoted by $\ket{\phi} \sim {\sf H}$. Due to the unitary invariance of the Haar measure, a Haar random state is suitable to check typical properties of quantum pure states.
Following this idea, we numerically and analytically investigate the pseudo entropy $S(\mathcal{T}^{\psi|\vv}_A)$ when $\vv$ and $\psi$ are independent Haar random states.

\begin{figure}[bt!]
\centering
\includegraphics[width=15cm]{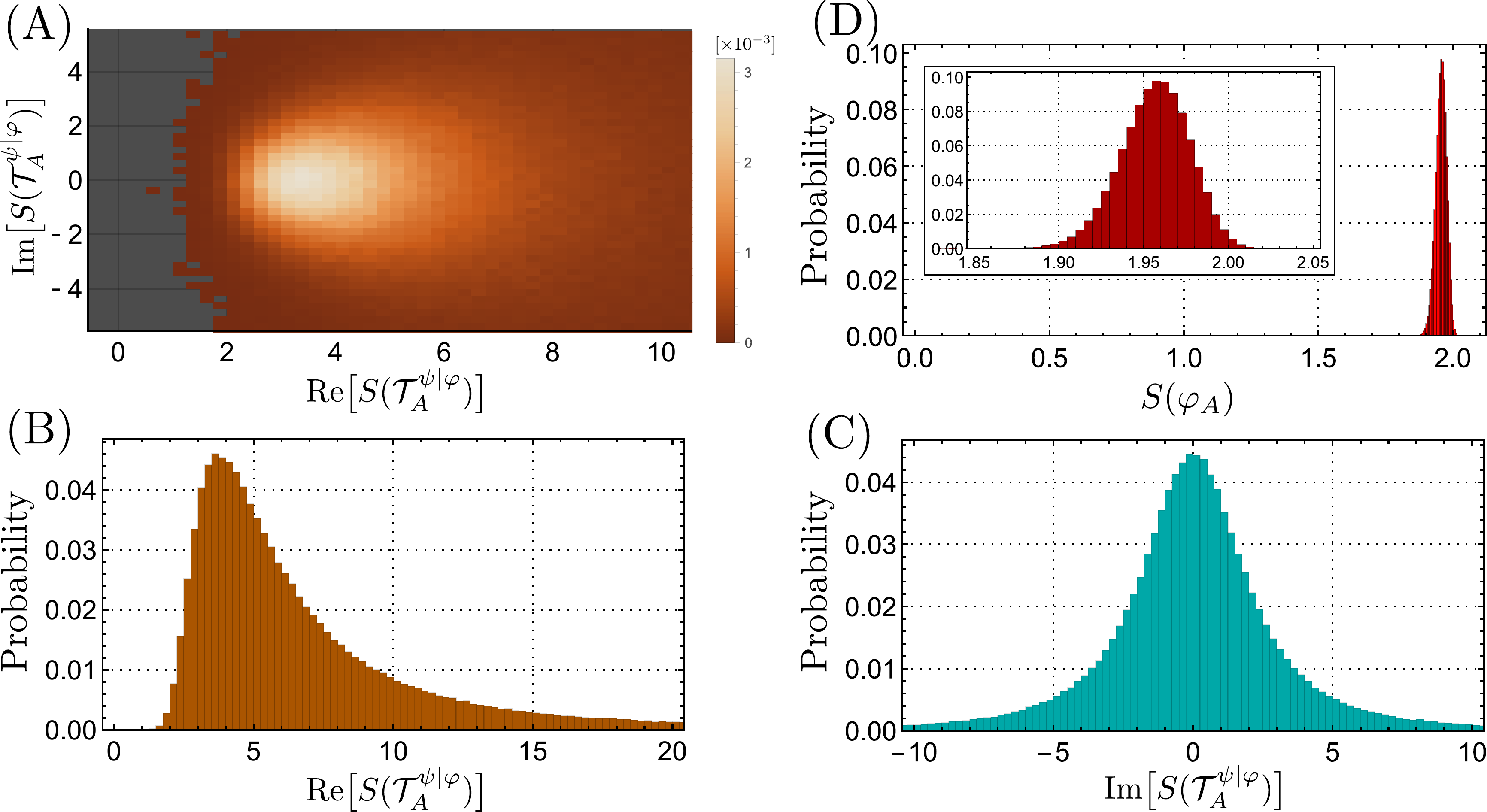}
\caption{Distributions of the pseudo entropy $S(\mathcal{T}^{\psi|\vv}_A)$ over Haar random states $\ket{\vv}$ and $\ket{\psi}$ (Panels A, B, and C) and that of the entanglement entropy $S(\vv_A)$ over $\ket{\varphi} \sim {\sf H}$ (Panel D). They are obtained for $\dim \CH_A = 8$ and $\dim \CH_B = 32$ with the number of sampling $655360$.
In Panel (A), we provide the distribution of the pseudo entropy. The color-bar on the right shows the corresponding probability for each color. As it is complex-valued in general, we plot the histogram over its real and imaginary parts. Note that the black part on the left-hand side implies that no instance was observed. In Panels (B) and (C), the real and imaginary parts of the pseudo entropy are plotted, respectively.
As a reference, we also provide the distribution of the entanglement entropy over a Haar random state in Panel (D), where the inset enlarges a non-trivial part of the distribution.
}\label{Fig:vNE}
\end{figure}

We first numerically provide in figure~\ref{Fig:vNE} the distribution of pseudo entropy $S(\mathcal{T}^{\psi|\vv}_A)$ over $\ket{\psi}, \ket{\varphi} \sim {\sf H}$. As a reference, we also provide the distribution of usual entanglement entropy $S(\vv_A)$ over $\ket{\varphi} \sim {\sf H}$ where $\varphi_A\equiv{\Tr_B}|\vv\rangle\langle\vv|$.
Despite the facts that pseudo entropy is in general complex-valued and that its absolute value can be arbitrarily large, the distribution centers around a moderately small real value. In the case of $\dim \CH_A = 8$ and $\dim \CH_B = 32$, the value is roughly $3.6$, which is larger than but comparable with the the maximum of the entanglement entropy, i.e. $\log \dim \CH_A \approx 2.08$. Also, the distribution seems to be symmetric about the real axis (see also Panel (C)). 
In Panel (D), we present the distribution of entanglement entropy over a Haar random state, where we clearly observe that the distribution highly concentrates around a nearly maximal value. This phenomena, i.e., most random states are highly entangled, has been repeatedly pointed out in the literature~\cite{L1978,Page1993,FK1994,S-R1995,S1996, HLW2006} and is a consequence of the \emph{concentration of measure phenomena}~\cite{HLW2006,L2001}, one of the generic properties of the Haar measure. Compared to the strong concentration of entanglement entropy for a random state, the distribution of pseudo entropy concentrates rather weakly. This is, however, likely to be due to the fact that there is no obvious upper bound on $|S(\mathcal{T}^{\psi|\vv}_A)|$, which is in contrast to finite range $[ 0, \log[ \dim \CH_A] ]$ of the entanglement entropy. We hence expect that concentration also occurs, though it may be weak, even for pseudo entropy. We will discuss this point later. We would like to note that no instances are found in region ${\rm Re}[S(\CT^{\psi|\vv}_A)]<0$ in figure \ref{Fig:vNE}. This implies that the probability for a transition matrix given by two Haar random states to satisfy ${\rm Re}[S(\CT^{\psi|\vv}_A)]<0$ is very small. However, this does not mean that ${\rm Re}[S(\CT^{\psi|\vv}_A)]\geq0$ holds for all transition matrices. For example, if we take $|\psi\rangle=(|0\rangle_A\otimes|0\rangle_B + |1\rangle_A\otimes|1\rangle_B)/\sqrt{2}$ and $|\vv\rangle=(2|0\rangle_A\otimes|0\rangle_B - |1\rangle_A\otimes|1\rangle_B)/\sqrt{5}$ where $\{|0\rangle_A,\cdots,|7\rangle_A\}$ gives a orthonormal basis of $\CH_A$ and $\{|0\rangle_B,\cdots,|31\rangle_B\}$ gives a orthonormal basis of $\CH_B$, then we can easily see that the eigenvalues of $\CT^{\psi|\vv}_A$ are given by $\{2,-1,0,\cdots,0\}$ and hence ${\rm Re}[S(\CT^{\psi|\vv}_A)] = -2\log2<0$.

The distribution of pseudo entropy can be more elaborated in terms of the Tsallis entropy instead of the von Neumann entropy. The Tsallis entropy is a generalization of the von Neumann entropy into a one-parameter family and is defined by $T_q (\rho) =(1- \tr [\rho^q])/(q-1)$ for $q \in \mathbb{R}$, where $\rho$ is normally a density matrix. Similarly to the R\'{e}nyi entropy, it converges to the von Neumann entropy in the limit of $q \rightarrow 1$. Using the Tsallis entropy, we define the pseudo Tsallis entropy $T_q(\CT_A^{\psi| \varphi})$ for the reduced transition matrix $\CT^{\psi|\vv}_A$. 

By the standard technique, we can show that the average of the pseudo Tsallis entropy over Haar random states $\ket{\psi}, \ket{\varphi} \sim {\sf H}$ exactly coincides with that of the entanglement Tsallis entropy over a Haar random state $\ket{\varphi} \sim {\sf H}$ for any $q \in \mathbb{N}^+\backslash\{1\}$, i.e. 
\begin{equation}
\mathbb{E}_{\ket{\varphi}, \ket{\psi} \sim {\sf H}} \bigl[ T_q(\CT_A^{\psi| \varphi}) \bigr] = \mathbb{E}_{\ket{\varphi} \sim {\sf H}} \bigl[ T_q(\varphi_A) \bigr], \label{Eq:AvT}
\end{equation}
where $\mathbb{E}$ represents the expectation over random states specified by the subscript, $\varphi_A = \Tr_{B} |\varphi \rangle \langle \varphi |$ is the reduced density matrix of $\ket{\varphi}$ in $A$. See Appendix~\ref{App:Haar} for the proof.
Recalling that the Tsallis entropy for quantum states takes the values between $0$ and $(1 - (\dim \CH_A)^{1-q})/(q-1) < 1$, we readily obtain from \eqref{Eq:AvT} that, although the pseudo Tsallis entropy generally takes complex values with arbitrarily large absolute values, its average remains between $0$ and $(1 - (\dim \CH_A)^{1-q})/(q-1)$. 
Note however that this does not imply that higher moments of the distributions also coincide, which can be explicitly checked by the same technique used in appendix~\ref{App:Haar}.  Thus, as far as we are concerned with the pseudo Tsallis entropy for random states, higher moments of the distributions are the factors that differentiate it from the entanglement Tsallis entropy.

\begin{figure}[tb!]
\centering
\includegraphics[width=16cm]{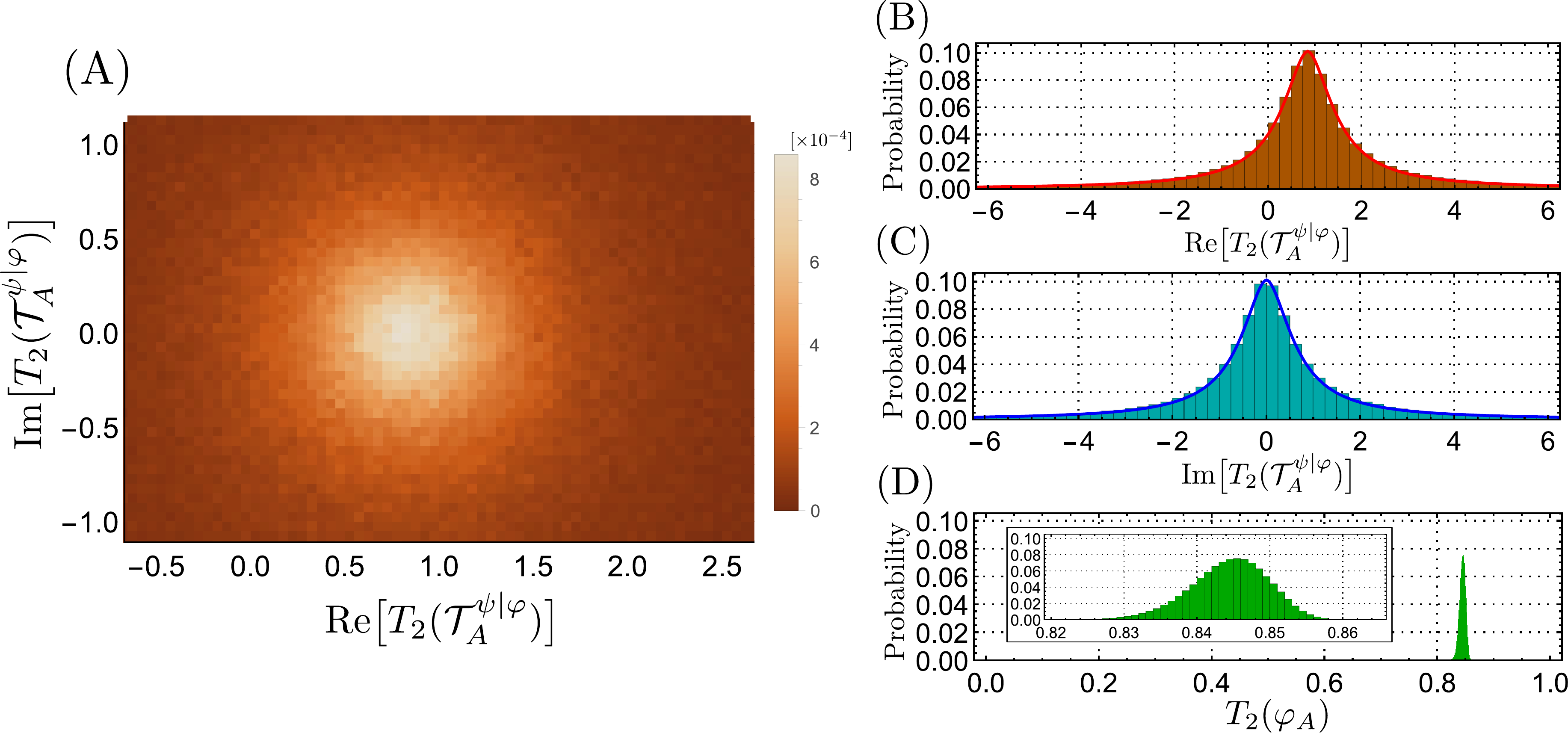}
\caption{Distributions of the pseudo Tsallis entropy $T_2(\mathcal{T}^{\psi|\vv}_A)$ over Haar random states $\ket{\vv}$ and $\ket{\psi}$ (Panels A, B, and C) and that of the entanglement Tsallis entropy $T_2(\vv_A)$ over $\ket{\varphi} \sim {\sf H}$ (Panel D). They are obtained for $\dim \CH_A = 8$ and $\dim \CH_B = 32$ with the number of sampling $655360$.
Similarly to Fig.~\ref{Fig:vNE}, the distribution of the pseudo entropy over its real and imaginary parts is given in Panel (A), that only of the real part in Panel (B), that of the imaginary part in Panel (C), and that of the entanglement Tsallis entropy over a Haar random state in Panel (D).
In Panels (B) and (C), we also plot inverse-polynomial functions by red and blue colours (see the main text).
}\label{Fig:Tsallis}
\end{figure}

To elucidate this point, we numerically demonstrate in figure~\ref{Fig:Tsallis} the distribution of the entanglement Tsallis entropy and that of pseudo Tsallis entropy over Haar random states. We especially consider $q=2$. It is clear that the shapes of the distributions are rather different: for the entanglement Tsallis entropy, the probability density function (PDF) is basically a Gaussian-shape flanked with rapidly decaying functions on its both sides (see Panel (D) and its inset). In contrast, the distribution of the pseudo Tsallis entropy has much heavier tails and resembles the Cauchy–Lorentz distribution rather than a Gaussian distribution (see Panel (B) and (C)). In fact, we numerically confirmed that the PDF $p \bigl({\rm Re}[T_2(\CT_A^{\psi| \varphi})]=t\bigr)$ for the real part of the pseudo Tsallis entropy with $q=2$ to take the value $t$ is given by
\begin{equation}
p \bigl({\rm Re}[T_2(\CT_A^{\psi| \varphi})]=t \bigr) \propto \frac{1}{(t- t_0)^{1.8}},  \text{\ \ where\ \ } t_0 := 1 - \frac{\dim \CH_A + \dim \CH_B}{ \dim \CH + 1}.
\end{equation}
We here point out that this PDF of the pseudo Tsallis entropy shall indicate that the values should be centered around its average $t_0 \in (0, 1)$, suggesting a weak concentration of the pseudo Tsallis entropy for random states.
We hence expect that, in the large dimension limit, it is atypical for the pseudo Tsallis entropy to take abnormal values such as complex numbers with large absolute values. 

We finally conclude this section with a comment on a theoretically interesting property of the entanglement distribution over a random state. It is known that the PDF for the entanglement entropy over a Haar random state has two singularities, which divides the distribution into three \emph{entanglement phases} with different entanglement spectra~\cite{G2007, FMPPS2008, PFPPS2010, NMV2010, PFGPPS2011,NMV2011, FFPPY2013, FPPSY2019}. 
It is natural to ask whether the PDF $p \bigl({\rm Re}[T_2(\CT_A^{\psi| \varphi})]=t \bigr)$ for the pseudo Tsallis entropy also has any singularity. This question can be analytically addressed using the technique of random matrix theory, which is the method used to analyze the PDF for the entanglement Tsallis entropy in great detail, but we will leave it as an open problem.


\section{Pseudo R\'{e}nyi Entropy in Free CFT}\label{sec:rfreecft}

In this section, we compute pseudo R\'{e}nyi entropy for a free CFT as a simple example of QFT. We consider a massless scalar field theory in two dimensional Euclidean space, whose coordinate is $w=x+i\tau$. We would like to calculate its pseudo-R\'{e}nyi entropy with respect to an interval subsystem $A=\{(x,\tau)\mid\tau=0,\ x_l\le x\le x_r\}$ by using the replica method we described in section \ref{sec:replica}. 

Before computing the pseudo R\'{e}nyi entropy, let us briefly review the ordinary calculation of R\'{e}nyi entropy of the vacuum state $\ket{0}$, the detail of which is described in \cite{CC04}. We define the density matrix $\rho_A^{(0)}=\Tr_B[|0\rangle\langle0|]$, where $B$ is the complement of $A$. 
In terms of the path integral as expressed in section \ref{sec:replica}, $\ket{0}$ and $\bra{0}$ are
\begin{align}
  \ket{0}=\imineq{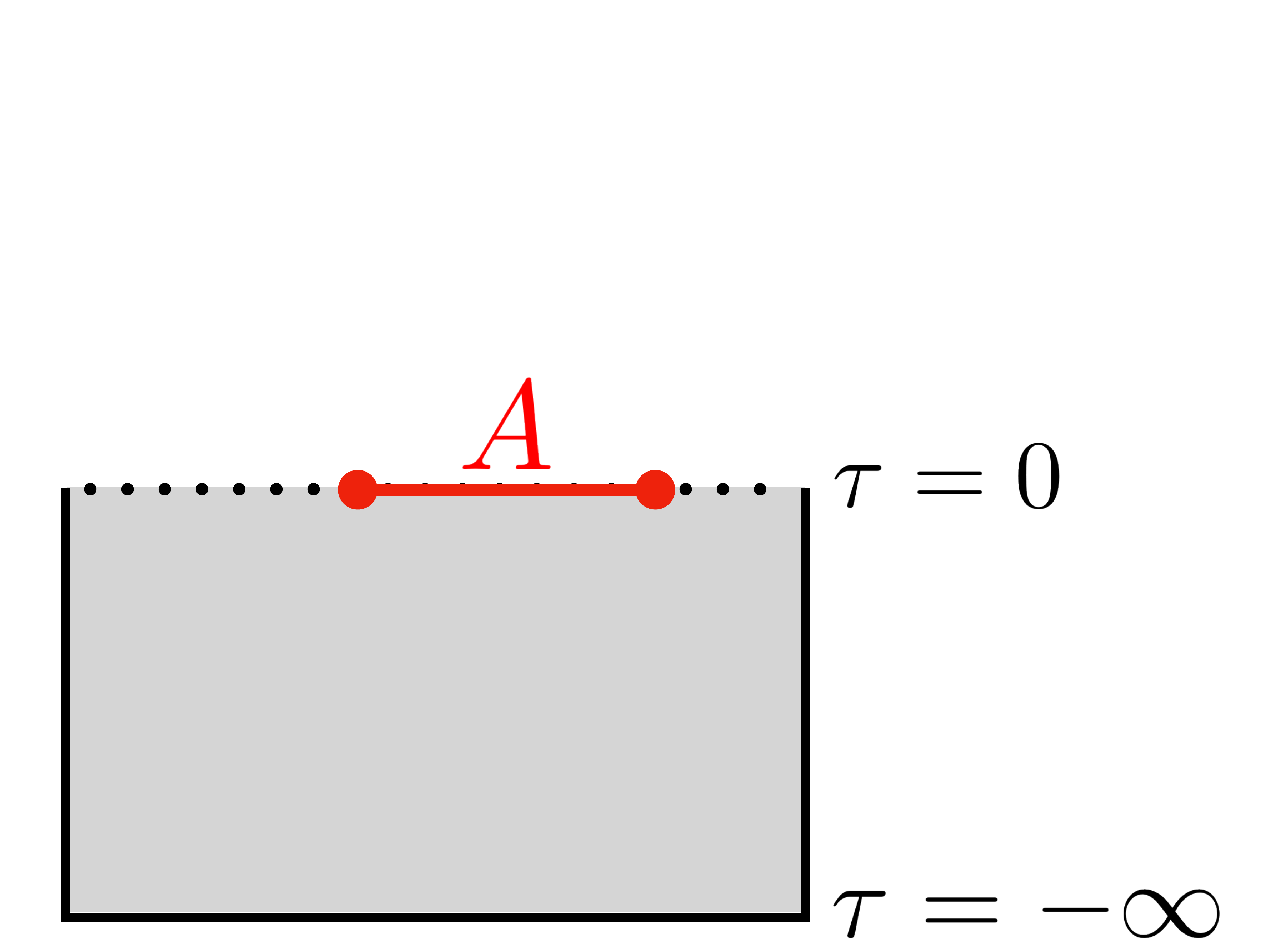}{20}, \ \bra{0}=\imineq{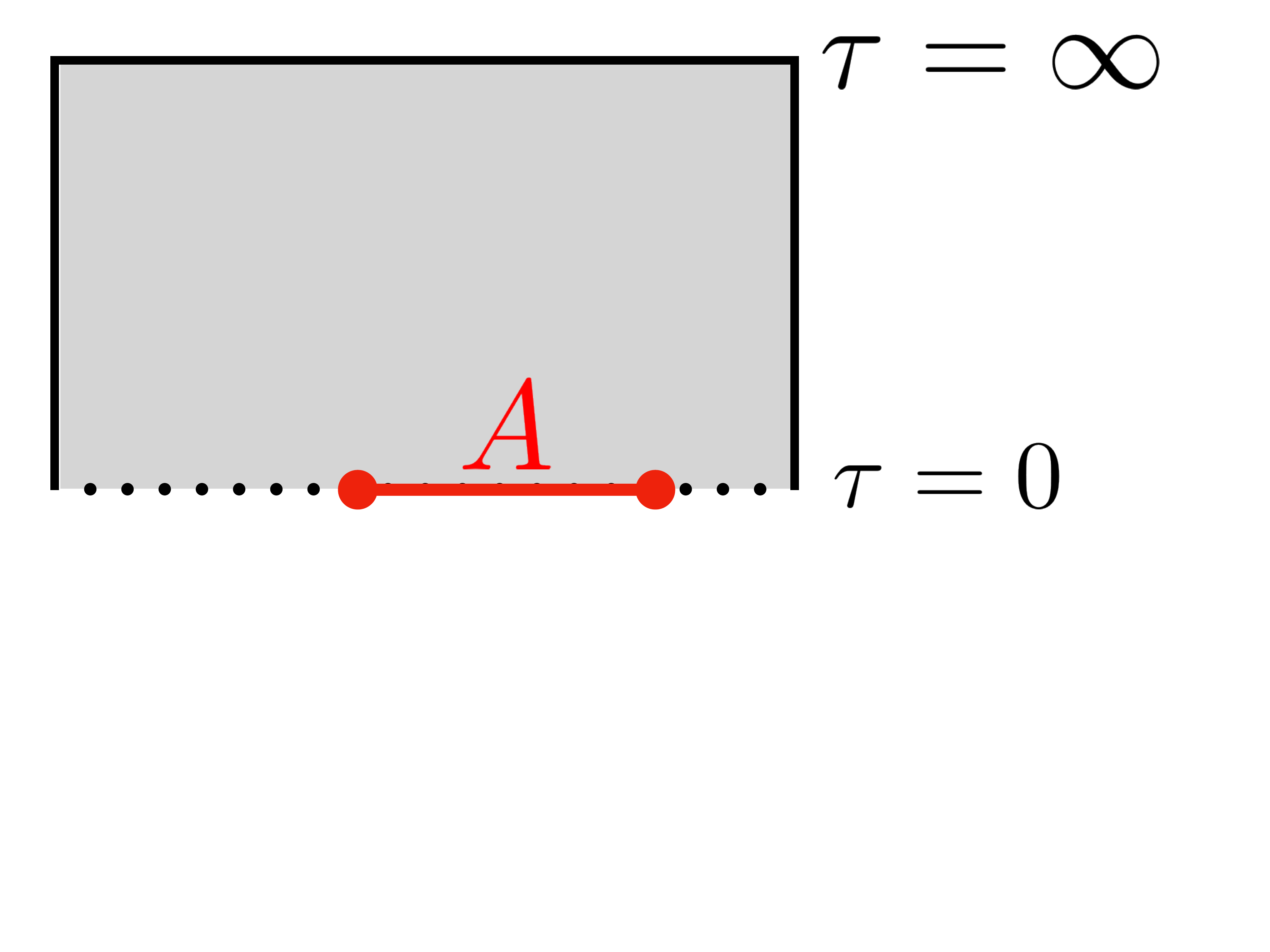}{20},
\end{align}
and tracing out on $B$ then 
\begin{align}
 \rho_A^{(0)}=\imineq{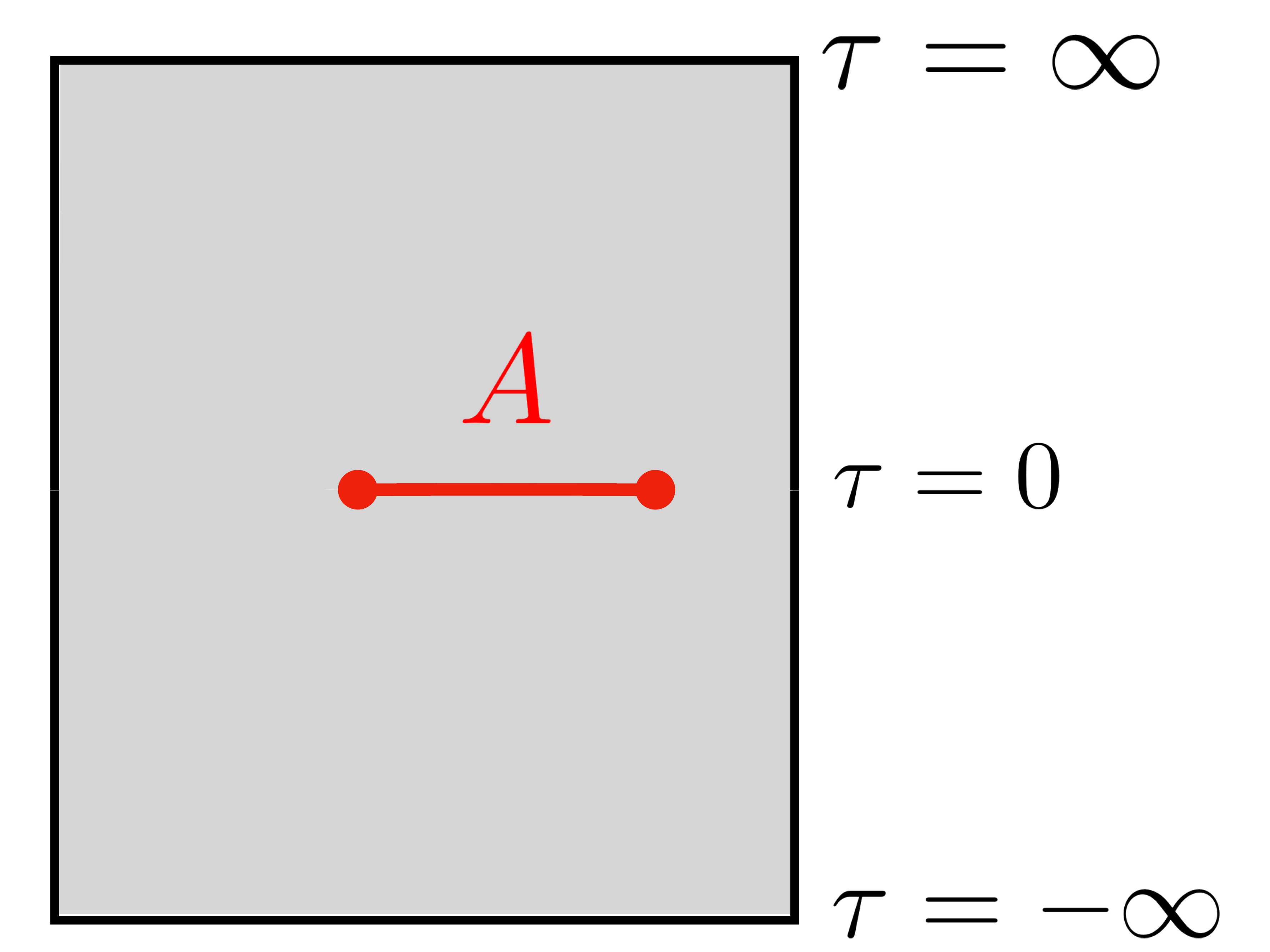}{20}.
\end{align}
For calculating the R\'{e}nyi entropy, we need $\Tr[(\rho_A)^n]$, which can be computed by replica method. We consider $n$ sheets and identify the upper side of $A$ on the $k$-th sheet and the lower side of $A$ on the $(k+1)$-th sheet, that is, 
\begin{align} \label{eq:nsheets}
  \Tr[(\rho_A^{(0)})^n]=\imineq{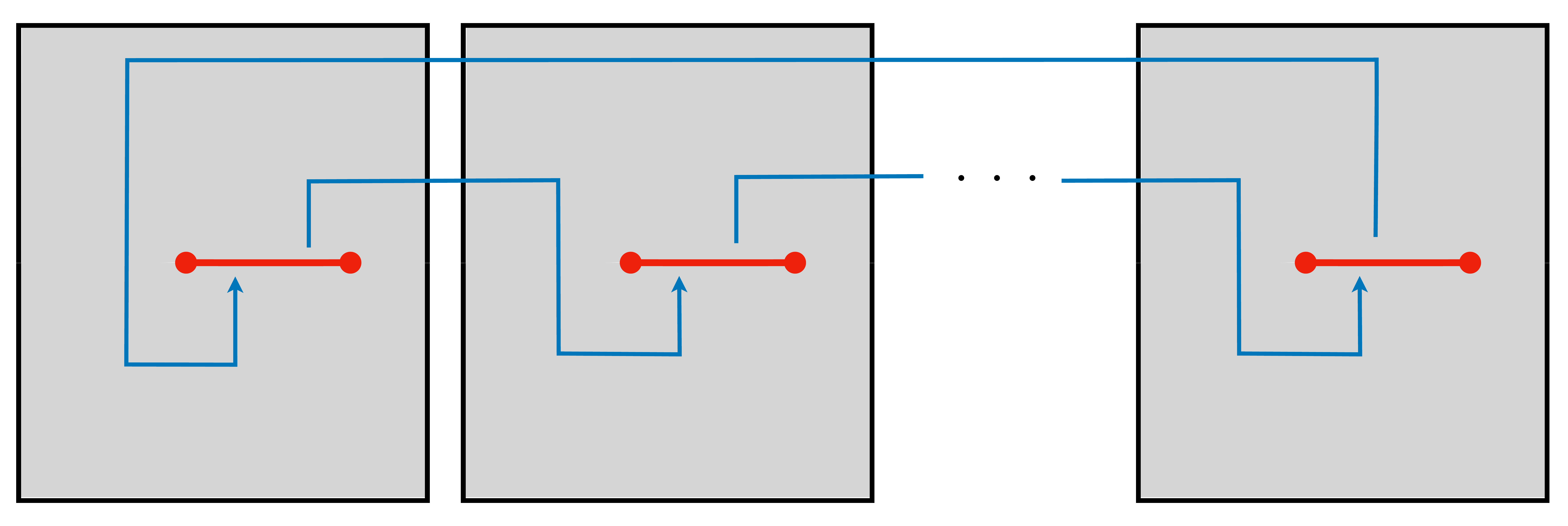}{20}.
\end{align}
The right-hand side of (\ref{eq:nsheets}) is the path integral on the manifold $\Sigma_n$ introduced in section \ref{sec:replica}. The identification between the $k$-th sheet and the $(k+1)$-th sheet corresponds to the multiplication as matrix, especially the $(n+1)$-th sheet and the first sheet to tracing out. By computing this path integral, the explicit expression of (\ref{eq:nsheets}) becomes 
\begin{equation}
  \Tr[(\rho_A^{(0)})^n]=c_n\left(\frac{x_r-x_l}{\epsilon}\right)^{-\frac{n^2-1}{6n}},
\end{equation}
where $\epsilon$ is the cutoff with the dimension of length and now the central charge is $1$. Also note that $c_n$ is a constant which depends on $n$ and cannnot be determined by the replica trick. Thus the $n$-th R\'{e}nyi entropy is 
\begin{equation}
  S^{(n)}(\rho_A^{(0)})=\frac{1}{6}\left(1+\frac{1}{n}\right)\log\frac{x_r-x_l}{\epsilon}+\frac{\log{c_n}}{1-n}.
\end{equation}
It is important that the vacuum state has the entanglement between $A$ and $B$, so we are interested in the variation of the pseudo R\'{e}nyi entropy from the ordinary R\'{e}nyi entropy of the vacuum state:
\begin{equation}
  \Delta S^{(n)}_A\equiv S^{(n)}(\mathcal{T}^{\psi|\vv}_A)-S^{(n)}(\rho^{(0)}_A).
\end{equation}
We will compute this quantity in the following sections.

\subsection{Example 1: Exciting the Same Space Point with Different Cutoffs} \label{sec:samepoints}
For the first example, we choose $\ket{\psi}$ and $\ket{\vv}$ as 
\begin{align} \label{eq:CFTstates}
  &\ket{\psi}=e^{-a H_{\rm CFT}}\mathcal{O}(x=b)\ket{0}, \nonumber \\
  &\ket{\vv}=e^{-a^\prime H_{\rm CFT}}\mathcal{O}(x=b^\prime)\ket{0}, 
\end{align}
where $b$ ($b^\prime$) is the excited space point and $a$ ($a^{\prime}$) is a cutoff to avoid UV divergence at $b=b'$. We can also regard $a$ ($a'$) as Euclidean inserting time. The inserted operator is given by
\begin{equation} \label{eq:belloperator1}
  \mathcal{O}=e^{\frac{i}{2}\phi}+e^{-\frac{i}{2}\phi},
\end{equation}
and its conformal dimension is $h=\bar{h}=1/8$. This operator is analogous to a Bell pair in a two qubit system.
Note that this operator creates a Bell pair at the inserted point \cite{Nozaki,HNTW}. This can be seen if we decompose the scalar field into the left moving mode and right moving one as $\phi=\phi_R+\phi_L$.
We can interpret the operator such that  $e^{\pm i\phi_L/2}|0\lb_L\sim|\pm\lb_L$ and  $e^{\pm i\phi_R/2}|0\lb_R\sim|\pm\lb_R$. This allows us to regard $(e^{\frac{i}{2}\phi}+e^{-\frac{i}{2}\phi})|0\lb_L |0\lb_R\sim|+\lb_L|+\lb_R+|-\lb_L|-\lb_R$, i.e. a Bell pair. If we chose  $\mathcal{O}=e^{\frac{i}{2}\phi}$ instead of (\ref{eq:belloperator1}), 
we would always get the trivial result $\Delta S^{(2)}_A=0$. 

Note that (\ref{eq:CFTstates}) is written in the Schr\"{o}dinger picture. When we perform the path integral, the corresponding operators are inserted at
\begin{align}
    (w_1,\bar{w}_1) \equiv (b-ia, b+ia), ~~~~
    (w_2,\bar{w}_2) \equiv (b^\prime+ia^\prime, b^\prime-ia^\prime)
\end{align}
respectively. 

\begin{figure}[H]
  \begin{center}
    \includegraphics[width=0.9\linewidth,pagebox=cropbox,clip]{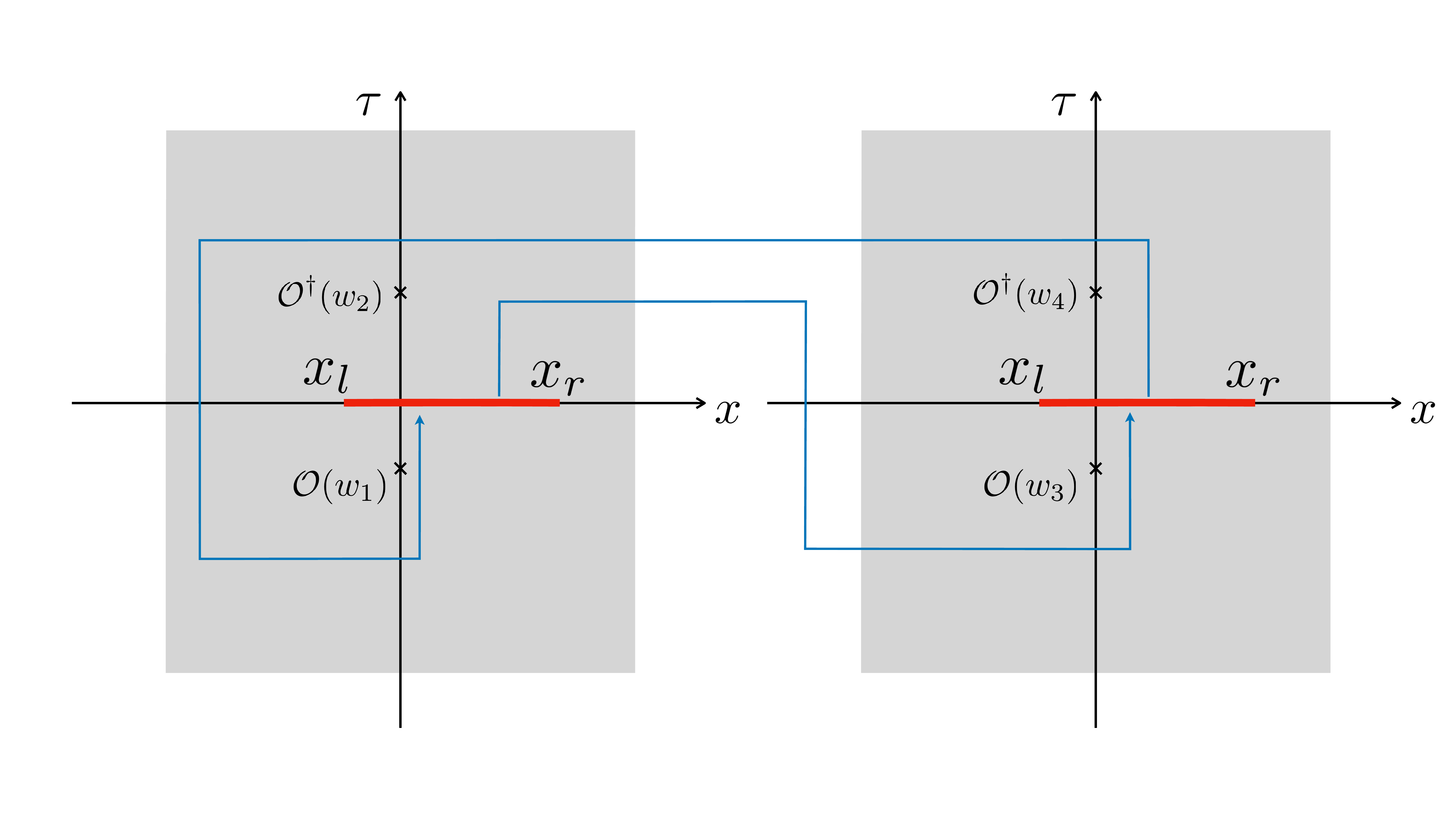}
    \caption{The manifold $\Sigma_2$ for the setting in section \ref{sec:samepoints}. The lower half expresses $\ket{\psi}$ and the upper half expresses $\bra{\vv}$, both of which is excited at the same space point but with different cutoffs.  The upper side of the subsystem in the first sheet is identified with the lower side of the subsystem in the second sheet. These correspond to matrix multiplication and tracing out respectively.}
    \label{fig:FREECFTA}
 \end{center}
\end{figure}
Now we consider the case where the same space points are excited with different cutoffs, i.e., we can parameterize as $w_1=-ia$ and $w_2=ia^{\prime}$, where $a,a^\prime>0$. Let us compute the variation of the second pseudo R\'{e}nyi entropy $\Delta S^{(2)}_A$ by replica method as in \cite{Nozaki,HNTW}. The manifold $\Sigma_2$ is depicted in figure \ref{fig:FREECFTA}. From the formula (\ref{localcpr}), the pseudo 2nd R\'{e}nyi entropy is 
\begin{equation}
  \Delta S^{(2)}_A=-\log\left(\frac{\langle\mathcal{O}(w_1,\bar{w}_1)\mathcal{O}^{\dagger}(w_2,\bar{w}_2)\mathcal{O}(w_3,\bar{w}_3)\mathcal{O}^{\dagger}(w_4,\bar{w}_4)\rangle_{\Sigma_2}}{(\langle\mathcal{O}(w_1,\bar{w}_1)\mathcal{O}^{\dagger}(w_2,\bar{w}_2)\rangle_{\Sigma_1})^2}\right).
\end{equation}
To compute the expectation values in the logarithm, we consider a conformal mapping 
\begin{equation} \label{eq:conformalmap1}
  \frac{w-x_l}{w-x_r}=z^2.
\end{equation}
This maps $\Sigma_2$ to a Riemann surface $\Sigma_1$. Let $z_i\in\Sigma_1\ (i=1,\ldots,4)$ be points mapped by (\ref{eq:conformalmap1}) from $w_i\in\Sigma_2\ (i=1,\ldots,4)$ respectively, then there are the relations
\begin{align}
  z_1=-z_3=\sqrt{\frac{w_1-x_l}{w_1-x_r}}, \\
  z_2=-z_4=\sqrt{\frac{w_2-x_l}{w_2-x_r}}.
\end{align}
In terms of $z_i$, $\Delta S^{(2)}_A$ turns out to be
\begin{equation} \label{eq:CFTRenyi1}
  \Delta S^{(2)}_A=\log\left(\frac{2}{1+|\eta|+|1-\eta|}\right),
\end{equation}
where $\eta$ is the cross ratio defined as 
\begin{equation}
  \eta=\frac{(z_1-z_2)(z_3-z_4)}{(z_1-z_3)(z_2-z_4)}.
\end{equation}
When both $a$ and $a'$ are real (i.e. the Euclidean time evolution), we always find $\Delta S^{(2)}_A\leq 0$. If we set $a=a'$, the pseudo 2nd R\'{e}nyi entropy reduces to the ordinary 2nd R\'{e}nyi entropy. Therefore $\Delta S^{(2)}_A=0$ when $a=a'$ \cite{Nozaki,HNTW}.

\begin{figure}[H]
    \centering
    \includegraphics[width=10cm]{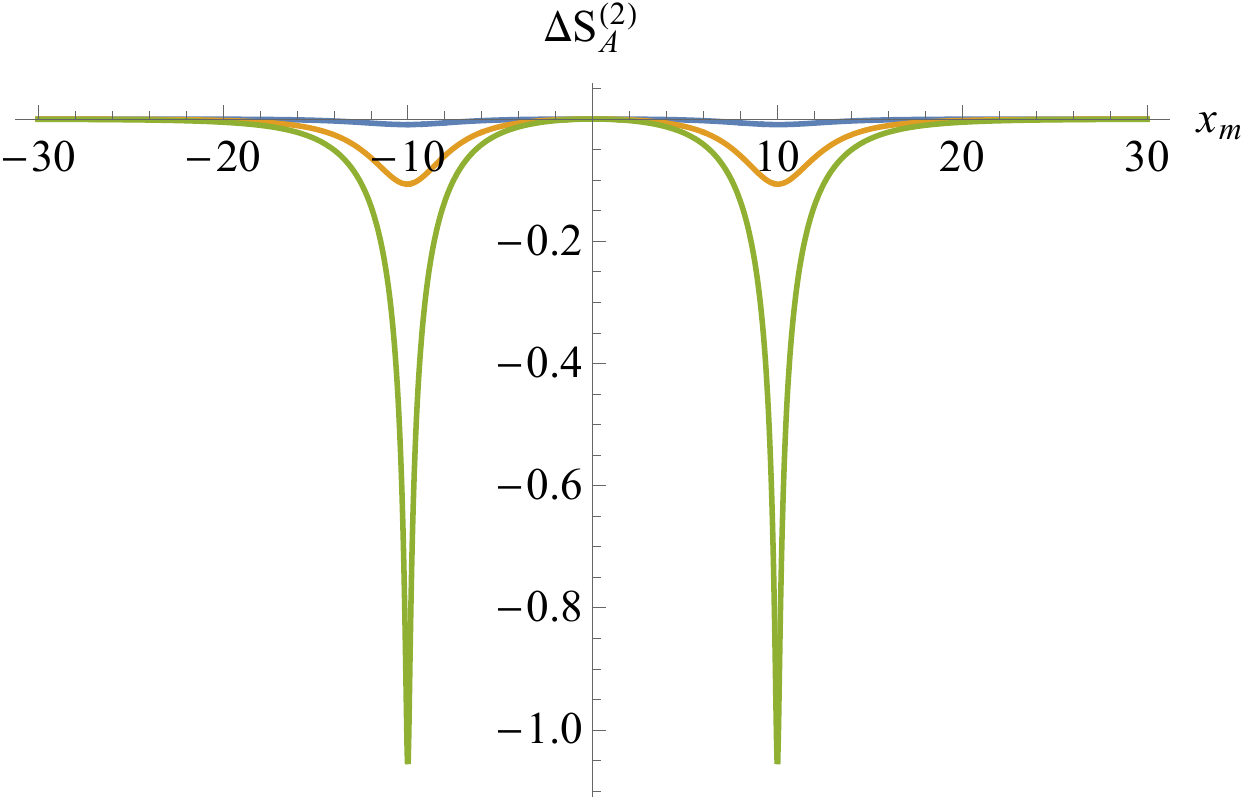}
    \caption{$\Delta S^{(2)}_A$ for subsystems centered at different space points with length $l=20$. Blue line: $a=4,a^\prime=6$, orange line: $a=2,a^\prime=8$, green line: $a=0.1,a^\prime=9.9$ case.}
    \label{fig:FREECFTB}
\end{figure}

Let us investigate the behavior of $\Delta S^{(2)}_A$ in an explicit example. Figure \ref{fig:FREECFTB} shows $\Delta S^{(2)}_A$ for subsystems centered at different points $x=x_m(\equiv\frac{x_l+x_r}{2})$ with length $l=20$, where $l\equiv x_r-x_l$. We can see that  $\Delta S^{(2)}_A$ sharply decreases when either of the operator insertion points become closer to the boundary of the interval $A$. This behavior can be understood from the fact that the pseudo entropy is reduced when we consider the entanglement swapping as we saw in section \ref{entwasps}. Indeed, when the boundary of $A$ gets close to one of the operators, the system experiences an entanglement swapping. On the other hand, if we insert an operator in the middle of the interval $A$, since the swapping does not occur near the boundaries of $A$, this does not contribute to $\Delta S^{(2)}_A$.

We can also analyze the real time evolution by setting the Euclidean times as $a=\delta+it_1$ and $a^\prime=\delta-it_2$, where $\delta(>0)$ is an infinitesimally small parameter which regularize the local quench. When $0<x_l<t_{1,2}<x_r$, we find $\Delta S^{(2)}_A=\log 2$, which is interpreted as the entanglement entropy for the Bell pair as in \cite{Nozaki,HNTW}.

\subsection{Example 2: Exciting Different Points with the Same Cutoff} \label{sec:sametime}
\begin{figure}[H]
  \begin{center}
    \includegraphics[width=0.9\linewidth]{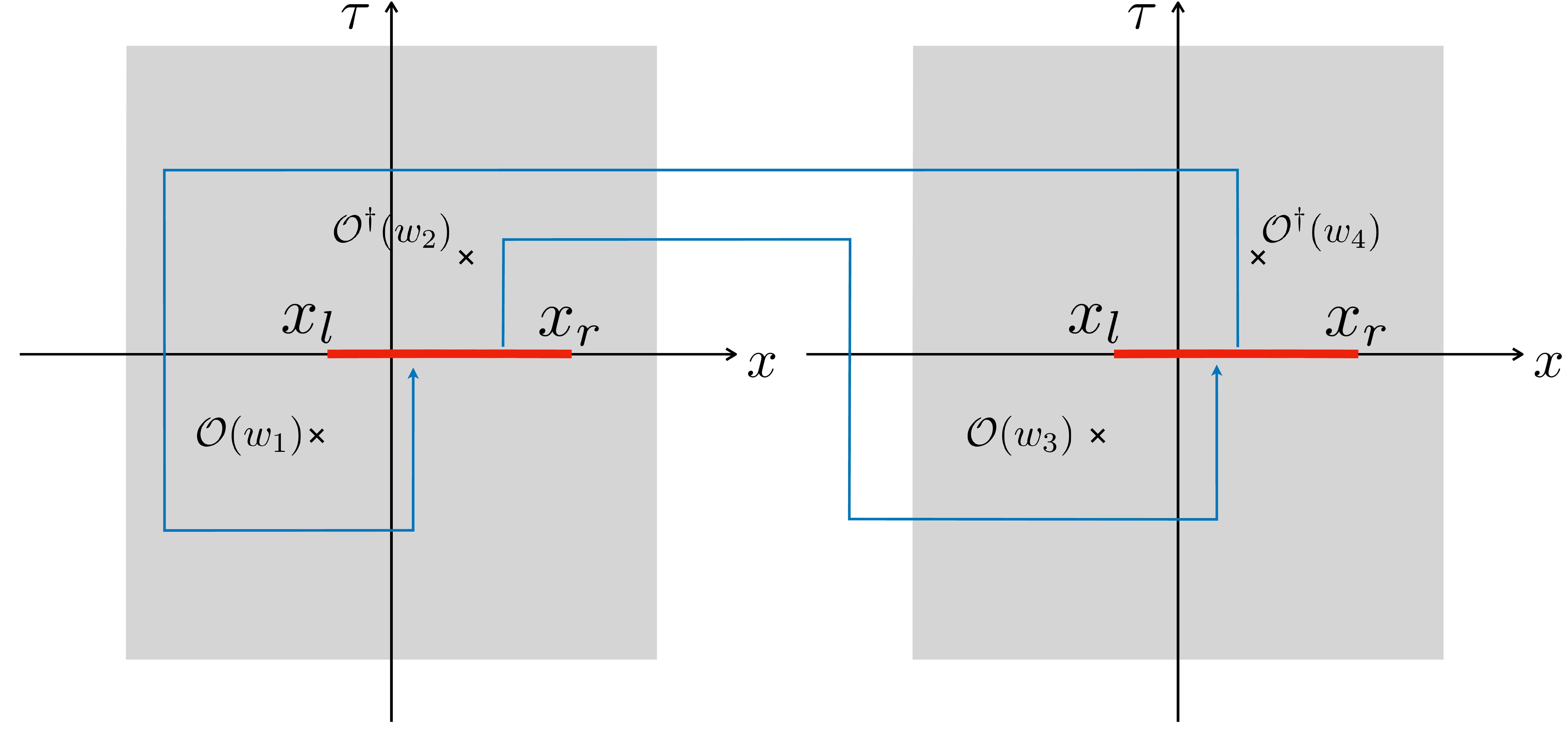}
  \end{center}
  \caption{The manifold $\Sigma_2$ for the setting in section \ref{sec:sametime}.}
  \label{fig:FREECFTD}
\end{figure}
We consider the same states as (\ref{eq:CFTstates}) and this time we excite two different space points with different cutoffs, i.e. $w_1=-d-ia$, $w_2=d+ia$ as depicted in figure \ref{fig:FREECFTD}. Since the difference from the previous case is only the location of the excitations, $\Delta S^{(2)}_A$ is the same as (\ref{eq:CFTRenyi1}).

\begin{figure}[H]
    \centering
    \includegraphics[width=10cm]{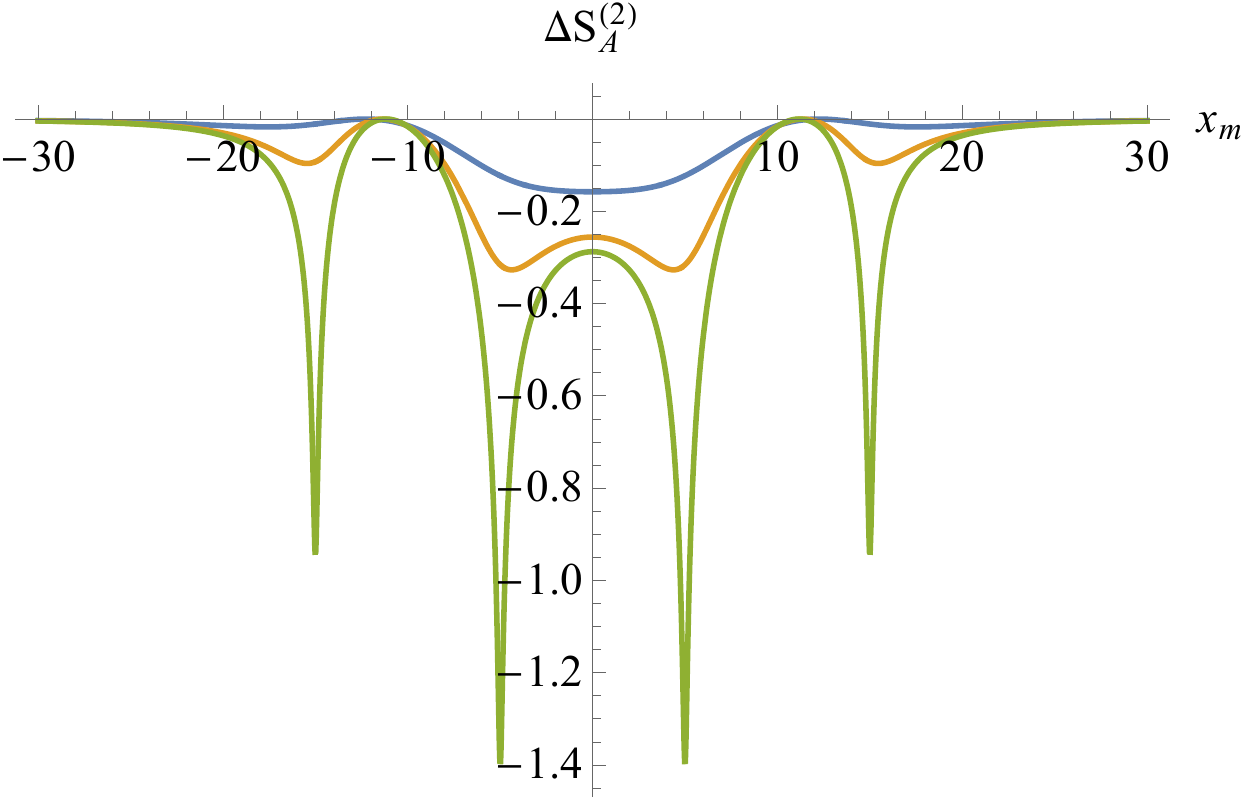}
    \caption{$\Delta S^{(2)}_A$ for subsystems centered at different points with length $l=20$. We set the points of excitation at $d=5$. Blue line: $a=5$, orange line: $a=2$, green line: $a=0.1$ case.}
    \label{fig:FREECFTC}
\end{figure}

Figure \ref{fig:FREECFTC} shows the behavior of $\Delta S^{(2)}_A$ as a function of the center of subsystems $x_m$. We again see that $\Delta S^{(2)}$ sharply decreases when an end point of the interval $A$ is close to either of the two operator insertion points, which can be interpreted as in the previous example.

\subsection{Example 3: Excitations by Different Operators}
In this section, we would like to consider an example of excitations at the same space points by different operators, which is similar to the setup in figure \ref{fig:FREECFTA}. We set $w_1=-ia$ and $w_2=ia'$, and let two states be 
\begin{align}
  &\ket{\psi}=e^{-a H_{\rm CFT}}\tilde{\mathcal{O}}(x=0)\ket{0}, \nonumber \\
  &\ket{\vv}=e^{-a^\prime H_{\rm CFT}}\mathcal{O}(x=0)\ket{0}, 
\end{align}
where $\mathcal{O}$ is same as defined in (\ref{eq:belloperator1}) and $\tilde{\mathcal{O}}$ is defined as 
\begin{align}
  \tilde{\mathcal{O}}=e^{\frac{i}{2}\phi}+e^{i\theta}e^{-\frac{i}{2}\phi}.
\end{align}
Its conformal dimension is also $h=\bar{h}=1/8$ and $\theta\in[-\pi,\pi]$. This is analogous to the states in 
(\ref{twoqttt1}). $\Delta S^{(2)}_A$ can be computed by replica method again, we have
\begin{equation} 
  \Delta S^{(2)}_A=-\log\left(\frac{\langle\tilde{\mathcal{O}}(w_1,\bar{w}_1)\mathcal{O}^{\dagger}(w_2,\bar{w}_2)\tilde{\mathcal{O}}(w_3,\bar{w}_3)\mathcal{O}^{\dagger}(w_4,\bar{w}_4)\rangle_{\Sigma_2}}{(\langle\tilde{\mathcal{O}}(w_1,\bar{w}_1)\mathcal{O}^{\dagger}(w_2,\bar{w}_2)\rangle_{\Sigma_1})^2}\right).
\end{equation}
Computing the expectation values in the logarithm by using the conformal mapping (\ref{eq:conformalmap1}), 
\begin{equation} \label{eq:CFTRenyi2}
  \Delta S^{(2)}_A=\log\left(\frac{1+\cos\theta}{\cos\theta+|\eta|+|1-\eta|}\right),
\end{equation}
which reduces to (\ref{eq:CFTRenyi1}) when $\theta=0$. From (\ref{eq:CFTRenyi2}), $\Delta S^{(2)}_A$ is always negative.
\begin{figure}[t]
  \begin{center}
    \includegraphics[width=7.5cm]{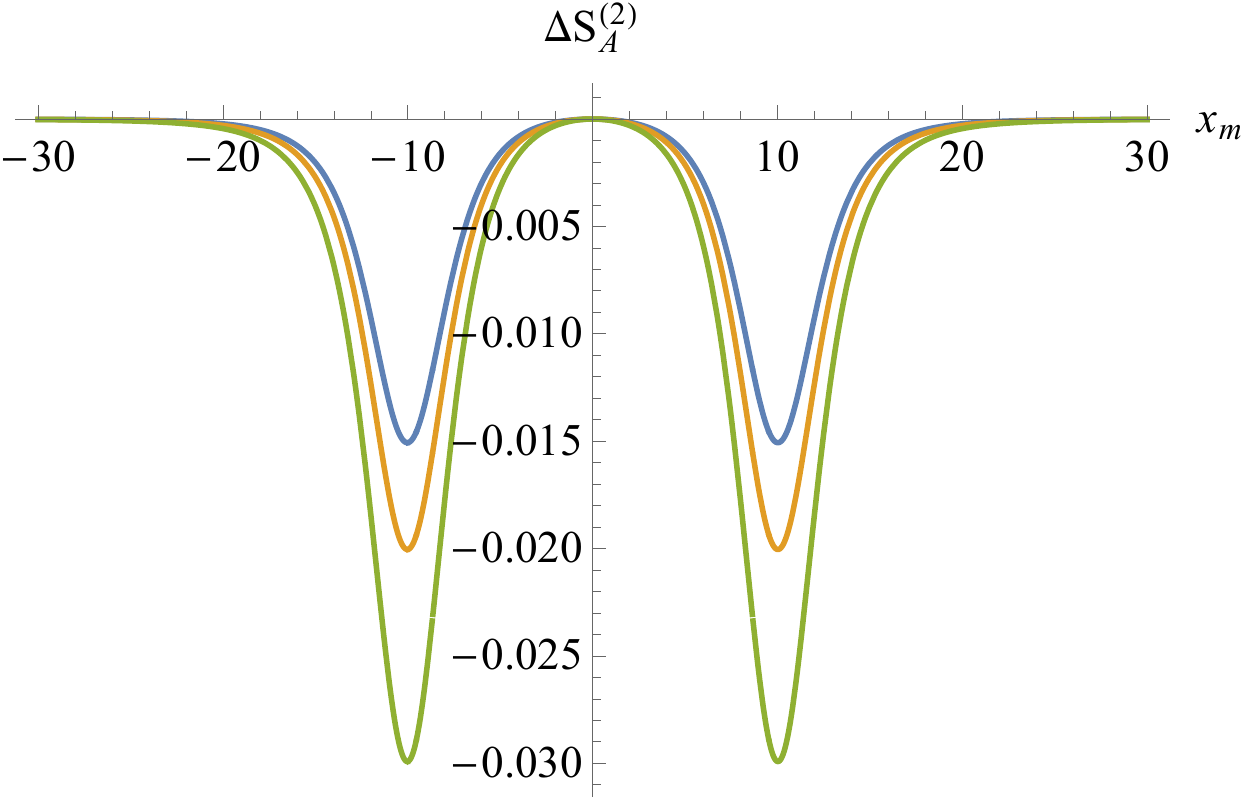}
    \includegraphics[width=7.5cm]{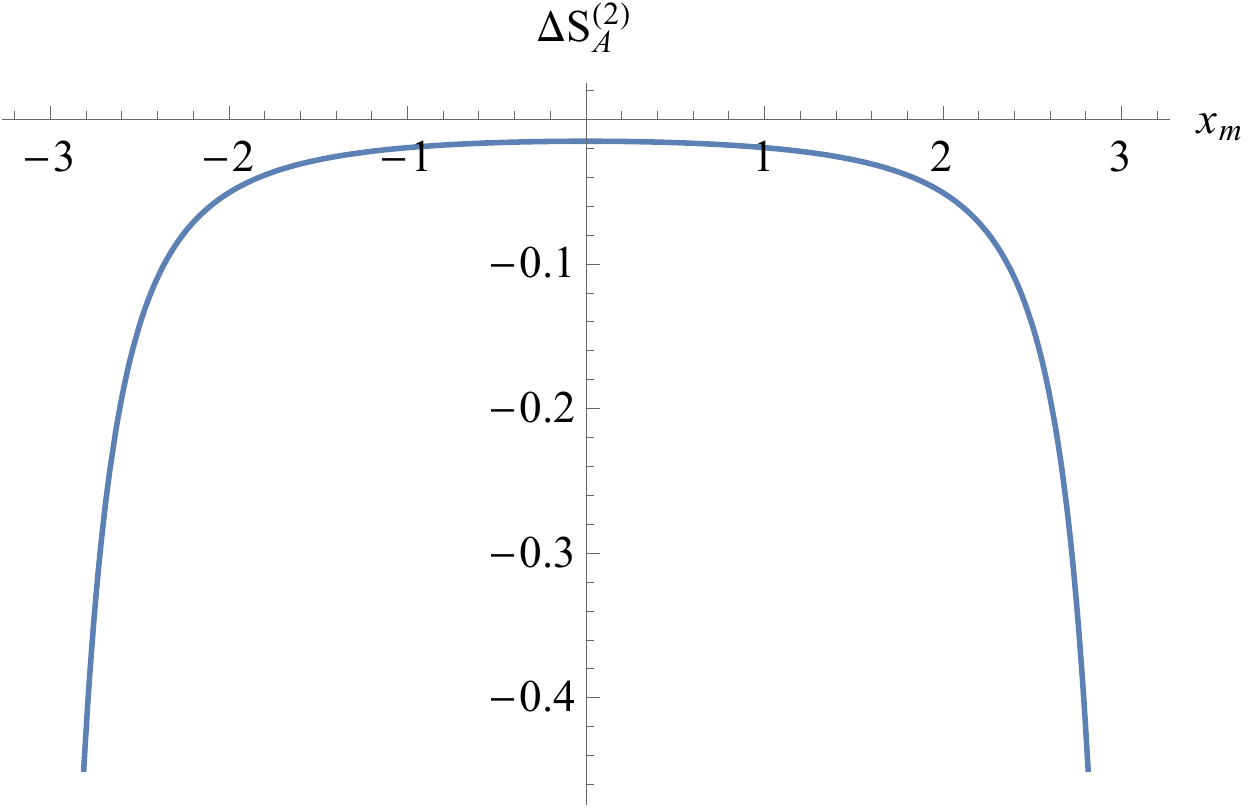}
  \end{center}
  \caption{The left figure shows the behavior of $\Delta S^{(2)}_A$ as functions of the center of the interval $A$ i.e. $x_m$. Here, we take the length of the subsystem to be $l=20$. We fix the points of excitations at $a=3,a'=5$. Blue line: $\theta=0$, orange line: $\theta=\pi/3$, green line: $\theta=\pi/2$ case. The right figure shows the behavior of $\Delta S^{(2)}_A$ as a function of the phase $\theta$ at $x_m=10$ and $l=20$. We also fix the points of excitations at $a=3,a'=5$. $\Delta S^{(2)}_A$ diverges at $\theta=\pm\pi$.}
  \label{fig:FREECFTE}
\end{figure}
Figure \ref{fig:FREECFTE} shows the behavior of $\Delta S^{(2)}_A$ when we change the phase $\theta$ of $\tilde{\mathcal{O}}$. The left panel shows $\Delta S^{(2)}_A$ as functions of the center of subsections $x_m$ and the right as a function of the phase $\theta$. We can see that $\Delta S^{(2)}_A$ becomes smaller when the ``difference" between $\mathcal{O}$ and $\tilde{\mathcal{O}}$ is larger. 

In the Lorentzian time evolution, we set $a=\delta+it$ and $a'=\delta-it$. If we take the limit $\delta\to0$ and assume the range $0<x_l<t<x_r$, which is equivalent to the limit $(\eta,\bar{\eta})\to(1,0)$, we obtain 
\begin{equation}
  \Delta S^{(2)}_A=\log\left(\frac{1+\cos\theta}{\cos\theta}\right).
\end{equation}
This agrees with the previous result (\ref{twoqtt}) of a two qubit system as expected.

Moreover, we can compute the pseudo 3rd R\'{e}nyi entropy by replica method. In this case, we consider the manifold $\Sigma_3$ composed by three sheets. Then we have 
\begin{equation}
  \Delta S^{(3)}_A=-\frac{1}{2}\log\left(\frac{\langle\tilde{\mathcal{O}}(w_1,\bar{w}_1)\mathcal{O}^{\dagger}(w_2,\bar{w}_2)\tilde{\mathcal{O}}(w_3,\bar{w}_3)\mathcal{O}^{\dagger}(w_4,\bar{w}_4)\tilde{\mathcal{O}}(w_5,\bar{w}_5)\mathcal{O}^{\dagger}(w_6,\bar{w}_6)\rangle_{\Sigma_3}}{(\langle\tilde{\mathcal{O}}(w_1,\bar{w}_1)\mathcal{O}^{\dagger}(w_2,\bar{w}_2)\rangle_{\Sigma_1})^3}\right).
\end{equation}
The expectation values are computed in a Riemann surface mapped by conformal mapping 
\begin{align}
  \frac{w-x_l}{w-x_r}=z^3,
\end{align}
then the new coordinates have the following relations:
\begin{align}
  z_1&=e^{-\frac{2}{3}\pi i}z_3=e^{\frac{2}{3}\pi i}z_5=\left(\frac{w_1-x_l}{w_1-x_r}\right)^{1/3}, \\
  z_2&=e^{-\frac{2}{3}\pi i}z_4=e^{\frac{2}{3}\pi i}z_6=\left(\frac{w_2-x_l}{w_2-x_r}\right)^{1/3}.
\end{align}
We define cross ratios 
\begin{equation}
  \eta_{ij}^{kl}=\frac{(z_i-z_j)(z_k-z_l)}{(z_i-z_k)(z_j-z_l)},
\end{equation}
where the subscripts and superscripts are $1\le i,j,k,l\le6$. In terms of $\eta_{ij}^{kl}$, $\Delta S^{(3)}_A$ can be computed as 
\begin{equation}
  \Delta S^{(3)}_A=\frac{1}{2}\log\left(\frac{\sqrt{2}(1+\cos\theta)^\frac{3}{2}}{\cos\frac{3}{2}\theta+3\cos\frac{\theta}{2}(|\eta_{14}^{32}|^2+|\eta_{56}^{14}|^2+|\eta_{32}^{56}|^2)}\right).
\end{equation}
\begin{figure}[t]
  \begin{center}
    \includegraphics[width=7.5cm]{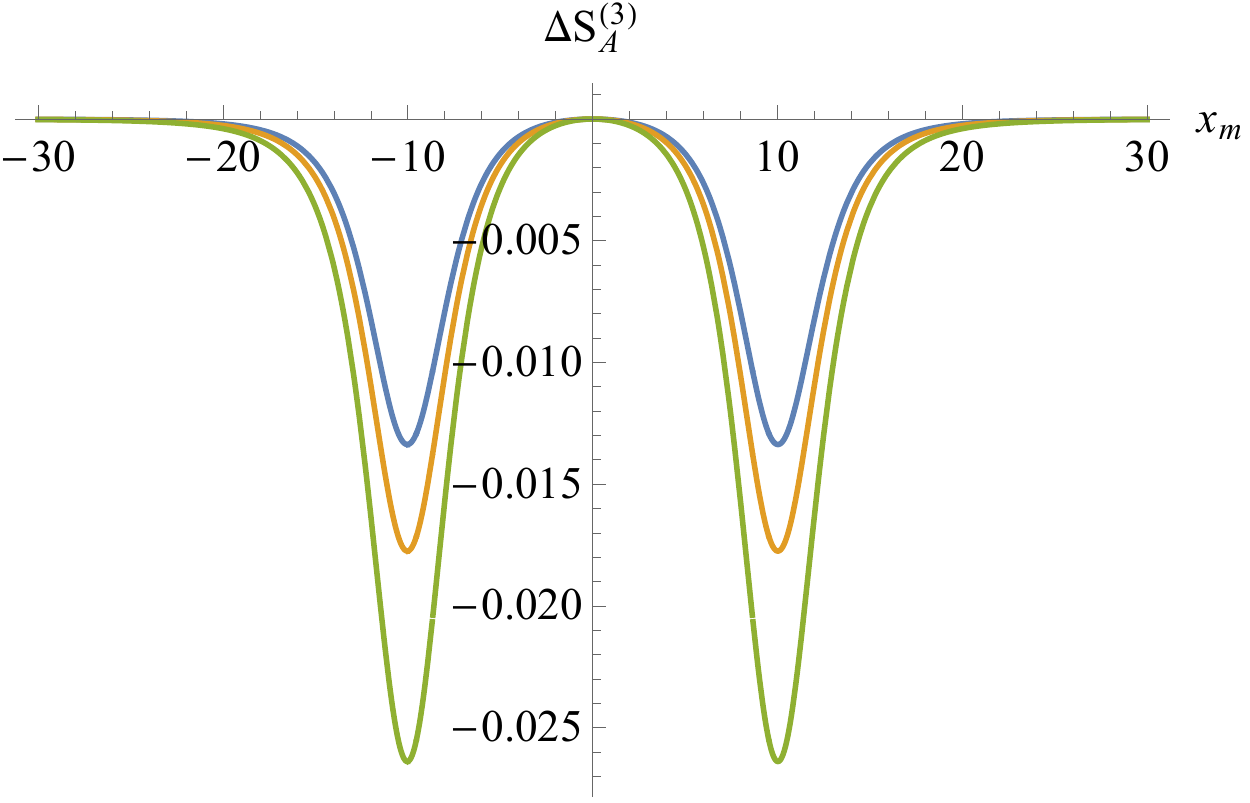}
    \includegraphics[width=7.5cm]{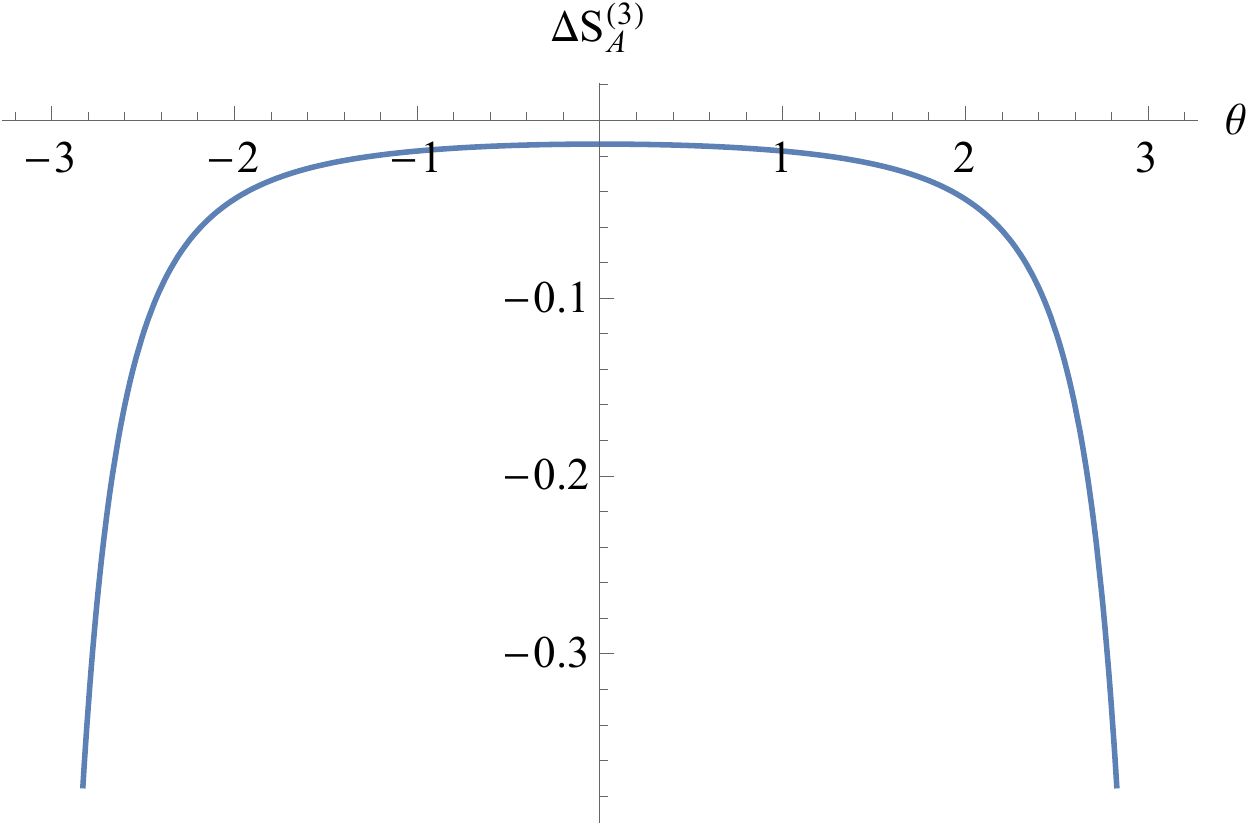}
    \caption{The left figure shows the behavior of $\Delta S^{(3)}_A$ as functions of the center of the interval $A$ i.e. $x_m$. Here, we take the length of the subsystem to be $l=20$. We fix the points of excitations at $a=3,a'=5$. Blue line: $\theta=0$, orange line: $\theta=\pi/3$, green line: $\theta=\pi/2$ case. The right figure shows the behavior of $\Delta S^{(3)}_A$ as a function of the phase $\theta$ at $x_m=10$ and $l=20$. We also fix the points of excitations at $a=3,a'=5$. $\Delta S^{(3)}_A$ diverges at $\theta=\pm\pi$.}
    \label{fig:FREECFTF}
  \end{center}
\end{figure}

The behavior of $\Delta S^{(3)}_A$ is similar to that of $\Delta S^{(2)}_A$. Figure \ref{fig:FREECFTF} shows the behavior of $\Delta S^{(3)}_A$, which are very similar to the graphs in figure \ref{fig:FREECFTE}. 

In the Lorenzian time evolution $0<x_l<t<x_r$ with the limit $\delta\to0$, we find 
\begin{equation}
  \Delta S^{(3)}_A=\frac{1}{2}\log\left(\frac{\sqrt{2}(1+\cos\theta)^{\frac{3}{2}}}{\cos{\frac{3}{2}\theta}}\right),
\end{equation}
which again reproduces the two qubit result (\ref{twoqttt}) at $n=3$.

\section{Aspects of Holographic Pseudo Entropy}\label{secHPE}

In this section we will explore properties of holographic pseudo entropy defined by the minimal area formula (\ref{eq:HEPE}) from various viewpoints including explicit examples.

\subsection{Nonnegativeness of Holographic Pseudo (R\'enyi) Entropy}\label{HEPEderive}

One surprising prediction of the holographic pseudo entropy is that this quantity is real and non-negative in the classical gravity dual calculation when we assume the bulk metric is real valued, in spite of the fact that the pseudo entropy $S(\CT^{\psi|\vv}_A)$ takes complex values for generic choices of $|\psi\lb$ and $|\vv\lb$. Moreover, a further holographic consideration shows that pseudo (R\'enyi) entropies are also real and non-negative in holographic computations:
\begin{align}
    ~^\forall n>0, ~S^{(n)}(\CT^{\psi|\vv}_A) \geq 0. 
    \label{eq:HolWEpos}
\end{align}
To see this, similar to \cite{Dong16}, we compute a refined version of the pseudo R\'enyi entropy holographically as
\begin{align}
    \tilde{S}^{(n)}(\CT^{\psi|\vv}_A) \equiv n^2\frac{\partial}{\partial n}\left(\frac{n-1}{n}S^{(n)}(\CT^{\psi|\vv}_A)\right) = \min_{\Gamma_A^{(n)}}\left(\frac{{\rm Area~of~}\Gamma^{(n)}_A }{4G_N}\right), 
    \label{eq:HolWRE}
\end{align}
where $\Gamma^{(n)}_A$ is a codimension-2 cosmic brane with tension
\begin{align}
    T_n = \frac{1}{4G_N}\frac{n-1}{n}.
\end{align}
and satisfies $\partial\Gamma_A^{(n)} = \partial A$. Therefore, $\tilde{S}^{(n)}(\CT^{\psi|\vv}_A)$ is supposed to be nonnegative for any $n>0$. Since the pseudo R\'enyi entropy is just given by the following integral \begin{align}
    S^{(n)}(\CT^{\psi|\vv}_A) = \frac{n}{n-1}\int_1^{n}dn'\frac{\tilde{S}^{(n)}(\CT^{\psi|\vv}_A)}{n'^2},
\end{align}
the non-negativity (\ref{eq:HolWEpos}) is shown. 

The inequalities (\ref{eq:HolWEpos}) give strong restrictions in the sense that holographic transition matrices belong to 
(at least) class $\mathscr{B}$. This can be regarded as a new characterization of holographic states.

\subsection{Subadditivity and Strong Subadditivity}

Since the holographic pseudo entropy is real and non-negative, we can think of possibilities of other inequalities.
It is clear to confirm the subadditivity expressed as 
\ba
 S(\CT^{\psi|\vv}_A)+S(\CT^{\psi|\vv}_B)- S(\CT^{\psi|\vv}_{AB})\geq 0.
\ea
Remember that the subadditivity is not always satisfied even if the eigenvalues of transition matrices are real and non-negative as we saw in section \ref{sec:vssa}.

Next, let us turn to the strong subadditivity 
\ba
 S(\CT^{\psi|\vv}_{AB})+S(\CT^{\psi|\vv}_{BC})- S(\CT^{\psi|\vv}_B)- S(\CT^{\psi|\vv}_{ABC})\stackrel{?}{\geq} 0
\label{ssah}
\ea
In the static background, we can prove the strong subadditivity immediately by restricting to a canonical time slice  \cite{Headrick:2007km}. In the Lorentzian signature, namely the covariant holographic entanglement entropy \cite{HRT}, we can employ the property that the minimal area on a time slice gets smaller as we deform the time slice away from the one which includes the true extremal surface, even if the all relevant extremal surfaces are not on the same time slice, as proved in \cite{Wall:2012uf}. 

However, in our Euclidean time-dependent setup, if we allow to deform the time slice to an arbitrary shape, which is dual to a quantum state via the Euclidean path-integral in CFT, we will immediately find that 
the strong subadditivity (\ref{ssah}) can be violated as follows\footnote{We would like to thank very much a referee of this article for pointing out this important fact.}. Consider a static AdS$_3$ for simplicity and take a canonical time slice. We define four points $(P,Q,R,S)$ which are lining up in this order on the time slice. If we take the subsystems $A=[P,Q],B=[Q,R]$ and $C=[R,S]$, then the strong subadditivity follows based on the argument of \cite{Headrick:2007km}. However, if we choose a distorted slice which passes through the points in the order $(P,Q,S,R)$, then it is obvious that the strong subadditivity is violated
because the strong subadditivity takes the opposite `wrong' form  $S(\CT^{\psi|\vv}_{AB})\!+\!S(\CT^{\psi|\vv}_{BC})\!\leq \! S(\CT^{\psi|\vv}_B)\!+\!S(\CT^{\psi|\vv}_{ABC})$. This problem might suggest that the notion of  strong subadditivity is not useful in Euclidean setups. 

Nevertheless, thinking of possible inequality analogous to the ordinary strong subadditivity in Lorentzian setups, we may focus on the strong subadditivity of 
holographic pseudo entropy by restricting the time slice to the canonical one on the AdS boundary, assuming 
that the AdS boundary is the flat space in this paper. Even under this restricted definition, we cannot 
derive the strong subadditivity of the holographic pseudo entropy because
the minimal area gets larger if we deform the time slice, which prohibits a similar proof \cite{Wall:2012uf}.
Let us mention, however,  that we did not find any violation of (\ref{ssah}) for explicit examples of classical gravity duals we studied. If we can consider a quantum mechanical superposition of several classical geometries, then as we will show in section \ref{subsec:linearity}, we will find examples which violate these inequalities as coefficients of linear combinations allow complex numbers.

\subsection{A Simple Example: Janus AdS/CFT}

As a simple example, we would like to analyze the holography for the Janus solutions which are dual to interface CFTs. We focus on a Janus solution in AdS$_3$/CFT$_2$ found in \cite{Bak:2007jm}.

\subsubsection{Holographic Pseudo Entropy in Janus AdS/CFT}
The Janus solution to three dimensional AdS gravity coupled to a scalar field:
\ba
I=\frac{1}{16\pi G_N}\int d^3 x\s{g}(R-\de_a\phi\de^a\phi+2),
\ea
is given by 
\ba
&& ds^2=d\rho^2+f(\rho)\frac{dx^2+dy^2}{y^2}, \no
&& \phi=\phi_0+\frac{1}{\s{2}}\log\left[\frac{1+\s{1-2\gamma^2}+\s{2}\gamma\tanh \rho}
{1+\s{1-2\gamma^2}-\s{2}\gamma \tanh \rho}\right],\no
&& f(\rho)=\frac{1}{2}\left(1+\s{1-2\gamma^2}\cosh 2\rho\right), \label{januss}
\ea
where $0\leq \rho\leq \infty$.
If we set $\gamma=0$, this solution is equivalent to the ordinary Poincare AdS$_3$ solution 
$ds^2=\frac{dz^2+d\tau^2+dx^2}{z^2}$.

If we write the exactly marginal operator (i.e. dimension two primary operator) dual to the bulk massless scalar field $\phi$ as $O(x)$, then the above Janus solution is dual to a two dimensional holographic CFT
deformed by $J_+\int dx^2O(x)$ in the region $\tau>0$ and by 
$J_-\int dx^2O(x)$ in the region $\tau<0$, where $J_\pm=\lim_{\rho\to\pm\infty}\phi(\rho)$.
These two deformations produce different states $\la \psi_+|$ and $|\psi_-\lb$, respectively and the total gravity partition function is equal to the inner product $\la \psi_+|\psi_-\lb$.

The time slice $\tau=0$ is equivalent to $\rho=0$ in (\ref{januss}).
By requiring the usual UV cutoff condition $g_{xx}=\frac{1}{\ep^2}$ we find the cutoff of the $y$ coordinate :
$y>\delta$ is given by 
\ba
\delta= \ep\cdot \s{f(0)}=\s{\frac{1+\s{1-2\gamma^2}}{2}}\cdot \ep.
\ea
When $A$ is a length $l$ interval, the holographic pseudo entropy 
is calculated as the length of geodesic and this leads to
\ba
S(\CT^{\psi_-|\psi_+}_A)=\frac{c}{3}\cdot \s{\frac{1+\s{1-2\gamma^2}}{2}}\cdot \log\frac{l}{\delta}
\simeq \frac{c}{3}\left(1-\frac{\gamma^2}{4}\right)\log\frac{l}{\ep}+\frac{c}{12}\gamma^2. \label{JanusWEE}
\ea
Therefore we find the holographic pseudo entropy 
is reduced by $O(\gamma^2)$ compared with the $\gamma=0$ result (i.e. the vacuum case $|\psi_+\lb=|\psi_-\lb=|0\lb$) due to the interface perturbation. This result is consistent with the perturbative analysis in (\ref{pertwee}).

\subsubsection{CFT Perturbation Analysis}

Let us analyze the previous Janus setup from the CFT viewpoint. In the replica calculation of pseudo entropy 
(look at figure \ref{fig:rep2d}), we write the exactly marginal perturbation from $\phi=\phi_-$ to $\phi=\phi_+$ 
as $\int_{\Sigma_{n(+)}}dx^2 JO(x)$, 
where $J=J_+-J_{-}$ and  $\Sigma_n$ is the $n-$sheeted Riemann surface. Also we call the  upper half of $\Sigma_n$ as $\Sigma_{n(+)}$. The flat complex plane $\mathbb{R}^2$ is denoted by $\Sigma=\Sigma_{n=1}$.

In the path-integral, the pseudo entropy 
is computed via the replica method as follows
\ba
S^{(n)}_A=\frac{1}{1-n}\log  \left[\frac{\la e^{\int_{\Sigma_{n(+)}} dx^2 JO(x)}\lb_{\Sigma_n}}
{\left(\la  e^{\int_{\Sigma_{1(+)}} dx^2 JO(x)}\lb_{\Sigma_1}\right)^n}\right]. \label{ssw}
\ea
If we take the difference from the entanglement entropy of $\phi=\phi_-$ theory and perform the perturbation w.r.t $J$ up to the quadratic order, we get
\ba
\Delta S^{(n)}_A=\frac{J^2}{1-n}\left[\int dx^2 \int dy^2 \la O(x)O(y) \lb_{\Sigma_n}
-n\int dx^2 \int dy^2 \la O(x)O(y) \lb_{\Sigma_1} \right]. \label{tpintgfh}
\ea
The part $[\ddd]$ is estimated as (we introduce the point splitting cut off $\ep$)
\ba
[\ddd]=a_1\log\frac{l}{\ep}+O(1), \label{logdig}
\ea
where $a_{1}>0$ are positive constant and also $a_{1}\propto n-1$ as these divergences arise when 
both $x$ and $y$ are close to one of two end points of $A$. The divergences arise when the two points $x$ and $y$ get closer to each other geometrically. 
Though the integrals of the two point function 
in (\ref{tpintgfh}) include $O(\frac{l^2}{\ep^2})$ terms, they cancel with each other as this quadratic divergences
are also equally present in the flat space $\Sigma_1=\mathbb{R}^2$.  We can explicitly confirm that a logarithmic term in (\ref{logdig}) arises via the non-trivial conformal map (\ref{xxxcmap}), which vanishes at $n=1$.

Thus we have the estimation (up to an $O(1)$ factor):
\ba
\Delta S_A\sim -J^2\log\frac{l}{\ep}.
\ea
For the Janus setup the scalar action is normalized such that $J^2\sim c\gamma^2$. Therefore we can reproduce the behavior (\ref{JanusWEE}).

Note also that the terms (\ref{ssw}) are in agreement with (\ref{coefre}) in the general argument of perturbation theory given in appendix \ref{pertep}. When two $O$s are on the same sheet of 
$\Sigma_2$, we have the second term in   (\ref{coefre}), while not on the same sheet, we get the third term in   (\ref{coefre}).



\subsection{Pseudo Entropy of Locally Excited States in Holographic CFT}\label{sec:hpelocp}

In this subsection we will calculate holographic pseudo entropy for locally excited states \cite{Nozaki,HNTW} in two dimensional CFTs.
In AdS$_3/$CFT$_2$, it is known that the dual geodesic lengths in a gravity dual with local excitations agree with those obtained from the two point functions of light operators in an excited state with heavy operators in holographic CFTs \cite{Fitzpatrick:2014vua}. For example, the holographic entanglement entropy under local quenches \cite{NNT} can be perfectly reproduced from the CFT calculations \cite{Hat}. Therefore, below, we will present results of the pseudo entropy computed from a two dimensional holographic CFT. These are guaranteed to agree with those computed from the gravity dual using the formula (\ref{eq:HEPE}) and thus we will omit the gravity dual calculation below. Refer also to appendix \ref{opepea} for general analysis of pseudo entropy and fidelity for local operator excited states.

In the former sections, we perform the replica trick, i.e. compute the correlation functions on a replica manifold $\Sigma_n$ to get the pseudo $n$-th R\'{e}nyi entropy. In CFT$_2$, the replica trick admits a so-called twist operator formalism \cite{CCD07}. In this formalism, instead of computing correlation functions on $\Sigma_n$, we compute the correlation functions of an $n$ replicated CFT with twist operators inserted on the edges of $A$. More specifically, in our case where scalar primaries $\CO$ with conformal dimension $(h_\CO,h_\CO)$ are inserted, we can compute the variation of the pseudo $n$-th R\'{e}nyi entropy from that in the vacuum state by using the twist operator formalism:
\begin{align}
    \Delta S^{(n)}_A &= S^{(n)} (\CT^{\psi|\vv}_A) - S^{(n)} (\Tr_{A^c}|0\rangle\langle0|) \nonumber\\
    &= \frac{1}{1-n}\log \frac{\braket{\CO_n(z_1,\bar{z}_1)\sigma_n(z_2,\bar{z}_2) \tilde{\sigma}_n(z_3,\bar{z}_3)\CO_n(z_4,\bar{z}_4)}}{\braket{\CO_n(z_1,\bar{z}_1)\CO_n(z_4,\bar{z}_4)}\braket{\sigma_n(z_2,\bar{z}_2) \tilde{\sigma}_n(z_3,\bar{z}_3)}}. 
\end{align}
Here, the correlation functions are those in the $n$ replicated CFT whose central charge is $nc$, $\CO_n \equiv \CO\otimes\CO\otimes\cdots\otimes\CO$ is the replicated primary whose conformal weight is $(nh_\CO,nh_\CO)$ and $\sigma_n$($\bar{\sigma}_n$) is the $n$-th twist operator with conformal weight
\begin{align}
    (h_n,\bar{h}_n)=\left(\frac{c}{24}\left(n-\frac{1}{n}\right),\frac{c}{24}\left(n-\frac{1}{n}\right)\right). 
\end{align}
If we focus on the pseudo entropy which is given by $n\rightarrow1$ limit, we can regard $\CO_n$ as the heavy operator and $\sigma_n$ as the light operator and use the heavy-light 4-point function  \cite{Fitzpatrick:2014vua} to evaluate $\Delta S^{(n)}_A$. As the result, we have 
\begin{align}
    \Delta S_A \equiv \lim_{n\rightarrow 1} \Delta S^{(n)}_A &= \lim_{n\rightarrow1} \frac{1}{1-n} \log \left|\frac{\eta^{\frac{1-\a}{2}}(1-\eta^\a)}{\a(1-\eta)}\right|^{-4h_n} \nonumber\\
    &= \lim_{n\rightarrow1} \frac{1}{1-n} \log \left|\frac{\eta^{\frac{1-\a}{2}}(1-\eta^\a)}{\a(1-\eta)}\right|^{\frac{c}{6}\frac{1-n^2}{n}} \nonumber\\
    &= \frac{c}{6}\log \left|\frac{\eta^{\frac{1-\a}{2}}(1-\eta^\a)}{\a(1-\eta)}\right|^2,
    \label{eq:EEOO}
\end{align}
where $\a \equiv \sqrt{1-24h_\CO/c}$ and 
\begin{align}
    \eta \equiv \frac{z_{12} z_{34}}{z_{13} z_{24}}
\end{align}
is the cross ratio.

\subsubsection{Exciting the Same Space Point with Different Cutoffs}
\begin{figure}[H]
    \centering
    \includegraphics[width=10cm]{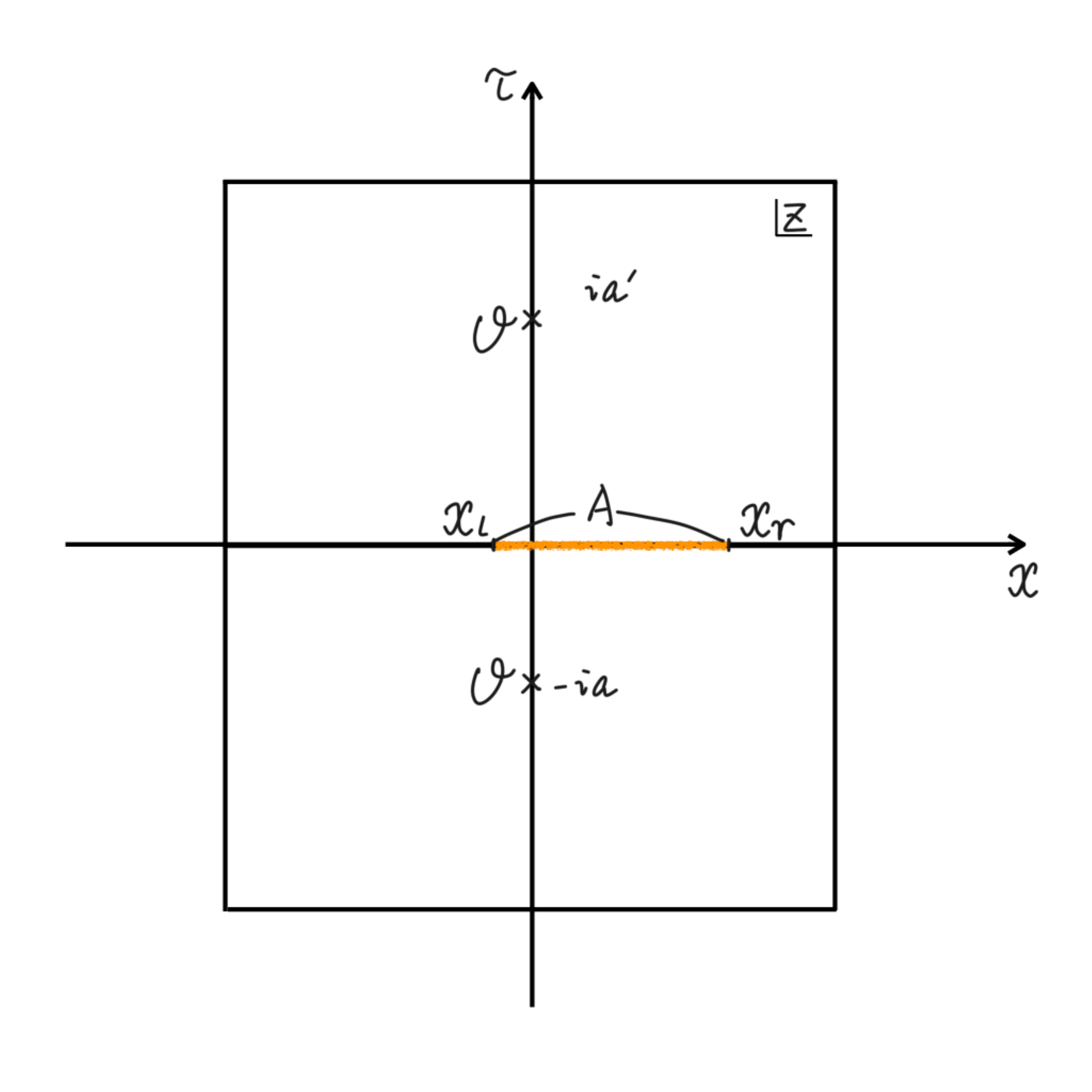}
    \caption{The Euclidean path integral setup. The orange line shows the subsystem $A$.}
    \label{fig:SE-SE-PI}
\end{figure}

\begin{figure}[H]
    \centering
    \includegraphics[width=10cm]{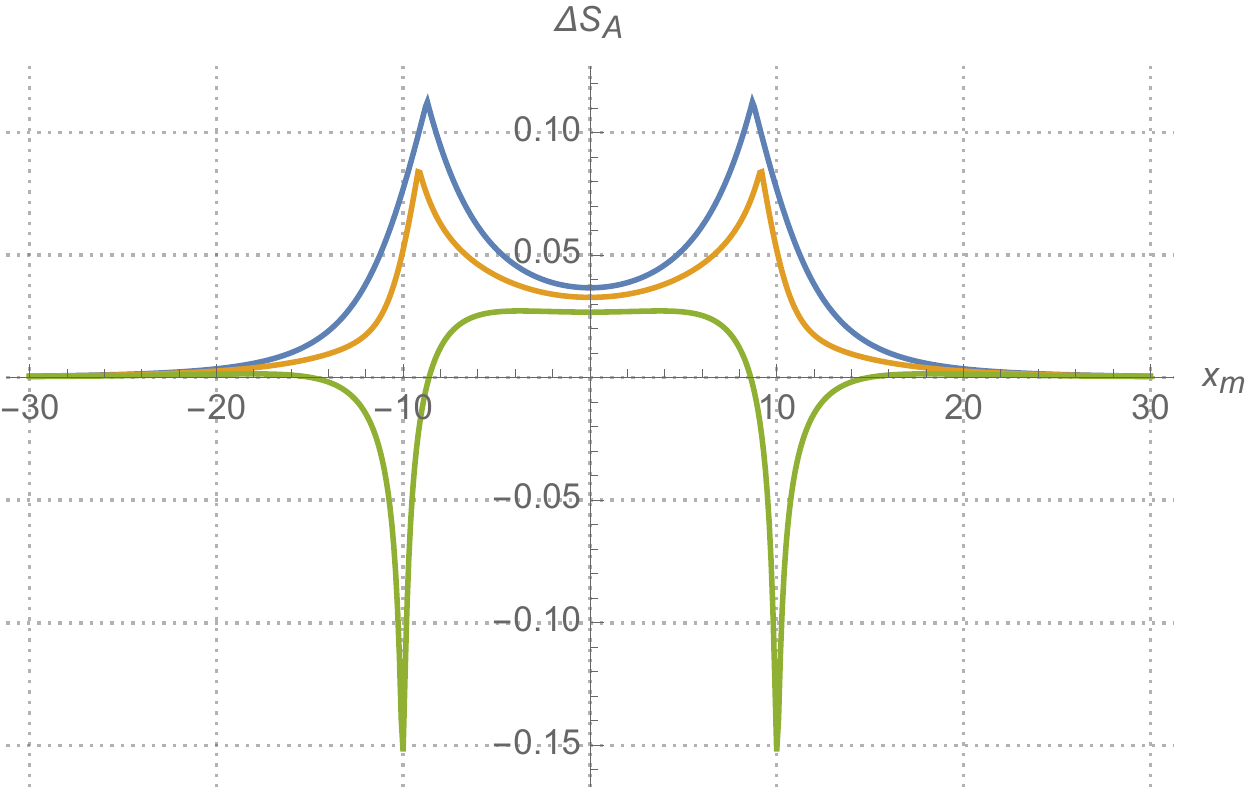}
    \caption{$\Delta S_A$ for subsystems centered at different points $x=x_m$ with length $l=20$. $\a$ is set to be $\a=1/2$. The blue line shows the $a=4$, $a'=6$ case, the orange line shows the $a=2$, $a'=8$ case and green line shows the $a=0.1$, $a'=9.9$ case.}
    \label{fig:SE-SE-VA-Hol}
\end{figure}

For the first example, let us consider the same setup as in section \ref{sec:samepoints}, i.e. we consider $\CT^{\psi|\vv}$ for two states 
\begin{align}
    \ket{\psi} &= e^{-aH_{\rm CFT}}\CO(x=0)\ket{0},\\
    \ket{\vv} &= e^{-a'H_{\rm CFT}}\CO(x=0)\ket{0}.
\end{align}
where $\ket{0}$ is the CFT ground state and $a$ ($a'$) are UV cutoffs introduced to avoid divergence. Here $\ket{\psi}$ is a state locally excited by a primary operators $\CO$ at $x=0$. We assume $\CO$ is a scalar primary with conformal weight $(h_\CO,h_\CO)$. The corresponding Euclidean path integral setup is shown in figure \ref{fig:SE-SE-PI}. Let us consider a connected subsystem $A=\{(x,\tau)|\tau=0,x_l\leq x\leq x_r\}$. 
In the current case
\begin{align}
    z_1 = -ia, ~z_2 = x_l, ~z_3 = x_r, ~z_4 =+ ia', \\
    \bar{z}_1 = +ia, ~\bar{z}_2= x_l, ~\bar{z}_3 = x_r, ~\bar{z}_4 =-ia'.
\end{align}
Figure \ref{fig:SE-SE-VA-Hol} shows $\Delta S_A$ for subsystems centered at different points $x=x_m$ with length $l=20$. 
From this plot, we can see that $\Delta S_A$ can be both positive and negative, in contrast with the behavior of free CFT in which $\Delta S_A^{(2)}$ is always negative (figure \ref{fig:FREECFTB}). On the other hand, if we focus on the case when one of the UV cutoff is very small (for example, the green line in figure \ref{fig:SE-SE-VA-Hol}), we can observe that $\Delta S_A$ sharply decreases when the edges of $A$ approach the excited space point. This is similar to $\Delta S_A^{(2)}$ in free CFT. 

Also note that, no matter how we choose the cutoff, $\Delta S_A$ at $x_m = 0$ is always positive. This can be easily understood from a holographic point of view. If we perform a conformal transformation and bring the two operator excitations to infinite past and infinite future respectively, then they give a Euclidean BTZ black hole microstate in the bulk with the two edges of $A$ lying on the same time slice \cite{Hat}. The geodesic connecting the two edges of $A$ should be larger in a BTZ black hole geometry than in a global AdS geometry. This observation immediately gives $\Delta S_A > 0$ at $x_m = 0$.

\subsubsection{Exciting Different Space Points with the Same Cutoff}
\begin{figure}[H]
    \centering
    \includegraphics[width=10cm]{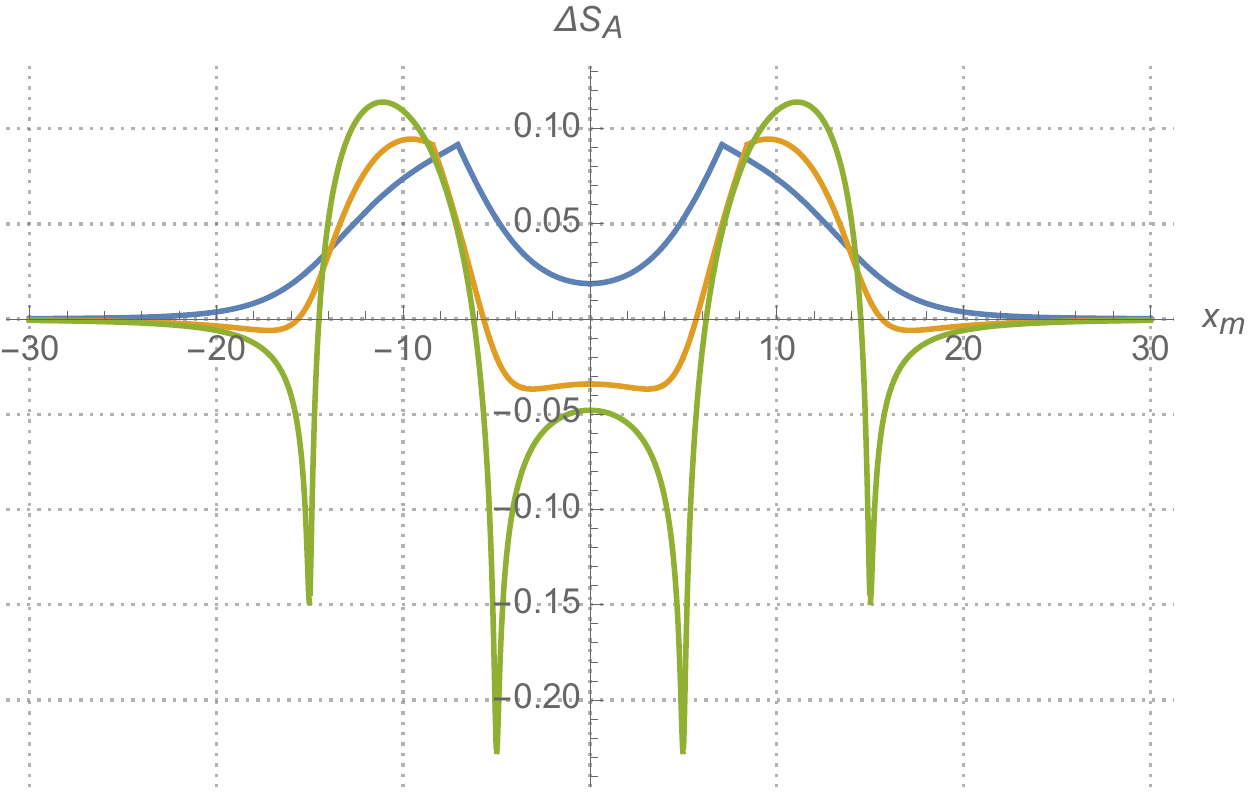}
    \caption{$\Delta S_A$ for subsystems centered at different points $x=x_m$ with length $l=20$, $d=5$, $\a=1/2$. The blue line shows the $a=5$ case, the orange line shows the $a=2$ case, and the green line shows the $a=0.1$ case.}
    \label{fig:SE-SE-HA-Hol}
\end{figure}

As one more example, let us consider the same setup as in section \ref{sec:sametime}, i.e. we consider $\CT^{\psi|\vv}$ for two states 
\begin{align}
    \ket{\psi} &= e^{-aH_{\rm CFT}}\CO(x=-d)\ket{0},\\
    \ket{\vv} &= e^{-aH_{\rm CFT}}\CO(x=+d)\ket{0}.
\end{align}
where $\ket{0}$ is the CFT ground state and $a$ is a UV cutoff to avoid divergence. Here $\ket{\psi}$ ($\ket{\vv}$) is a state locally excited by a primary operators $\CO$ on $x=-d$ ($x=+d$). 
\begin{figure}[H]
    \centering
    \includegraphics[width=10cm]{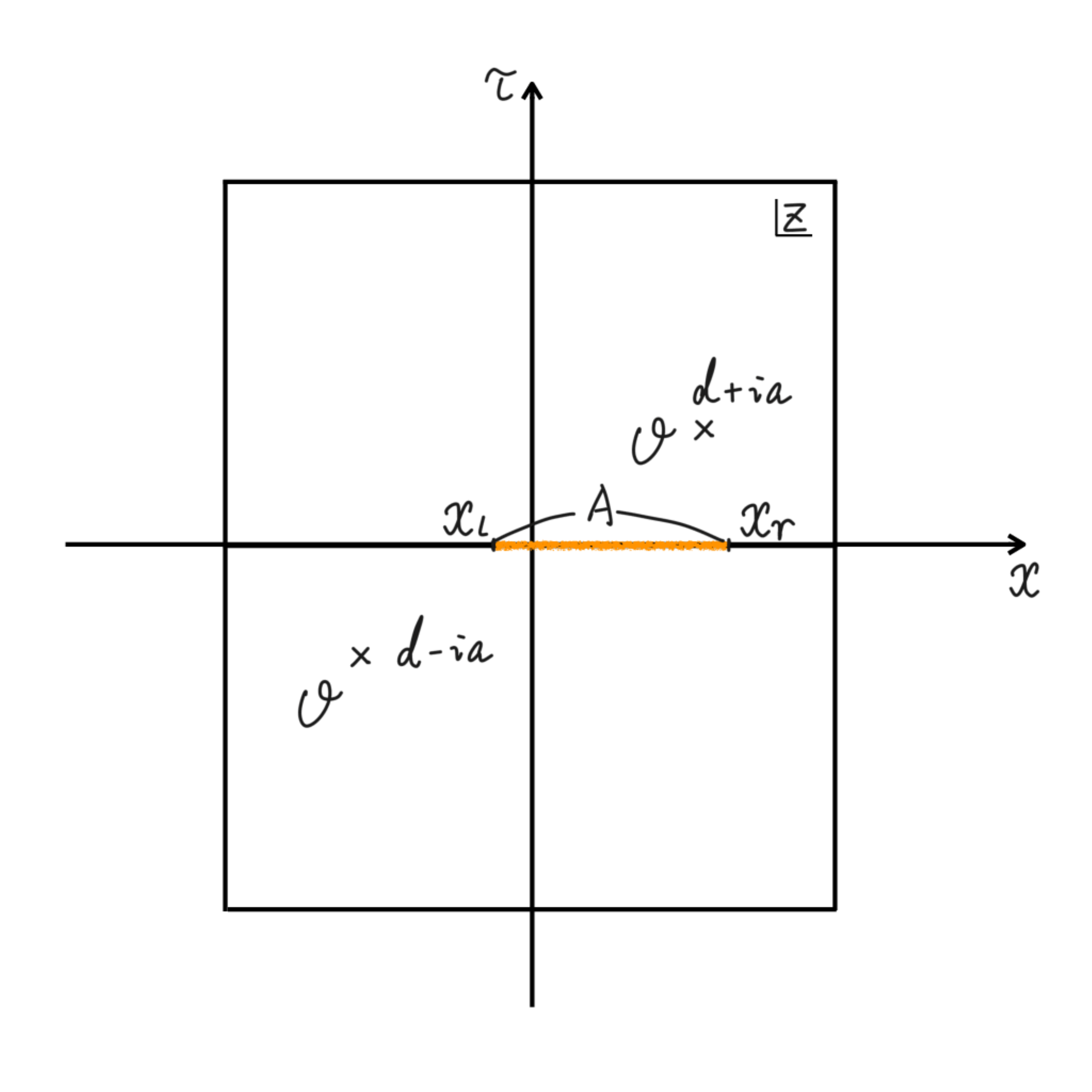}
    \caption{The Euclidean path integral setup. The orange line shows the subsystem $A$.}
    \label{fig:SE-SE-PI2}
\end{figure}
In this case
\begin{align}
    z_1 = -d-ia, ~z_2 = x_l, ~z_3 = x_r, ~z_4 =d+ ia, \\
    \bar{z}_1 = -d+ia, ~\bar{z}_2= x_l, ~\bar{z}_3 = x_r, ~\bar{z}_4 =d-ia.
\end{align}
Figure \ref{fig:SE-SE-HA-Hol} shows $\Delta S_A$ for subsystems centered at different points $x=x_m$ with length $l=20$.

From this plot, we can see that if the cutoff is small, then $\Delta S_A$ sharply decreases when one edge of $A$ gets close to one of the excitations. This is very similar to $\Delta S_A^{(2)}$'s behavior observed in the free CFT (figure \ref{fig:FREECFTC}). However, also in this case, we can see that $\Delta S_A$ can be both positive and negative. 

Also note that, when the cutoff $a$ is small enough, $\Delta S_A < 0$ at $x_m = 0$. This behavior can be easily understood from a holographic point of view. Performing a conformal transformation and bring the two operator excitations to infinite past and infinite future, then we will get a BTZ black hole microstate in the bulk. The two edges of $A$ lie on the same space slice and the geodesic connecting the two edges of $A$ should be shorter than that in a global AdS geometry. This observation immediately gives $\Delta S_A < 0$ at $x_m = 0$ when the cutoff $a$ is small enough. 

As a concluding remark, $\Delta S_A$ in two local operator excited states in a holographic CFT can be both positive and negative while $\Delta S_A^{(2)}$ in the corresponding setup in the free CFT is always negative. This behavior in holographic CFT can be understood from the bulk side. If we bring the two heavy operators to infinite past and infinite future, they produce a BTZ black hole in the bulk. Whether $\Delta S_A$ is positive or negative roughly depends on where $A$ lies in this black hole geometry. As we have already seen above, if $A$ lies on a time slice, then $\Delta S_A$ becomes positive. If $A$ lies on a space slice, them $\Delta S_A$ becomes negative. However, we do not know how to interpret this behavior from a CFT point of view while the $\Delta S_A^{(2)}\leq0$ behavior in the free CFT can be thought as coming from entanglement swapping.

\subsubsection{Doubly Excited States and the Ground State}
\begin{figure}[H]
    \centering
    \includegraphics[width=10cm]{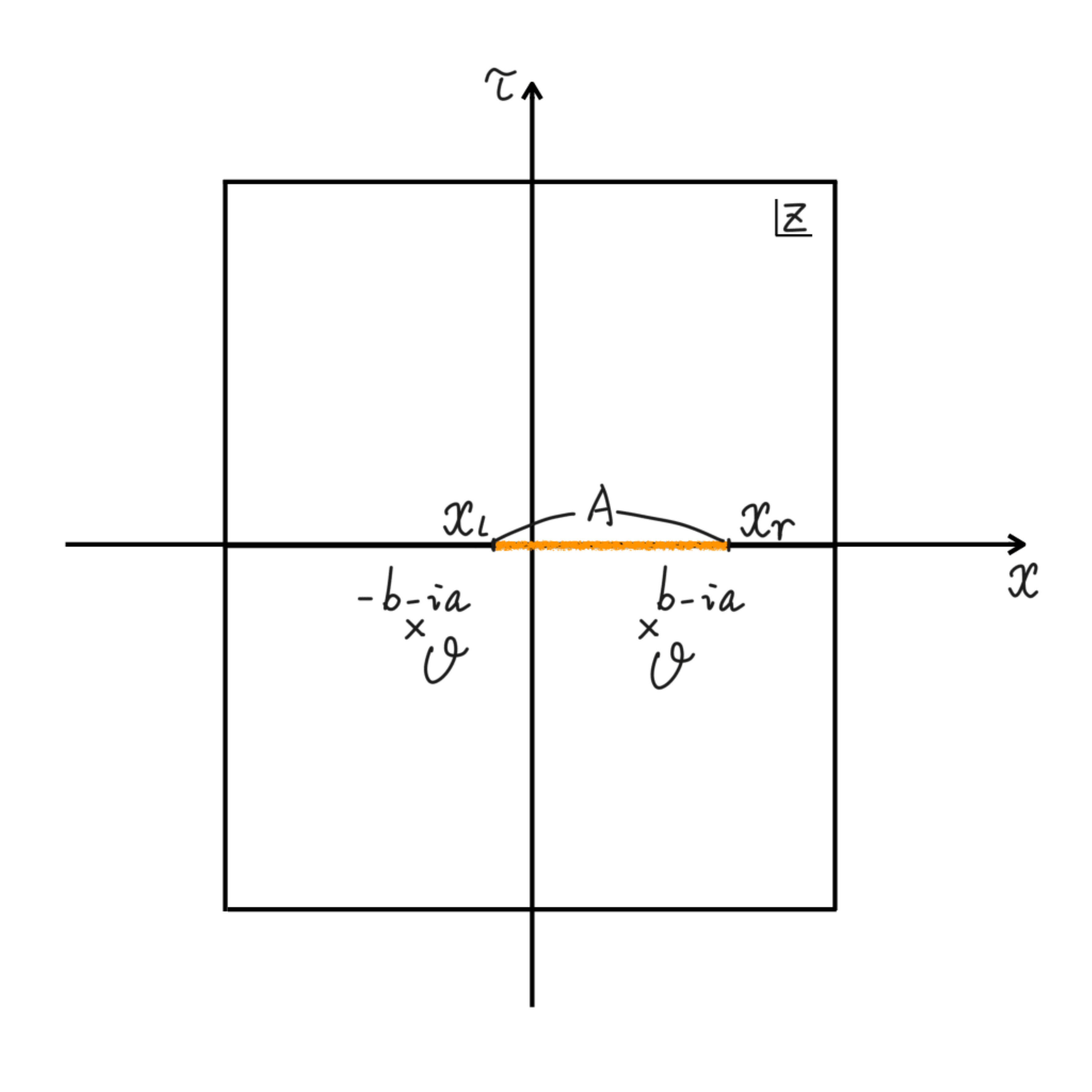}
    \caption{The Euclidean path integral setup. The orange line shows the subsystem $A$.}
    \label{fig:DE-Vac-PI}
\end{figure}

We can also consider the following setup in a similar manner:
\begin{align}
    \ket{\psi} &= e^{-aH_{\rm CFT}}\CO(x=-b)\CO(x=b)\ket{0},\label{eq:DEState}\\
    \ket{\vv} &= \ket{0}.\label{eq:GState}
\end{align}
Here, $\ket{0}$ is the CFT ground state and $a$ is a UV cutoff to avoid divergence. $\ket{\psi}$ is a state\footnote{The dynamics of such a state is known as a double local quench, which is studied in \cite{CNSTW19,KM19}.} locally excited by two identical primary operators $\CO$ on two different space points $x=-b$ and $x=b$. We assume $\CO$ is scalar primary with conformal weight $(h_\CO,h_\CO)$. The corresponding Euclidean path integral setup is shown in figure \ref{fig:DE-Vac-PI}. Let us consider a connected subsystem $A=\{(x,\tau)|\tau=0,x_l\leq x\leq x_r\}$. 

\begin{figure}[H]
    \centering
    \includegraphics[width=10cm]{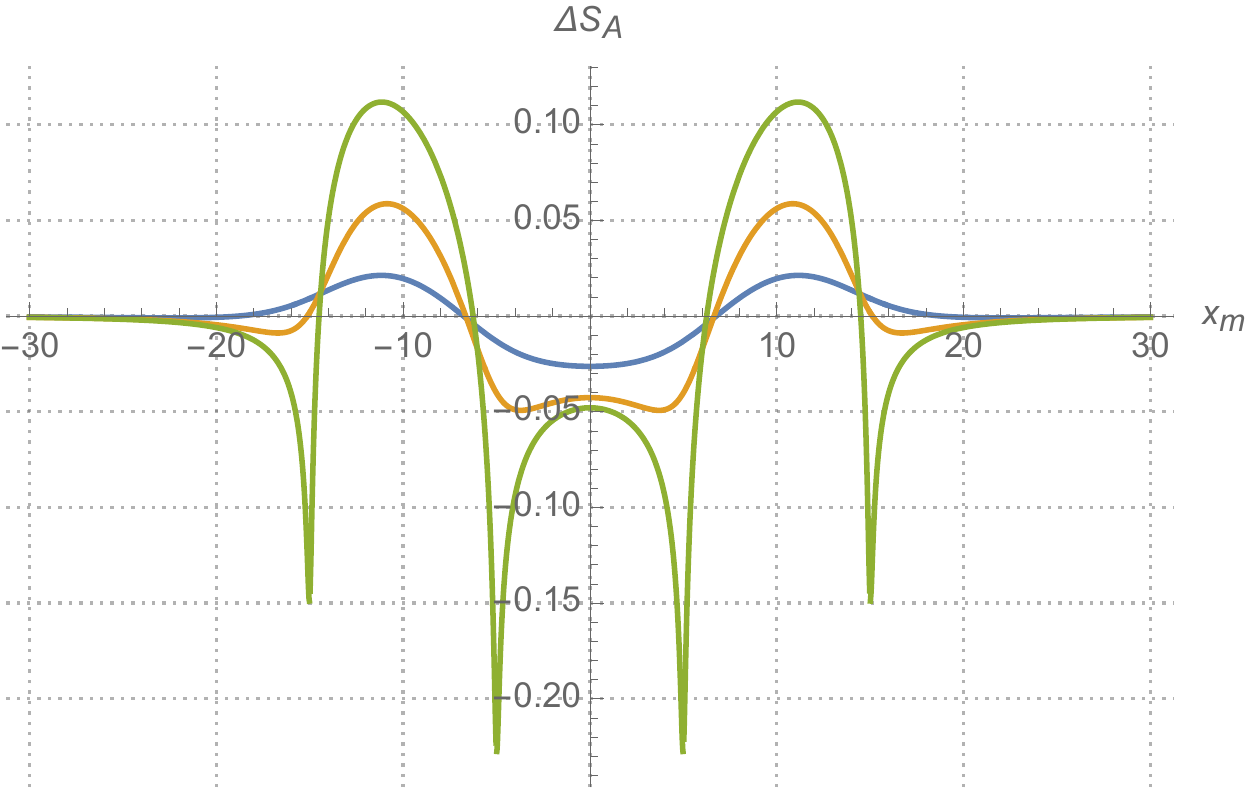}
    \caption{$\Delta S_A$ for subsystems centered at different points $x=x_m$ with length $l=20$, $b=5$, $\a=1/2$. The blue line shows the $a=5$ case, the orange line shows the $a=2$ case, and the green line shows the $a=0.1$ case.}
    \label{fig:DE-Vac-Hol}
\end{figure}

Here, we are considering the states given by (\ref{eq:DEState}) and (\ref{eq:GState}). In this case,
\begin{align}
    z_1 = -b-ia, ~z_2 = x_l, ~z_3 = x_r, ~z_4 = b-ia, \\
    \bar{z}_1 = -b+ia, ~\bar{z}_2= x_l, ~\bar{z}_3 = x_r, ~\bar{z}_4 = b+ia.
\end{align}
Figure \ref{fig:DE-Vac-Hol} shows $\Delta S_A$ for subsystems centered at different points $x=x_m$ with length $l=20$.

\subsection{Pseudo Entropy for Boundary States in Holographic CFTs}\label{sec:HolBCFT}
\begin{figure}[H]
    \centering
    \includegraphics[width=10cm]{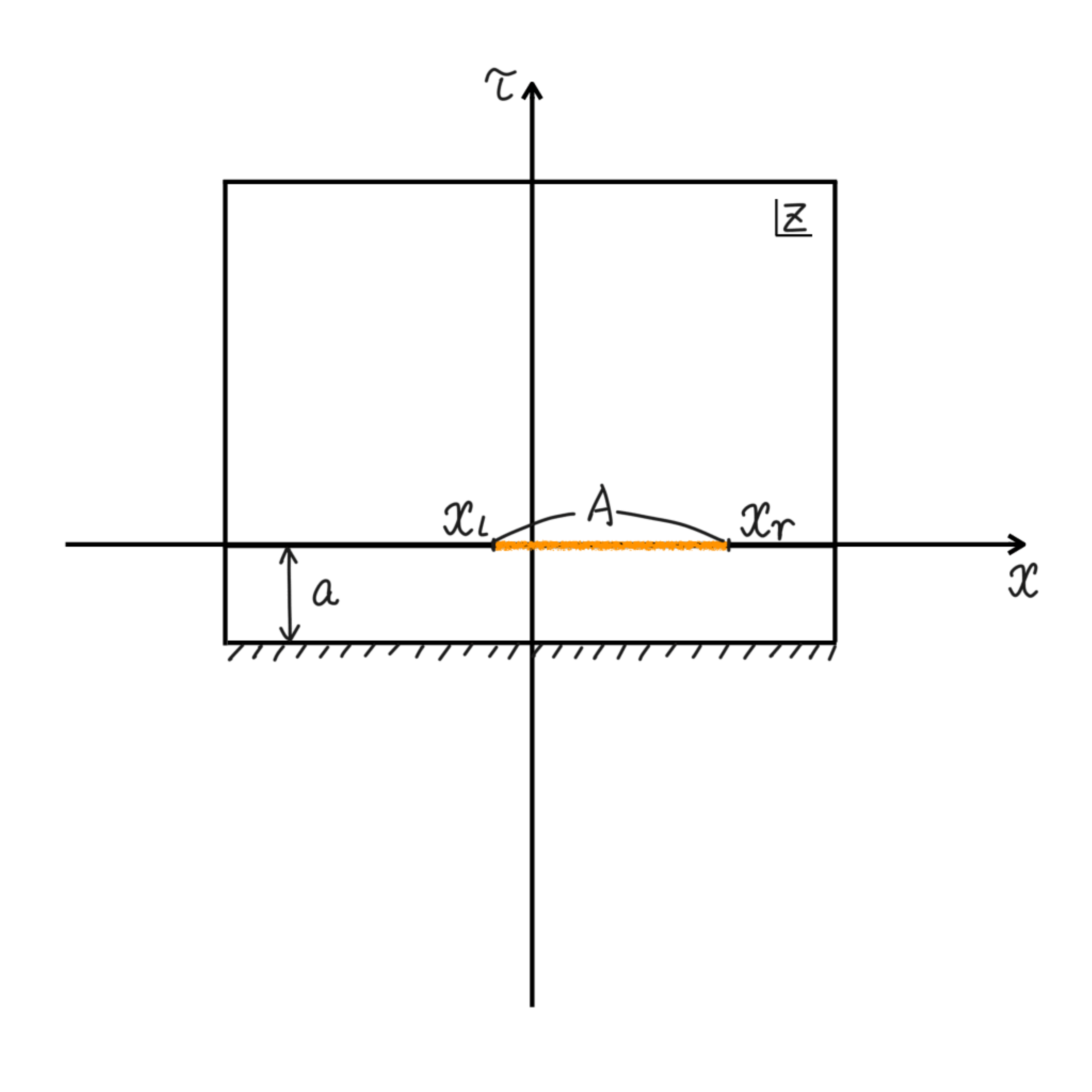}
    \caption{The Euclidean path integral setup. The orange line shows the subsystem $A$.}
    \label{fig:BS-GS-PI}
\end{figure}

From now on, we will focus on the case where one of the two states is given by the boundary state
(or Cardy state). Such a state is employed to describe e.g. global quantum quenches \cite{Calabrese:2005in}.
As the most simple example, we can consider $\CT^{\psi|\vv}$ for two states 
\begin{align}
    \ket{\psi} &= e^{-aH_{\rm CFT}}\ket{B},\label{eq:BState}\\
    \ket{\vv} &= \ket{0}.
\end{align}
where $a$ is a UV cutoff to avoid divergence and $\ket{B}$ is a CFT boundary state. Figure \ref{fig:BS-GS-PI} shows the corresponding Euclidean path integral setup.

\begin{figure}[H]
    \centering
    \includegraphics{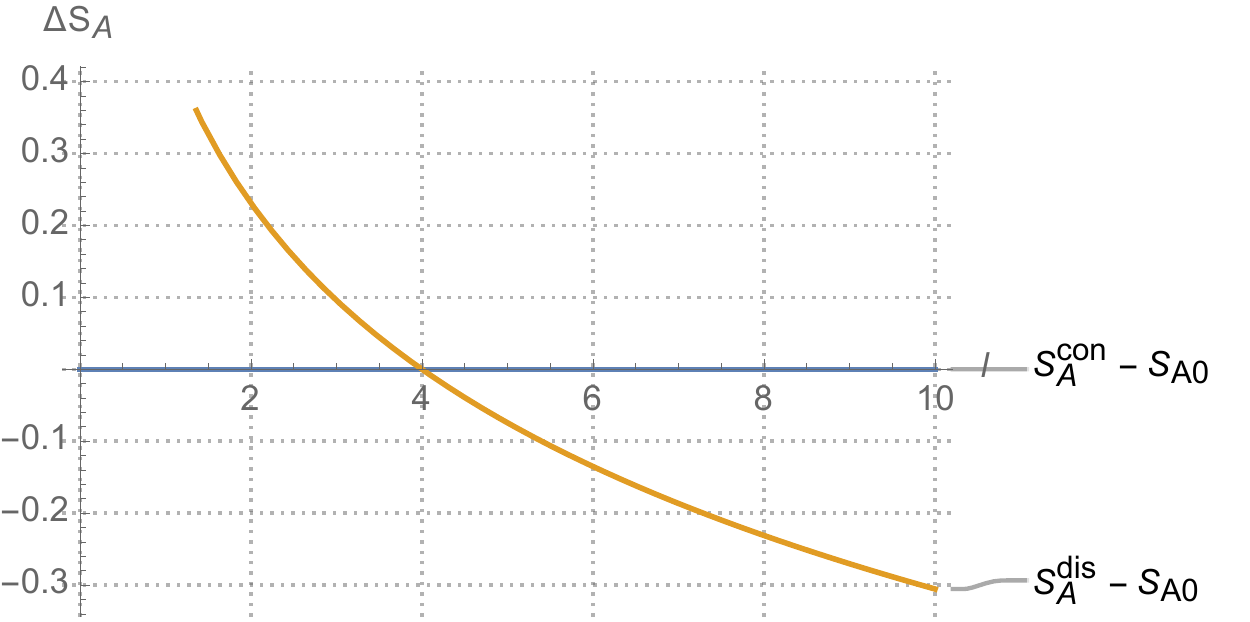}
    \caption{$\Delta S_A$ for a subsystem $A$ with different length. Here, we set $a=2$.}
    \label{fig:BS-GS}
\end{figure}

The minimal surface whose area computes holographic entanglement entropy can end on the end-of-the-world brane in AdS/BCFT \cite{AdSBCFT}. Therefore, in the zero tension $T=0$ case, we have
\begin{align}
    &S_A = \min\{S_A^{con},S_A^{dis}\} \\
    &S_A^{con} = \frac{c}{3} \log\frac{l}{\ep} \\
    &S_A^{dis} = \frac{c}{3} \log\frac{2a}{\ep}
\end{align}
where $l$ is the length of the subsystem $A$ and $\epsilon$ is the UV cutoff corresponding to the lattice distance. Figure \ref{fig:BS-GS} shows the $\Delta S_A$ for a subsystem $A$ with different length.

Notice also in the limit $a\to 0$, we obtain $S^{dis}_A$ by setting $a\sim \ep$. This is consistent with 
(\ref{propa}) because the boundary state does not have any entanglement under a spacial decomposition
\cite{Miyaji:2014mca}.

\subsection{Linearity of Holographic Pseudo Entropy}\label{subsec:linearity}
For CFT states dual to semiclassical geometries (holographic states), we expect the entanglement entropy can be approximated as an expectation value of the so-called area operator \cite{Faulkner:2013ana,Almheiri:2016blp, Harlow:2016vwg}. Let us consider the linear combination of holographic states,
\begin{equation}
\ket{\Psi_i}=\sum^M_{k=1}\alpha_{ik}\ket{\psi_{ik}},
\end{equation}
and its reduced density matrix $\rho^{\Psi_i}_{A}=\mathrm{Tr}_{\bar{A}}\ket{\Psi_i}\hspace{-1mm}\bra{\Psi_i}$. 
Then, the entanglement entropy gives
\begin{equation}
S(\rho^{\Psi_i}_{A})\simeq \dfrac{\bra{\Psi_i}\frac{\hat{\mathcal{A}}}{4G_N}\ket{\Psi_i}}{\braket{\Psi_i|\Psi_i}}=\sum^M_{k=1}|\alpha_{ik}|^2\dfrac{\mbox{A}(\Gamma^{\psi_{ik}}_A)}{4G_N}, \label{eq:linearity}
\end{equation}
where $\mbox{A}(\Gamma^{\psi_{ik}}_A)$ corresponds to the area of the minimal surface (homologous to $A$) in a geometry dual to a holographic state $\ket{\psi_{ik}}$. Note that $M$ is supposed to be a small number compared with $e^{\mathcal{O}{(G_N^{-1})}}$ \cite{Almheiri:2016blp}. We also assumed that $\braket{\psi_{ik}|\psi_{i\ell}}=\delta_{k\ell}$ and $\sum_{k}|\alpha_{ik}|^2=1$. 
We expect a version of the linearity of pseudo entropy for transition matrix $\mathcal{T}^{\Psi_i|\Psi_j}_A=\mathrm{Tr}_{\bar{A}}\ket{\Psi_i}\hspace{-1mm}\bra{\Psi_j}/\braket{\Psi_j|\Psi_i}$,
\begin{equation}
S(\mathcal{T}^{\Psi_i|\Psi_j}_A)\simeq \dfrac{\bra{\Psi_j}\frac{\hat{\mathcal{A}}}{4G_N}\ket{\Psi_i}}{\braket{\Psi_j|\Psi_i}}=\frac{1}{\sum_k\alpha^\ast_{jk}\alpha_{ik}}\sum_k \alpha^\ast_{jk}\alpha_{ik}\dfrac{\mbox{A}(\Gamma^{{\psi_{jk}|\psi_{ik}}}_A)}{4G_N}, \label{eq:linearity2}
\end{equation}
where $\mbox{A}(\Gamma^{{\psi_{jk}|\psi_{ik}}}_A)$ corresponds to the minimal surface (homologous to $A$) in a geometry dual to an inner product $\braket{\psi_{jk}|\psi_{ik}}$. 
It means that the holographic pseudo entropy computes a weak value of the area operator. 
Therefore, if one chooses coefficients $c_{ik}$ and $c_{jk}$ appropriately, like examples of qubit systems, one can easily obtain complex values even from holographic states. We leave a proof of \eqref{eq:linearity2} for linear combinations of so-called heavy states in appendix \ref{app:linearity}. 

\begin{figure}[t]
 \begin{center}
 \resizebox{140mm}{!}{
 \includegraphics{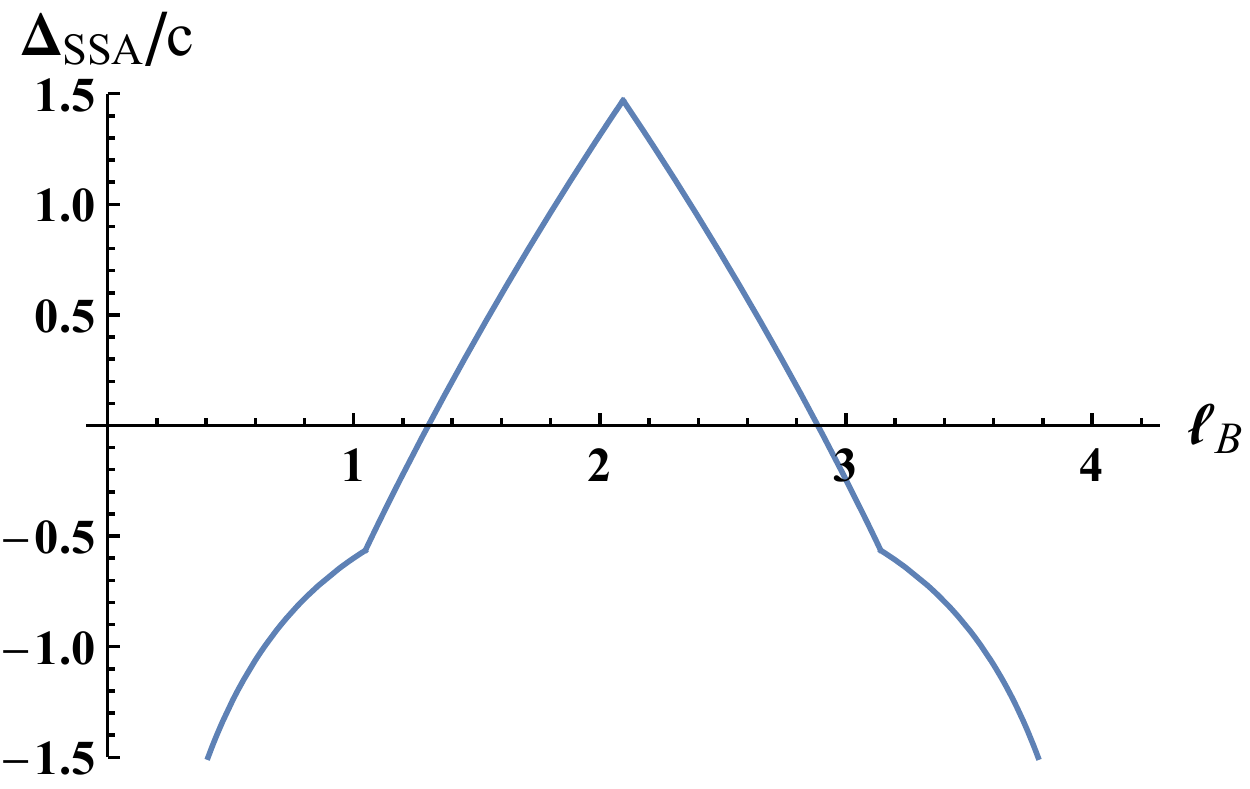}\hspace{1cm}
 \includegraphics{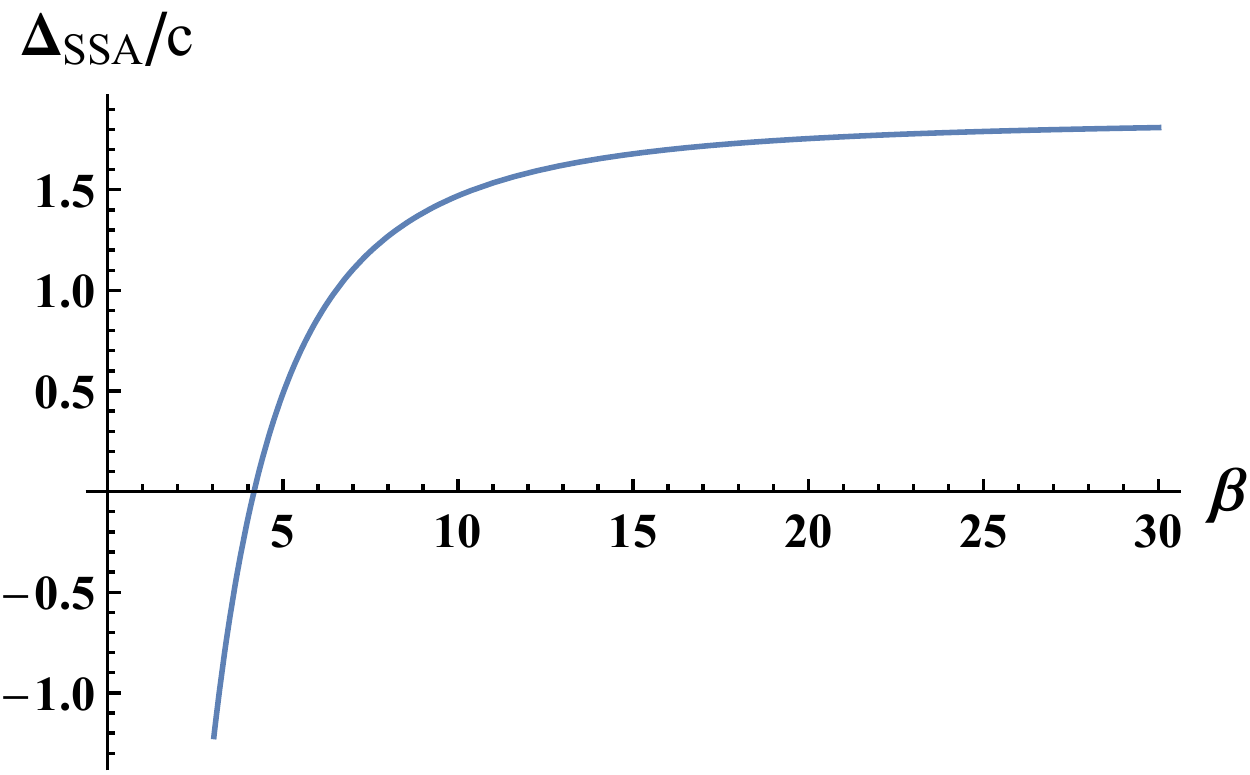}}
 \end{center}
 \caption{Plots for $\Delta_{SSA}$ (with the unit of the central charge $c$) defined in \eqref{eq:dssa}. The regime $\Delta_{SSA}>0$ shows the breaking of strong subadditivity. For simplicity, we consider adjacent intervals: $A=[0,\pi/3], B=[\pi/3,\ell_B], $and $C=[\ell_B,\ell_B+\pi/3]$. Left panel: $\ell_B$ dependence with $\beta_{H_1}=1$ and $\beta_{H_2}=10$. Note that this state satisfies both positivity and subadditivity of pseudo entropy. Right panel: $\beta_{H_2}$ dependence with $\ell_B=2\pi/3$ and $\beta_{H_1}=1$. }
 \label{fig:ssa_heavystates}
\end{figure}
Here we just illustrate an example of the linear combination of holographic states which does break the strong subadditivity for pseudo entropy \eqref{ssah} even when we restrict to a canonical time slice for the definition of subsystems, while it holds the positivity.
We focus on the two-dimensional holographic CFT on a cylinder with periodicity $2\pi$ and take a given connected subsystem size as $\ell_A$.  Let us consider a superposition of two heavy states,
\begin{equation}
\ket{\Psi_\pm}=\dfrac{1}{\sqrt{5}}(\ket{\mathcal{O}_{H_1}}\pm2\ket{\mathcal{O}_{H_2}}),
\end{equation}
where each $\mathcal{O}_{H_i}$ has the conformal dimension $h_{H_i}=\bar{h}_{H_i}=\mathcal{O}(c)\geq \frac{c}{24}$. 

The linearity relation \eqref{eq:linearity2} leads
\begin{equation}
S(\mathcal{T}^{\Psi_+|\Psi_-}_{A})=\dfrac{5}{3}\left(-\dfrac{\mbox{A}(\Gamma^{H_1}_A)}{4G_N}+4\dfrac{\mbox{A}(\Gamma^{H_2}_A)}{4G_N}\right),
\end{equation}
where
\begin{align}
\mathcal{T}^{\Psi_+|\Psi_-}_{A}&=\mathrm{Tr}_{\bar{A}}\frac{\ket{\Psi_+}\hspace{-1mm}\bra{\Psi_-}}{\braket{\Psi_-|\Psi_+}},\\
\dfrac{\mbox{A}(\Gamma^{H_i}_A)}{4G_N}&=\dfrac{c}{6}\log\left(\dfrac{\beta_{H_i}}{\pi\epsilon}\sinh\left[\dfrac{\pi}{\beta_{H_i}}\mathrm{min}(\ell_A,2\pi-\ell_A)\right]\right), \\
\beta_{H_i}&=\dfrac{2\pi}{\sqrt{\frac{24h_{H_i}}{c}-1}}.
\end{align}
Each area term describes the area of the minimal surfaces for (a micro state of) BTZ blackhole with the inverse temperature $\beta_{H_i}$\footnote{Strictly speaking, there is another phase for these pure states \cite{Kusuki:2019rbk, Kusuki:2019evw}. In this paper, however, we neglect this phase just for simplicity.}. Remarkably, this linear combination of holographic states leads the breaking of strong subadditivity for pseudo entropy, while we maintain the positivity $S(\mathcal{T}^{\Psi_+|\Psi_-}_{A})\geq0$, for example. To see this, let us define
\begin{equation}
\Delta_{SSA}=S(\mathcal{T}^{\Psi_i|\Psi_j}_{ABC})+S(\mathcal{T}^{\Psi_i|\Psi_j}_{B}) - S(\mathcal{T}^{\Psi_i|\Psi_j}_{AB})+S(\mathcal{T}^{\Psi_i|\Psi_j}_{BC}). \label{eq:dssa}
\end{equation}
We plotted $\Delta_{SSA}$ for our example in figure \ref{fig:ssa_heavystates}.


\section{Generalizations for Operator and Mixed States}\label{sec:MixedGen}
Finally, let us consider a possible generalization of the pseudo entropy for mixed states. 
So far, we have seen that the holographic pseudo entropy $S(\mathcal{T}^{\psi|\vv}_A)$ computes the minimal surfaces on the geometry associated with $\braket{\vv|\psi}$. Based on this observation, it is natural to introduce some generalizations for mixed states which are holographically equivalent to entanglement wedge cross sections \cite{Takayanagi:2017knl, Nguyen:2017yqw} in the dual geometry. There are several proposals for mixed state measures dual to this cross-section \cite{Takayanagi:2017knl, Nguyen:2017yqw,Kudler-Flam:2018qjo,Tamaoka:2018ned,Dutta:2019gen,Umemoto:2019jlz,Levin:2019krg}, but here we define a generalization based on a canonical purification in \cite{Dutta:2019gen}.

\subsection{Operator States for Transition Matrices}
The entanglement entropy for a given purified density matrix can be regarded as a specific example of so-called operator entanglement entropy \cite{Bandyopadhyay:2005,Prosen:2007}
(see also \cite{Hosur:2015ylk,Nie:2018dfe} for recent applications).
Therefore, we first discuss its transition matrix and pseudo entropy counterpart. 
Let us start from an operator,
\begin{align}
X=\sum_{i,j}[X]_{ij}\ket{i_{A}}\hspace{-1mm}\bra{j_{A}},
\end{align}
which is not necessarily to be Hermitian, $X^\dagger\neq X$. 
We can define a corresponding operator state\footnote{To avoid an unnecessarily ambiguity from unitary transformations, we further assume the above basis $\ket{i_{A}}$ are product states. },
\begin{align}
\ket{X}=\sum_{i,j}[X]_{ij}\ket{i_{A}}\hspace{-1mm}\ket{j^\ast_{A^\ast}}. \label{eq:TXXd}
\end{align}
where $\ket{j^\ast}$ is a complex conjugate of $\ket{j}$. Then, one can define any transition matrices for operator states and compute entanglement entropy for such transition matrices. 
From the next section, we are particularly interested in a specific case that our transition matrix\footnote{In  \eqref{eq:top}, we abused the notation of the transition matrix. In below, the operators assigned in the arguments of transition matrices will be assumed to be corresponding operator states. } is given by
\begin{align}
\mathcal{T}^{X|X^{\dagger}}&=\dfrac{\ket{X}\hspace{-1mm}\bra{X^{\dagger}}}{\braket{X^{\dagger}|X}}, \label{eq:top}
\end{align}
where $X$ will be given by a transition matrix between two holographic states. 
Note that if we started from a Hermitian operator $X$, this reduces to the usual operator entanglement for $X$. 

Before discussing the transition matrix as \eqref{eq:top}, it is rather natural to consider the ordinary operator entanglement for a single reduced transition matrix itself. In the rest part of this subsection, as a warmup exercise, we discuss entanglement entropy for a given operator state dual to a reduced transition matrix. The reader who is familiar with the reflected entropy may skip to the next subsection. 

Starting from a given reduced transition matrix $\mathcal{T}^{\psi|\vv}_{A}$, we can introduce a series of operator states,
\be
\ket{(\mathcal{T}^{\psi|\vv}_{A})^\frac{m}{2}}=\sum_{i,j}[(\mathcal{T}^{\psi|\vv}_{A})^{\frac{m}{2}}]_{ij}\ket{i_A}\ket{j^\ast_{A^\ast}},
\ee
where $m$ is assumed to be an even integer. We can then introduce reduced {\it density matrices} associated with these operator states,
\be
\rho^{(m)}_A=\mathrm{Tr}_{A^\ast}\ket{(\mathcal{T}^{\psi|\vv}_{A})^\frac{m}{2}}\bra{(\mathcal{T}^{\psi|\vv}_{A})^\frac{m}{2}}. \label{eq:rdmt}
\ee
Note that we are loose about the normalization which will be taken into account later. In what follows, we focus on $m\rightarrow1$ limit where $m$ is an analytic continuation of an even integer.  

If $\mathcal{T}^{\psi|\vv}_{A}$ is a physical state, nothing is what we did, just purifying and undoing a given state $\mathcal{T}^{\psi|\vv}_A$. Therefore, in such cases, the pseudo entropy for $\mathcal{T}^{\psi|\vv}_{A}$ and the entanglement entropy for \eqref{eq:rdmt} (with $m\rightarrow1$) does match. A particularly interesting example is the class $\mathscr{E}$ in two qubit systems (see around \eqref{twoqubitf}) where we have a nice interpretation of the pseudo entropy as counting of Bell pairs. We stress that the transition matrices are not hermitian in general, hence the resulting {\it physical state} $\rho^{(m)}_A$ is not the original $(\mathcal{T}^{\psi|\vv}_A)^m$ in general. We can easily confirm this discrepancy between $\rho^{(m)}_A$ and $(\mathcal{T}^{\psi|\vv}_A)^m$ from the two qubit examples which are out of class $\mathscr{D}$. 

\begin{figure}[t]
 \begin{center}
 \resizebox{80mm}{!}{
 \includegraphics{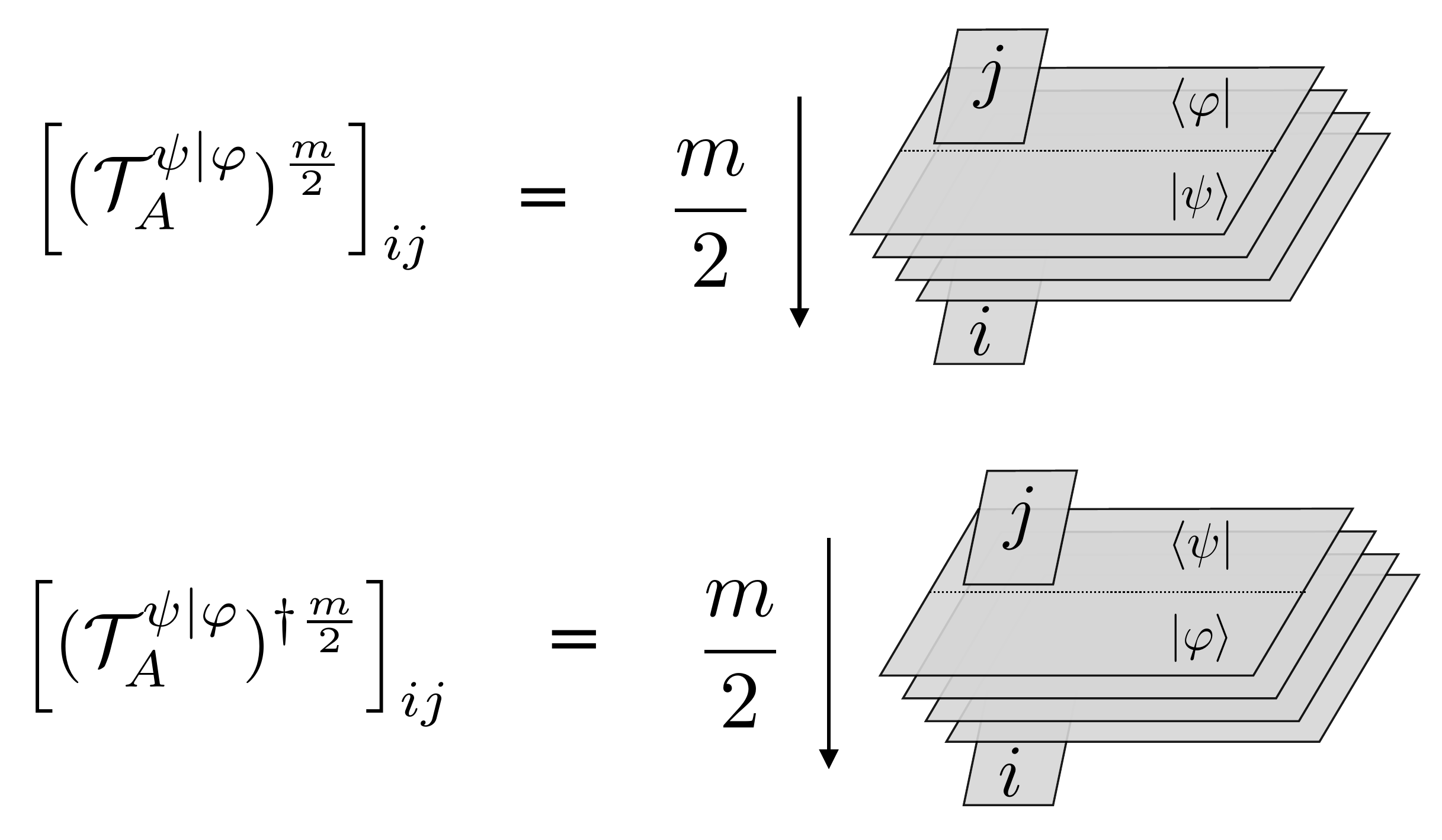}}
  \resizebox{80mm}{!}{
 \includegraphics{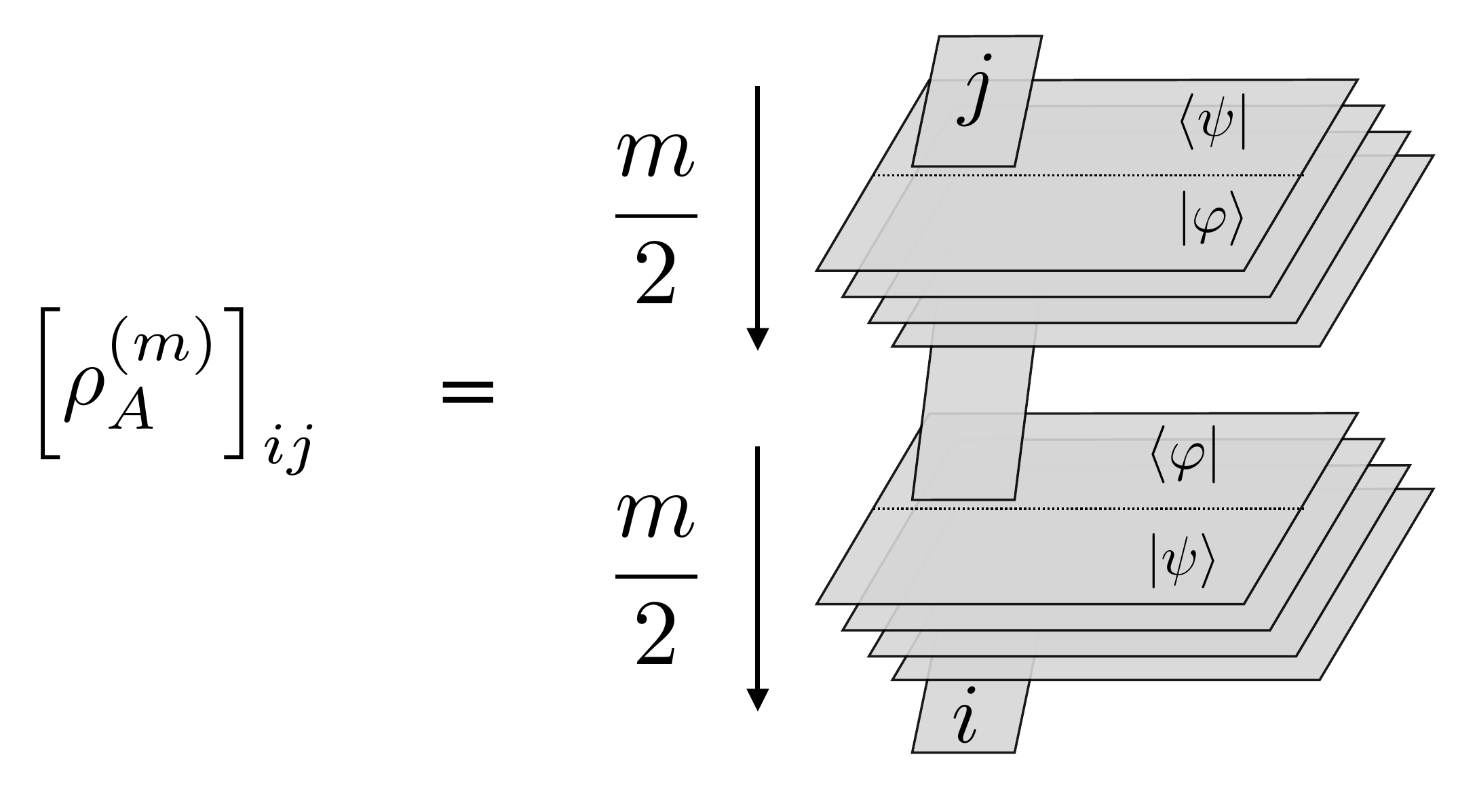}}
 \end{center}
 \caption{Left: replica manifolds for matrix elements of $(\mathcal{T}^{\psi|\vv}_{A})^{\frac{m}{2}}$ and $(\mathcal{T}^{\psi|\vv}_{A})^{\dagger \frac{m}{2}}$. Right: One for $\rho^{(m)}_A$. Here $m$ is assumed to be an even integer. }\label{fig:oet_replica}
\end{figure}

We can compute $n$-th R\'{e}nyi entropy for $\rho^{(m)}_A$ by using the standard replica trick. Figure \ref{fig:oet_replica} shows the replica manifold for $[(\mathcal{T}^{\psi|\vv}_{A})^{\frac{m}{2}}]_{ij}, [(\mathcal{T}^{\psi|\vv}_{A})^{\dagger \frac{m}{2}}]_{ij},$ and $[\rho^{(m)}_A]_{ij}$. In this case, the global structure of the replica manifold is the same as the ordinary $n$-th R\'{e}nyi entropy. Therefore, we can in principle reuse the replica trick calculations discussed in the previous sections. One main difference is that now we have two replica numbers $m$ and $n$ both of which will be taken to be one. From the next subsection, we move to the pseudo generalization of the reflected entropy, where we will use the similar replica trick described above with an extra complication due to topology of a replica manifold\footnote{But with one simplification: since $\mathcal{T}^{\psi|\vv}_{A}$ is not Hermitian, there is no obvious $\mathbb{Z}_2$ reflection symmetry in the $n$-th R\'{e}nyi entropy for $\rho^{(m)}_A$. The lack of this symmetry causes difficulties to compute the replica partition function and to identify its bulk dual. We would like to come back this point in the future. In the next subsection, we construct a pseudo version of the reflected entropy such that we artificially hold this symmetry. It indeed makes our computation simpler. }. 

\subsection{Pseudo Reflected Entropy}
Based on the pseudo entropy for the operator states introduced in the previous subsection, let us define ``pseudo reflected entropy'' as
\begin{equation}
S_R(\mathcal{T}^{\psi|\vv}_{AB})=\lim_{m\rightarrow1}\lim_{n\rightarrow1} S^{(n,m)}_R(\mathcal{T}^{\psi|\vv}_{AB}),
\end{equation}
where
\begin{equation}
S^{(n,m)}_R(\mathcal{T}^{\psi|\vv}_{AB})=S^{(n)}(\mathcal{T}^{(\mathcal{T}^{\psi|\vv}_{AB})^{\frac{m}{2}}|(\mathcal{T}^{\psi|\vv\,\dagger}_{AB})^{\frac{m}{2}}}_{AA^\ast}).
\end{equation}
As like the original reflected entropy, we assumed that $m$ is an analytic continuation of an even integer. 

In particular, if $B=\bar{A}$, 
\begin{equation}
\mathcal{T}^{\psi|\vv}_{A\bar{A}}\equiv\mathcal{T}^{\psi|\vv}=\dfrac{\ket{\psi}\hspace{-1mm}\bra{\vv}}{\braket{\vv|\psi}},
\end{equation}
the pseudo reflected entropy $S_R(\mathcal{T}^{\psi|\vv}_{A\bar{A}})$ reduces to the double of pseudo entropy,
\begin{equation}
S_R(\mathcal{T}^{\psi|\vv}_{A\bar{A}})=S(\mathcal{T}^{\psi|\vv}_A)+S((\mathcal{T}^{\psi|\vv}_{A})^T)=2S(\mathcal{T}^{\psi|\vv}_A). \label{eq:pre_pure}
\end{equation}
In this sense, the pseudo reflected entropy gives a generalization of the pseudo entropy for mixed states. 
We will argue that holographic pseudo reflected entropy gives (double of) the entanglement wedge cross section on the geometry dual to $\braket{\vv|\psi}$,
\begin{equation}
S_R(\mathcal{T}^{\psi|\vv}_{AB})=2E_W(\mathcal{T}^{\psi|\vv}_{AB}). 
\end{equation}
We again stress that this definition is different from the usual operator entanglement for the transition matrix $(\mathcal{T}^{\psi|\vv})^{\frac{1}{2}}$. For example, taking Hermitian conjugate for one of two $X$ in \eqref{eq:TXXd} is crucial to obtain the desired pure state limit in \eqref{eq:pre_pure}. 

\subsection{Replica Trick}
\begin{figure}[t]
 \begin{center}
 \resizebox{120mm}{!}{
 \includegraphics{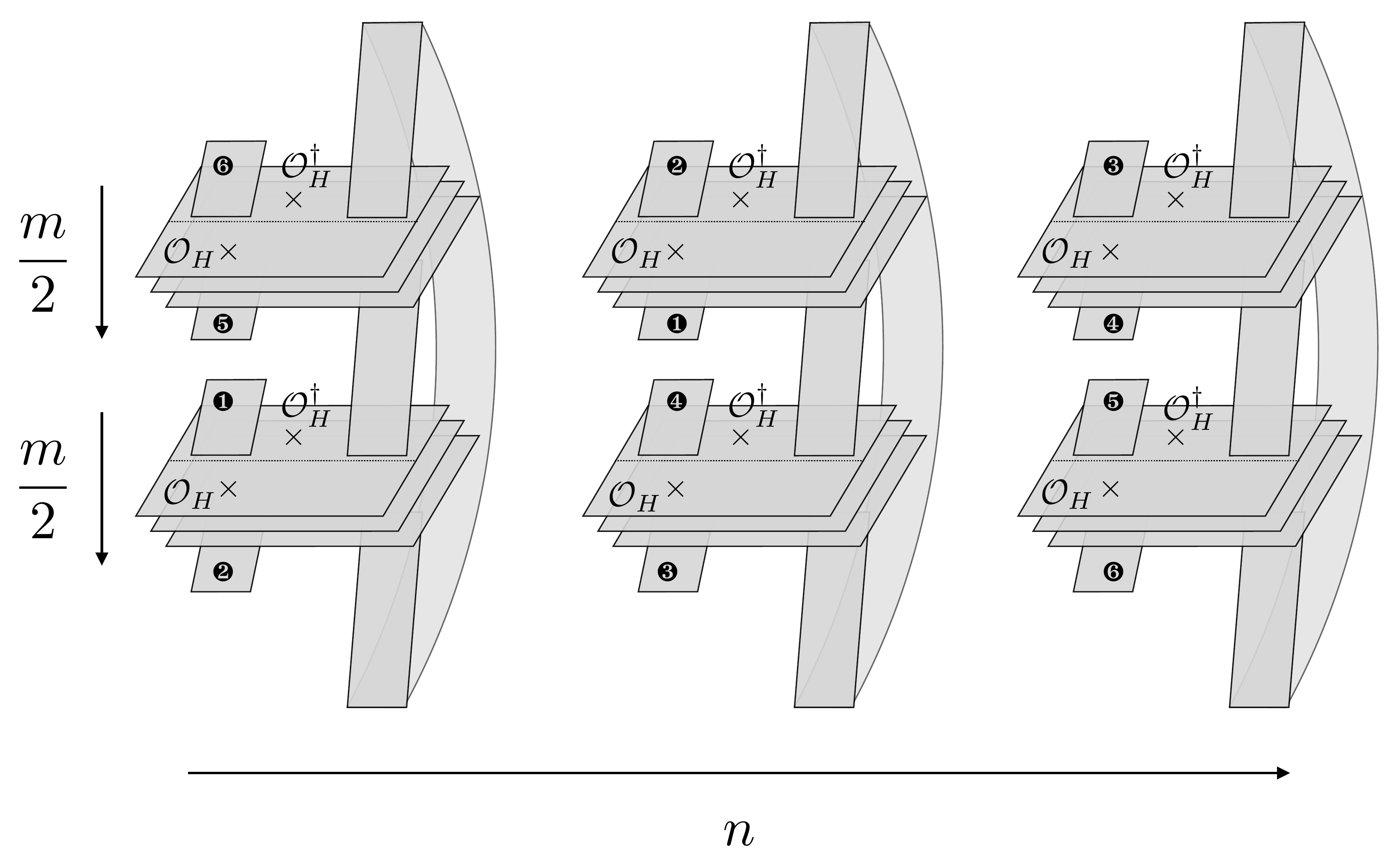}}
 \end{center}
 \caption{Replica manifold for computing $Z_{m,n}$. (The figure shows the $n=3$ case.) Each fixed $n$ corresponds to $\mathcal{T}^{(m)}_{AA^\ast}\equiv\mathrm{Tr}_{BB^\ast}\mathcal{T}^{X|X^\dagger}$ with $X=(\mathcal{T}^{\psi|\vv}_{AB})^{\frac{m}{2}}$,\, ($\ket{X_{AA^\ast BB^\ast}}=\ket{(\mathcal{T}^{\psi|\vv}_{AB})^{\frac{m}{2}}}$). We glue each label in the figure (each circled number with black background) such that we obtain $\mathrm{Tr}(\mathcal{T}^{(m)}_{AA^\ast})^n$.  It can be rewritten as the correlator of twist operators as \eqref{eq:twist_wre}.}\label{fig:re_replica}
\end{figure}
One of the main advantage of reflected entropy is that it can be computed using the correlation function of certain twist operators via the replica trick \cite{Dutta:2019gen, Kusuki:2019rbk, Kusuki:2019evw}.
This is also the case for the pseudo reflected entropy. For concreteness, let us consider a reduced transition matrix obtained from two excited states,
\begin{equation}
\mathcal{T}^{\psi|\vv}_{AB}=\mathrm{Tr}_{\overline{AB}}\left(\frac{\ket{\psi}\hspace{-1mm}\bra{\vv}}{\braket{\vv|\psi}}\right),
\end{equation}
where
\begin{equation}
\ket{\psi}=\mathcal{O}_{H}(w_5,\bar{w}_5)\ket{0},\; \ket{\vv}=\mathcal{O}_{H}(w_0,\bar{w}_0)\ket{0}.
\end{equation}
We will specify the conformal dimension later. 
Here we took subregions as $A=[w_1,w_2]$, $B=[w_3,w_4]$ and set
\begin{align}
w_1&=\bar{w}_1=u_1,\, w_2=\bar{w}_2=v_1,\, w_3=\bar{w}_3=u_2,\, w_4=\bar{w}_4=v_2, \\
w_0&=a_2+ib_2,  \bar{w}_0=a_2-ib_2, w_5=a_1-ib_1, \bar{w}_5=a_1+ib_1,
\end{align}
Based on \cite{Dutta:2019gen}, it is straightforward to construct the replica partition function for a series of canonically purified states. One can define the R\'{e}nyi version of the reflected entropy as,
\begin{align}
S^{(n,m)}_R&=\dfrac{1}{1-n}\log \dfrac{Z_{m,n}}{(Z_{m,1})^n}, \label{eq:renyi_wre}
\end{align}
where
\begin{align}
Z_{m,n}&=\mathrm{Tr}(\mathcal{T}^{(m)}_{AA^\ast})^n, \\
\mathcal{T}^{(m)}_{AA^\ast}&=\mathrm{Tr}_{BB^\ast}\left[\frac{\ket{(\mathcal{T}^{\psi|\vv}_{AB})^{\frac{m}{2}}}\hspace{-1mm}\bra{(\mathcal{T}^{\psi|\vv\,\dagger}_{AB})^\frac{m}{2}}}{\braket{(\mathcal{T}^{\psi|\vv\,\dagger}_{AB})^{\frac{m}{2}}|(\mathcal{T}^{\psi|\vv}_{AB})^{\frac{m}{2}}}}\right]. \label{eq:TXXdtt}
\end{align}
The $Z_{m,n}$ corresponds to a partition function on the manifold described in figure \ref{fig:re_replica}. 
Notice that each replica sheet represents the same one thanks to the Hermitian conjugate in \eqref{eq:TXXdtt}. From \eqref{eq:renyi_wre}, one can obtain
\begin{equation}
S_{R}(\mathcal{T}^{\psi|\vv}_{AB})=\lim_{m\rightarrow1}\lim_{n\rightarrow1}\dfrac{1}{1-n}\log \dfrac{Z_{m,n}}{(Z_{m,1})^n},
\end{equation}
In particular, here we took an analytic continuation of an even integer $m$.  
We can rewrite the $Z_{m,n}$ in terms of the correlation function of twist operators,
\begin{align}
Z_{m,n}&=\braket{0|\mathcal{O}^{\otimes mn\dagger}_H(w_0,\bar{w}_0) \sigma_{g_A}(u_1)\sigma_{g_A^{-1}}(v_1)\sigma_{g_B}(u_2)\sigma_{g_B^{-1}}(v_2)\mathcal{O}^{\otimes mn}_H(w_5,\bar{w}_5)|0}, \label{eq:twist_wre}
\end{align}
where we have twist operators whose chiral conformal dimensions are given by $h_{\sigma_{g_A}}=h_{\sigma_{g_A^{-1}}}=h_{\sigma_{g_B}}=h_{\sigma_{g_B^{-1}}}=\frac{nc}{24}\left(m-\frac{1}{m}\right)\equiv nh_m$, so as anti-chiral ones. These twist operators are essentially different from the ones appeared in the entanglement entropy because the global structure of the replica manifold is totally different. 
As a consequences of this difference, we have a particularly important OPE channel,
\begin{equation}
\sigma_{g_A^{-1}}\sigma_{g_B}\sim\sigma_{g_Bg_A^{-1}}+\cdots,
\end{equation}
where $\sigma_{g_Bg_A^{-1}}$ has the conformal dimension $h_{\sigma_{g_Bg_A^{-1}}}=\bar{h}_{\sigma_{g_Bg_A^{-1}}}=\frac{2c}{24}\left(n-\frac{1}{n}\right)\equiv 2h_{n}$. Notice that this is the same correlation function as one for the reflected entropy \cite{Kusuki:2019rbk, Kusuki:2019evw} except for the position of two operators $\mathcal{O}_H$ and $\mathcal{O}^\dagger_H$. 
\subsection{Pseudo Reflected Entropy in Holographic CFT}
Let us focus our analysis on the holographic CFT${}_2$. We specify two excited states in transition matrix with heavy states whose conformal dimensions are given by $h_{H}=\bar{h}_{H}=\mathcal{O}(c)\geq \frac{c}{24}$. 
In holographic CFT${}_2$, the correlation functions are well-approximated by the single conformal block. The $6$-point correlation functions of our interest are often referred as heavy-heavy-light-light-light-light (HHLLLL) correlation functions, whose conformal blocks under the large $c$ limit are studied in literature. 
We apply a conformal map from cylinder to plane,
\begin{equation}
z=e^{-i\tilde{w}}, \;\tilde{w}=\dfrac{w-w_5}{w-w_0},
\end{equation}
so that we get familiar forms of HHLLLL blocks in literature (refer to \cite{Banerjee:2016qca,Hirai:2018jwy} where the related conformal blocks are presented explicitly). 

\begin{figure}[t]
 \begin{center}
 \resizebox{100mm}{!}{
 \includegraphics{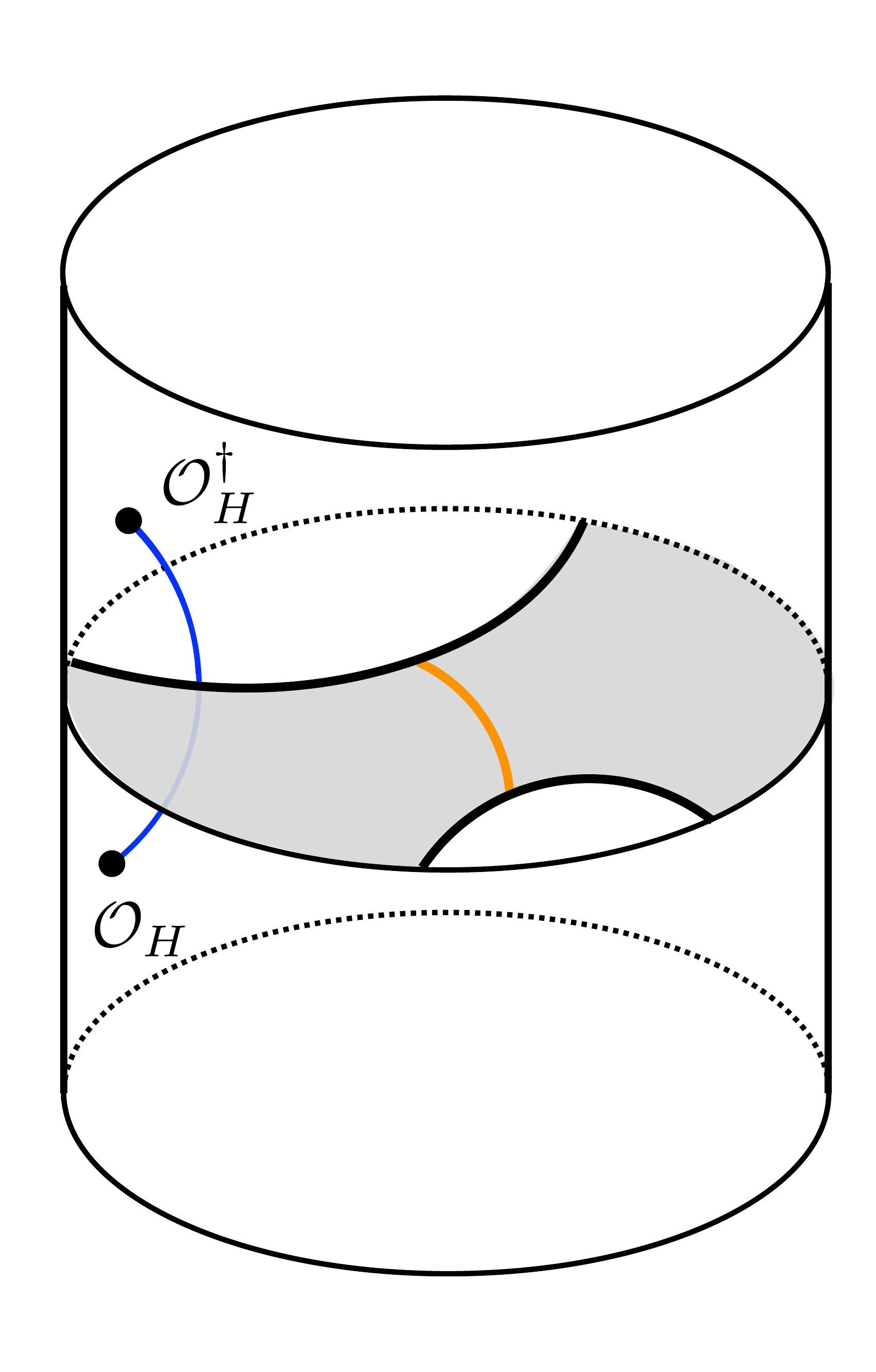}\includegraphics{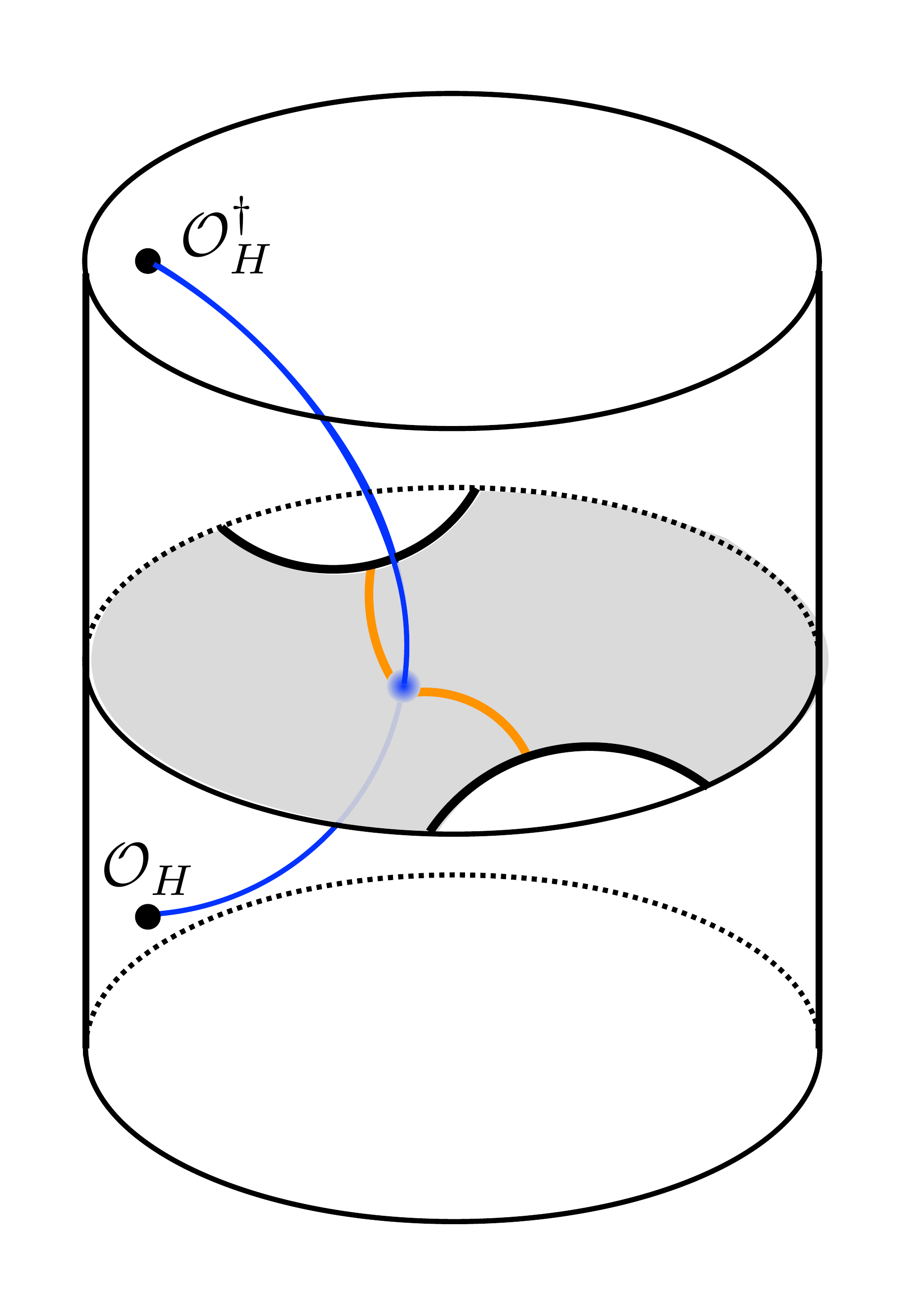}}
 \end{center}
 \caption{Two typical phases for the entanglement wedge cross section in the geometry dual to $\braket{\vv|\psi}$. Left: a connected cross section. Right: a disconnected cross section. Here we skip intermediate phases between left and right panels where we have transition of the entanglement wedge itself. }\label{fig:hwre}
\end{figure}

We have essentially two classes of non-zero entanglement wedge cross section. We can classify them with respect to whether the resulting cross sections have disconnected pieces or not. We will call these classes as disconnected or connected phases. 
The simplest cross section in the connected phase accords with one in the BTZ blackhole (see left panel of figure \ref{fig:hwre}),
\begin{equation}
S_{R}(\mathcal{T}^{\psi|\vv}_{AB})=\dfrac{c}{6}\log\coth\left(\dfrac{\alpha x}{4}\right)+(\textrm{anti-holomorphic}),
\end{equation}
where the cross ratios $x, \bar{x}$ are given by
\begin{equation}
x=\frac{(\tilde{w}_1-\tilde{w}_2)(\tilde{w}_3-\tilde{w}_4)}{(\tilde{w}_1-\tilde{w}_3)(\tilde{w}_2-\tilde{w}_4)}=\frac{(u_1-v_1)(u_2-v_2)}{(u_1-u_2)(v_1-v_2)}=\bar{x},
\end{equation}
and
\begin{equation}
\alpha=\frac{2\pi}{\sqrt{\frac{24h_H}{c}-1}}.
\end{equation}
Therefore, this cross section does not depend on the location of heavy states in the Euclidian manifold. It is rather natural---because except for back reaction of the geometry, the cross section has no direct connection with the geodesics created by the heavy operators. 

For sufficiently large subsystems, we reach the disconnected phase, where the cross section consists of two disconnected contributions both of which end at  the horizon of blackhole  (see right panel of figure \ref{fig:hwre}),
\begin{equation}
S_{R}(\mathcal{T}^{\psi|\vv}_{AB})=\dfrac{c}{6}\log\coth\left(\dfrac{\alpha x_{14}}{4}\right)+\dfrac{c}{6}\log\coth\left(\dfrac{\alpha x_{23}}{4}\right)+ (\textrm{anti-holomorphic}).  
\end{equation}
where
\begin{equation}
x_{14}=\dfrac{(w_1-w_4)(w_0-w_5)}{(w_1-w_0)(w_2-w_0)},\; x_{23}=\dfrac{(w_2-w_3)(w_0-w_5)}{(w_2-w_0)(w_3-w_0)},
\end{equation}
so as anti-holomorphic ones, $\bar{x}_{14}$ and $\bar{x}_{23}$. These results are consistent with the entanglement wedge cross section as is obvious from the related conformal blocks.

\section{Conclusions and Discussions}\label{sec:Conclusions}

In this paper, we introduced a novel quantity called pseudo entropy motivated by a natural question in AdS/CFT: what is the CFT quantity dual to the minimal codimension-2 surface in the bulk AdS which ends on the boundary when there is no time reversal symmetry. A Euclidean path integral with time reversal symmetry in CFT gives a density matrix, while that without time reversal symmetry can be regarded as a transition matrix. A minimal codimension-2 surface anchored on a time reversal symmetric time slice gives the entanglement entropy of a reduced density matrix while that does not lie on such a time slice gives the pseudo entropy of a reduced transition matrix. In this sense, pseudo entropy is a straightforward generalization of entanglement entropy. 

We studied general properties of pseudo entropy in section \ref{sec:PEbasics} and then focused on plenty of explicit examples in qubit systems, free CFT, and holographic setups in section \ref{sec:qubitpr}, \ref{sec:rfreecft}, \ref{secHPE}. In section 6, we introduced the pseudo reflected entropy based on the pseudo entropy for the operator state. In particular, we argued it is equivalent to double of the entanglement wedge cross section in the dual geometry. 

Pseudo entropy, introduced in the present paper with the above motivation and calculations, is a new fundamental quantity deﬁned for any quantum systems. Therefore we expect this to open new directions to study quantum information theory, condensed matter physics and high energy theory. Below we summarize our conclusions on these analysis in more details as well as future problems from a comprehensive point of view.

\paragraph{Pseudo Entropy in Qubit Systems and Operational Meaning}~\par
We classified the structure of two qubit transition matrices with the relation to pseudo entropy in figure \ref{fig:venn2qubit}. In general, the transition matrices $\CT^{\psi|\vv}_A$ are not Hermitian and are not regular quantum states, which lead to complex values of pseudo entropy in general. 
Nevertheless, this quantity shares several properties common also to entanglement entropy. We also uncovered several useful behaviors of pseudo entropy in qubit systems. In two qubit systems, we find that the pseudo entropy always gets increased under local unitary transformations. However, this is only a special property of two qubit systems, and we find the pseudo entropy in general gets reduced under local unitary transformations in larger systems especially when the entanglement swapping occurs. We observed this kind of reductions of pseudo entropy in a number of our examples in CFTs.

We then focus on the class $\mathscr{B}$, i.e. the case where the pseudo (R\'enyi) entropies are all real and non-negative. 
Our CFT examples in this paper fit into this class.  In particular, when the reduced transition matrices $\CT^{\psi|\vv}_A$ and  $\CT^{\psi|\vv}_B$ are Hermitian and positive semi-definite (i.e. the class $\mathscr{E}$), we showed that the pseudo entropy can be interpreted as the averaged number of Bell pairs included in intermediate states between the initial state and final state. {This allows us to quantitatively investigate the amount of virtual entanglement involved in the post-selection process. It will be of interest to consider what physical phenomena follow from the virtual entanglement behind a given process.
It is also important to know how to measure pseudo entropy in real experiments, and to provide a similar interpretation of pseudo entropy for general transition matrices}.

We have also investigated the pseudo entropy when both $\ket{\psi}$ and $\ket{\vv}$ are chosen uniformly at random from the whole Hilbert space. It was observed that the pseudo entropy seems to weakly concentrate around a moderately small real value. This may indicate that, despite the facts that pseudo entropy is in general complex-valued and that its absolute value can be arbitrarily large, it is atypical for states to take such abnormal values of pseudo entropy, especially in the large dimension limit.

\paragraph{Holographic Pseudo Entropy}~\par
Our main motivation of considering pseudo entropy is the holographic duality. For a generic transition matrix and a generic factorization of the Hilbert space, pseudo entropy takes complex values. However, as we have seen in section \ref{HEPEderive}, holographic pseudo $n$-th R\'{e}nyi entropy is nonnegative for any $n>0$, assuming a single classical gravity background with a real valued bulk metric. This gives rather strong restrictions since this implies that, no matter what spatial factorization we take for the whole Hilbert space, the reduced transition matrix should belong to class $\mathscr{B}$. 

Moreover, one may also go further to expect that the reduced transition matrices in holographic setups have only real and nonnegative eigenvalues, i.e. they belong to class $\mathscr{C}$. One way to check this expectation would be to evaluate weak values of observables other than the area operator. This is also left as a future problem. 

Nevertheless, holographic transition matrix generically do not give Hermitian reduced transition matrices and hence do not belong to $\mathscr{D}$. Indeed, it is straightforward to explicitly check
$\Tr\left[\left(\CT^{\psi|\vv}_A\right)^2\right] \neq \Tr\left[\CT^{\psi|\vv}_A\left(\CT^{\psi|\vv}_A\right)^\dagger\right]$
and hence $\CT^{\psi|\vv}_A \neq \left(\CT^{\psi|\vv}_A\right)^\dagger$.

As one of the simplest examples,we compute the holographic pseudo entropy explicitly for the Janus solution in AdS$_3/$CFT$_2$, which is dual to two quantum states given by two different exactly marginal deformations. We showed that the coefficient of the logarithmic divergence of the pseudo entropy is decreased by the Janus perturbation and that the perturbative analysis in the dual field theory correctly reproduces this behavior. 

Also, we found that transition matrices between two linear combinations of holographic states are even out of class $\mathscr{A}$ in general as we have seen in section \ref{subsec:linearity}. In the AdS/CFT, this is dual to a quantum mechanical superposition of two different classical gravity duals. We showed that in this general case, the holographic pseudo entropy is equal to the weak value of the area operator and this possesses the linearity property. 

Since holographic pseudo entropy is a codimension-2 surface in the bulk, we also expect that it is possible to account for the holographic pseudo entropy as Hayward term or gravity edge modes \cite{TT19,BMZ20} if we focus on the fixed-area contribution and whether the basis for fixed area-states is universal or not.

\paragraph{Comparisons of Pseudo Entropy in Free and Holographic CFTs}~\par

We found the class $\mathscr{B}$ property 
(i.e. all  pseudo R\'{e}nyi entropies take non-negative real values)
arises naturally in general field theories
when we prepare the two states $|\psi\lb$ and $|\vv\lb$ by Euclidean path-integrals. 
It is obvious that the transition matrices constructed by Euclidean path-integrals with real valued 
action and external sources at least belong to the class $\mathscr{A}$ because the transition matrices themselves are also real valued. In addition, ${\Tr \left(\CT^{\psi|\vv}_A\right)^n}>0$ for any $n>0$ in this special case. This means that the pseudo R\'{e}nyi entropies are real valued. Moreover, we found in our explicit examples that the resulting pseudo entropy is non-negative. Indeed, the pseudo entropy typically follows the area law as in the standard entanglement entropy in field theories 
\cite{BKLS,Sr}. In the two dimensional massless free scalar CFT, we calculated pseudo R\'{e}nyi entropies when  $|\psi\lb$ and $|\vv\lb$ are (different) locally excited states and found real valued pseudo entropy. In this free theory example, we always find  that the pseudo entropy gets smaller than that for the ground state. In particular, the pseudo entropy is reduced sharply when the positions of the local operator excitations get closer to the boundaries of the subsystem $A$.  We interpret this behavior as the reduction of pseudo entropy in the presence of the entanglement swapping. 

We studied the pseudo entropy for similar locally excited states in holographic CFTs. We again found the reduction of pseudo entropy when the positions of the local operators are close to the boundaries of the subsystem. Moreover, as opposed to the results in free CFTs, we observed that the holographic pseudo entropy can also increase when the local excitation is close to the center of the subsystem. This positive contribution seems to be a special feature of holographic CFTs, which is missing in free CFTs. It is a long-standing question what kinds of quantum states admit classical geometry duals and there are many known restrictions characterizing these states such as the monogamy of mutual information \cite{mono}, holographic entropy cone \cite{BNOSSW15} and holographic entropy arrangement \cite{HRR18}. For the behaviors of  entanglement entropy under local excitations, it is also known that holographic results show a characteristic growth \cite{NNT,Hat}, which is largely different from the free or integrable CFT results \cite{Nozaki,HNTW}.
Our study suggests that pseudo entropy can be used as another novel restriction. Also, since the calculation of pseudo entropy involves two quantum states, it is clear that this restriction is independent of those known ones. It will be an intriguing future direction to pursue this further.

One can also study pseudo entropy in alternative setups. One possibly interesting class is the confomal field theory on a manifold with boundaries (BCFT). We have already studied one of the simplest examples in section \ref{sec:HolBCFT} while there are also other setups such as joining quenched state \cite{CC07}, splitting quenched state \cite{STW}, etc. left to explore.

It will also be intriguing to explore condensed matter applications. Pseudo entropy is expected to provide a new quantum order parameter which depends both the initial state and ﬁnal state. Therefore an important future direction is to analyze how pseudo entropy probes quantum properties of excited states in quantum many-body systems under topological orders or quantum phase transitions.

\paragraph{Subadditivity and Strong Subadditivity}~\par
In section \ref{sec:vssa}, we have seen that even a transition matrix which gives reduced transition matrices whose eigenvalues are real and nonnegative, i.e. a transition matrix in class $\mathscr{C}$ for any factorization, do not always satisfy subadditivity and strong subadditivity. On the other hand, we showed that holographic transition matrices for a single classical gravity background satisfy subadditivity and no evidence suggests that they do not satisfy strong subadditivity. While we leave the confirmation of strong subadditivity of holographic pseudo entropy as a future problem, we would like to note that these give another restrictions on holographic states which are independent of those discussed in the former paragraph. 

\paragraph{Mixed State Generalization}~\par
As a mixed state generalization of the pseudo entropy, 
we introduced the pseudo entropy generalization of reflected entropy,
motivated by the idea of so-called operator
entanglement, based on the transition matrix between two operator states. As
a concrete example, we defined the pseudo reflected entropy where two
operator states are given by a reduced transition matrix and its Hermitian
conjugate. We argued that the pseudo reflected entropy is, as the name
indicates, holographically equivalent to double of the entanglement wedge
cross section in the dual geometry. Notice that such generalizations are
also complex-valued in general and hence constraint holographic states. We
leave extensive study for more general operator states as future work.
Besides the pseudo entropy generalization, we also analyzed 
the ordinary operator entanglement for a given transition matrix, whose 
holographic interpretation was left as a future problem.

\paragraph{Pseudo Complexity?}~\par
Motivated by holographic proposals of computational complexity \cite{Susskind,Brown}, 
it is also intriguing to explore if we can define a complexity measure for transition matrices, which 
might be called as pseudo complexity.\footnote{Please distinguish this from the pseudo-complexity introduced in \cite{Sa} (see also further disucssions in \cite{Sb}), which is a coarse-grained version of 
complexity defined for a single quantum state.  On the other hand, our pseudo complexity is a fined grained complexity defined for two different quantum states.} As one simple way to define such a quantity, we can employ the canonical purification of any given (reduced) transition matrix $\ket{(\mathcal{T}^{\psi|\vv}_{A})^\frac{1}{2}}$, explained in section \ref{sec:MixedGen} and apply known definitions of complexity in quantum field theories e.g. \cite{Chapman:2017rqy,Jefferson:2017sdb,Caputa:2017urj,Caputa:2018kdj,Chen:2020nlj, Erdmenger:2020sup,Flory:2020eot}. However, such a computational complexity for a pure transition matrix   $\mathcal{T}^{\psi|\vv} \equiv \frac{|{\psi}\rangle\langle{\vv}|}{\braket{\vv|\psi}}$ is simply given by the sum of complexity for the state 
$|\psi\lb$ and $\la\vv|$ as the canonical purification is just a direct product of them. This sum rule looks too simple compared with any expected gravity duals such as the minimal volume or gravity action in the gravity duals where non-trivial back reactions should be taken into account, depending on the choices of 
$|\psi\lb$ and $\la\vv|$. In this sense, other approaches to CFT duals of volumes, 
which are based on the information metric \cite{MIyaji:2015mia} and its suitable modification \cite{Belin:2018fxe}, seem to be able to capture the above mentioned gravitational back reactions even for pure transition matrices. This is because the information metric and its generalization is based on two point functions of the form $\la \vv|O_1O_2|\psi \lb$. More details of possible definitions of pseudo complexity and their gravity duals will be left as a future problem.

\section*{Acknowledgements}

We are grateful to Dmitry Ageev, Koji Hashimoto, Norihiro Iizuka, Masahiro Nozaki, Hirosi Ooguri, Noburo Shiba and Tomonori Ugajin for useful discussions and comments.
YN is supported by JST, PRESTO Grant Number JPMJPR1865, Japan.
TT and KT are supported by the Simons Foundation through the ``It from Qubit'' collaboration.  
TT is supported by Inamori Research Institute for Science and World Premier International Research Center Initiative (WPI Initiative) 
from the Japan Ministry of Education, Culture, Sports, Science and Technology (MEXT). 
TT is also supported by JSPS Grant-in-Aid for Scientific Research (A) No.16H02182 and 
by JSPS Grant-in-Aid for Challenging Research (Exploratory) 18K18766.
KT is also supported by JSPS Grant-in-Aid for Research Activity start-up No.19K2344.
ZW is supported by the ANRI Fellowship and Grant-in-Aid for JSPS Fellows No.20J23116.
TT would like to thank the organizers and participants of the online conference ``Frontiers of holographic duality" at Steklov Mathematical Institute and the East Asian String Webinar Series at Osaka U. where this work was presented.

\appendix
\section{Expressing Pseudo (R\'{e}nyi) Entropy in Matrix Form}\label{sec:REandvNE}

Throughout this paper, we study pseudo (R\'{e}nyi) entropy of reduced transition matrices which is a straightforward generalization of R\'{e}nyi entropy and von Neumann entropy of density matrices. In section 2, we firstly define the pseudo $n$-th R\'{e}nyi entropy of $\CT^{\psi|\vv}_A$ for $n\in\mathbb{N}^+\backslash\{1\}$ in a matrix form which can be alternatively written down with eigenvalues\footnote{Recall that $\log(z)$ is used to denote the principal value of logarithmic function with $-\pi<\operatorname{Im}[\log (z)] \leq \pi$.}
\begin{align}
    S^{(n)}(\mathcal{T}^{\psi|\vv}_A) \equiv \frac{1}{1-n}\log\mbox{Tr}\left[(\mathcal{T}^{\psi|\vv}_A)^n\right] = \frac{1}{1-n} \log \left[\sum_j\lambda_j(\CT^{\psi|\vv}_A)^n\right]
\end{align}
where $\lambda_j(M)$ are the eigenvalues of $M$.\footnote{We will use $\lambda_j$ instead for simplicity when there is no ambiguity on which matrix we are considering.} Then we use the eigenvalue form to extend the definition of pseudo $n$-th R\'{e}nyi entropy to $n\in\mathbb{R}^+\backslash\{1\}$ as\footnote{$a^n \equiv e^{n\log(a)}$ for $a\in \mathbb{C}\backslash\{0\}$, $n\in\mathbb{R}^+$, and $0^n \equiv \lim_{a\rightarrow0}a^n$ = 0 for $n\in\mathbb{R}^+$.} 
\begin{align}
    S^{(n)}(\CT^{\psi|\vv}_A) \equiv \frac{1}{1-n} \log \left[\sum_j\lambda_j(\CT^{\psi|\vv}_A)^n\right], 
\end{align}
and take the $n\rightarrow1$ limit to define pseudo entropy\footnote{$0\log0 \equiv \lim_{a\rightarrow0}a\log a = 0$.}
\begin{align}
    S(\CT^{\psi|\vv}_A) &\equiv \lim_{n\rightarrow1}S^{(n)}(\CT^{\psi|\vv}_A)  \nonumber\\
    &= \lim_{n\rightarrow1} \frac{1}{1-n} \log \left(\sum_j\lambda_j^n\right) \nonumber \\
    &= \lim_{\varepsilon\rightarrow0} \frac{1}{\varepsilon} \log \left(\sum_j\lambda_j^{1-\varepsilon}\right) \nonumber \\
    &= \lim_{\varepsilon\rightarrow0} \frac{1}{\varepsilon} \log \left(\sum_j e^{(1-\varepsilon)\log\lambda_j}\right) \nonumber \\
    &= \lim_{\varepsilon\rightarrow0} \frac{1}{\varepsilon} \log \left[\sum_j\lambda_j \left(1-\varepsilon\log{\lambda_j}+\mathcal{O}(\varepsilon^2)\right) \right]\nonumber \\
    &= \lim_{\varepsilon\rightarrow0} \frac{1}{\varepsilon} \log \left(1-\varepsilon\sum_j\lambda_j\log\lambda_j+\mathcal{O}(\varepsilon^2)\right)\nonumber \\
    &= -\sum_{j} \lambda_j(\CT^{\psi|\vv}_A) \log \left[\lambda_j(\CT^{\psi|\vv}_A)\right].
\end{align}
Thinking about the special case when we are considering the pseudo (R\'{e}nyi) entropy of a density matrix $\rho$, then it simply reduce to R\'{e}nyi entropy or von Neumann entropy and admits the following expression in matrix form:
\begin{align}
    &S^{(n)}(\rho) = \frac{1}{1-n}\log\left(\Tr\rho^n\right), \label{eq:RErhoM}\\
    &S(\rho) = -\Tr\left(\rho\log\rho\right). \label{eq:vNErhoM}
\end{align}
A natural question is whether pseudo (R\'{e}nyi) entropy of general reduced transition matrices admits a similar expression in matrix form:
\begin{align}
    &S^{(n)}(\CT^{\psi|\vv}_A) \stackrel{?}{=} \frac{1}{1-n}\log\left(\Tr\left(\CT^{\psi|\vv}_A\right)^n\right), \\
    &S(\CT^{\psi|\vv}_A) \stackrel{?}{=} -\Tr\left(\CT^{\psi|\vv}_A\log\CT^{\psi|\vv}_A\right). 
\end{align}
The answer is yes, while we have to clarify some subtle points on the definition of matrix functions and their traces. To see this, let us start by reviewing the expression (\ref{eq:RErhoM}) and (\ref{eq:vNErhoM}). 

\paragraph{R\'{e}nyi entropy and von Neumann entropy revisted}~\par
When we talk about functions of a $d\times d$ Hermitian matrix $\rho$, what are we talking about? For an arbitrary function $f(x)$ defined on $\mathbb{R}$, the corresponding function of Hermitian matrices $f(\rho)$ is defined as 
\begin{align}
    f(\rho) \equiv U {\rm diag}\left(f(\lambda_1),f(\lambda_2),\cdots,f(\lambda_d)\right)U^\dagger 
    \label{MFHermitianDef}
\end{align}
where 
\begin{align}
    \rho = U {\rm diag}\left(\lambda_1,\lambda_2,\cdots,\lambda_d\right) U^\dagger
\end{align}
is an eigenvalue decomposition of $\rho$. Here, we note that $f(\rho)$ is unique even though the way of decomposition is not. A crucial property is that, for any unitary $V$,
\begin{align}
    f(V\rho V^{\dagger}) = V f(\rho) V^{\dagger}.
    \label{MFHermitianProp}
\end{align}
This then allows us to have 
\begin{align}
    &S^{(n)}(\rho) = \frac{1}{1-n}\log\left(\sum_j\lambda_j^n\right) =  \frac{1}{1-n}\log\left(\Tr\rho^n\right), \\
    &S(\rho) = -\sum_j \lambda_j \log\lambda_j = -\Tr\left(\rho\log\rho\right) . 
\end{align}

\paragraph{Pseudo (R\'{e}nyi) entropy in matrix form}~\par
When we want to write pseudo (R\'{e}nyi) entropy in matrix form, we should extend the domain of definition of matrix functions to general matrices. We cannot use diagonalization to define it like (\ref{MFHermitianDef}) since not all the matrices are diagonalizable. However, we would like to have a definition of matrix functions which satisfies the following properties. 
\begin{itemize}
    \item $f(M)$ should reduce to (\ref{MFHermitianDef}) when $M$ is Hermitian. 
    \item It should satisfy an extension of (\ref{MFHermitianProp}). 
    \item It should be consistent with conventional definitions for simple functions such as positive integer power functions defined by matrix multiplication, negative integer power functions defined by inverse matrices of positive ones, fractional power functions defined by inverse functions of integer ones, exponential function defined by infinite power series, logarithmic function defined by the inverse function of exponential function, etc. 
    \item The power function $f(M)=M^n$, if exists, should be continuous with respect to $n$. This is desirable because we are now concerned at pseudo R\'{e}nyi entropy which is supposed to be continuous with respect to $n$.
\end{itemize} 
There are many definitions of matrix functions, and many of them are equivalent \cite{Higham08}. Now, we would like to pick up a definition which is convenient to perform calculations. Recall that any $d\times d$ matrix can be brought into a Jordan normal form
\begin{align}
    P^{-1} M P=\left(\begin{array}{ccc}{J_{m_{1}}\left(\lambda_{1}\right)} & {} & {0} \\ {} & {\ddots} & {} \\ {0} & {} & {J_{m_{p}}\left(\lambda_{p}\right)}\end{array}\right) \equiv J \equiv {\rm diag}\left(J_{1}, J_{2}, \cdots, J_{p}\right)   
\end{align}
where 
\begin{align}
    J_{k} = J_{m_{k}}(\lambda_k)=\left(\begin{array}{ccccc}{\lambda_k} & {1} & {} & {} & {0} \\ {} & {\lambda_k} & {1} & {} & {} \\ {} & {} & {\ddots} & {\ddots} & {} \\ {} & {} & {} & {\lambda_k} & {1} \\ {0} & {} & {} & {} & {\lambda_k}\end{array}\right)
\end{align}
is an $m_k\times m_k$ Jordan block. Note that the $\sum_{k=1}^p m_k = d$ diagonal elements of $J$ are the $d$ eigenvalues of $M$ and a specific order is chosen to label the eigenvalues.

For a function $f(z)$ defined on $\mathbb{C}$ and any matrix $M$ with a Jordan decomposition given above, if $f(\lambda_k), f^{\prime}(\lambda_k), \cdots, f^{(m_k-1)}(\lambda_k)$ exist for any $k$, then the corresponding matrix function can be defined as \cite{Higham08}
\begin{align}
    f(M) \equiv P f(J) P^{-1} = P {\rm diag}\left(f(J_{1}), f(J_{2}), \cdots, f(J_{p})\right) P^{-1}
\end{align}
where 
\begin{align}
    f\left(J_{k}\right) \equiv \left(\begin{array}{cccc}
    f\left(\lambda_{k}\right) & f^{\prime}\left(\lambda_{k}\right) & \dots & \frac{f^{\left(m_{k}-1\right)}\left(\lambda_{k}\right)}{\left(m_{k}-1\right) !} \\
    & f\left(\lambda_{k}\right) & \ddots & \vdots \\
    & & \ddots & f^{\prime}\left(\lambda_{k}\right) \\
    & & & f\left(\lambda_{k}\right)
    \end{array}\right).
\end{align}
Note that $f(M)$ is unique even though the way of Jordan decomposition is not. It is easy to check that this definition satisfies the properties listed above. Moreover, it satisfies 
\begin{align}
    f(XMX^{-1}) = Xf(M)X^{-1}
\end{align}
for any non-singular matrix $X$. Therefore, matrix functions given by this definition, once exist, immediately lead us to 
\begin{align}
    &S^{(n)}(M) = \frac{1}{1-n} \log \left[\sum_j\lambda_j(M)^n\right] = \frac{1}{1-n}\log\Tr M^n, \\
    &S(M) = -\sum_j\lambda_j(M)\log\left[\lambda_j\left(M\right)\right] = -\Tr\left(M \log M \right). 
\end{align}
Now all we need to do is to check whether the matrix function $M^n$ and $M\log M$ are defined in our case. For both pseudo R\'{e}nyi entropy and pseudo entropy, for a Jordan block $J_k$ with $\lambda_k \neq 0$, $(J_k)^n~(n\in\mathbb{R}^+)$ and $J_k\log J_k$ are well-defined. However, for a Jordan block $J_k$ with $\lambda_k=0$, things are different since the $1,2,...,m_k$-th order derivative of $z^n$ and $z\log z$ do not always exist at $z=0$. We have to do something to cope with $f(J_m(0))$. 

This problem can be resolved by performing the following identifications:
\begin{align}
    &\Tr \left(J_m(0)\right)^n \equiv \lim_{z\rightarrow0} \Tr \left(J_m(z)\right)^n = 0, \\
    &\Tr \left(J_m(0)\log J_m(0)\right) \equiv \lim_{z\rightarrow0} \Tr J_m(z)\log J_m(z) = 0.
\end{align}
These may seem cunning, but they are just generalizations of the treatment
\begin{align}
    0\log0 \equiv \lim_{z\rightarrow0} z\log z = 0
\end{align}
which we have already encountered when defining von Neumann entropy. With the above identifications in mind, now we can write our pseudo (R\'{e}nyi) entropy in matrix form.
\begin{align}
    &S^{(n)}(M) = \frac{1}{1-n}\log \left(\Tr M^n\right), \\
    &S(M) = -\Tr\left(M \log M \right). 
\end{align}

\section{Thermofield Double States}\label{tfdpe}
Consider two thermofield double states at different inverse temperatures $\beta_1$ and $\beta_2$:
\ba
&& |\psi_1\lb=\frac{1}{\s{Z(\beta)}}\sum_n e^{-\frac{\beta_1}{2} E_n}|n\lb_A |n\lb_B,\no
&& |\psi_1\lb=\frac{1}{\s{Z(\beta')}}\sum_n e^{-\frac{\beta_2}{2} E_n}|n\lb_A |n\lb_B ,
\ea
where the two identical CFTs are called $A$ and $B$.

The transition matrix reads 
\ba
\mathcal{T}^{\psi_1|\psi_2}_A=\frac{1}{Z\left(\frac{\beta_1+\beta_2}{2}\right)}\sum_n e^{-\frac{\beta_1+\beta_2}{2} E_n}|n\lb_A \la n|.
\ea
Finally the pseudo entropy is found as 
\ba
S(\mathcal{T}^{\psi_1|\psi_2}_A)=S_{th}\left(\frac{\beta_1+\beta_2}{2}\right),
\ea
where $S_{th}(\beta)$ is the standard thermal entropy at the inverse temperature $\beta$.
In $d$ dimensional CFTs, the thermal entropy behaves as $S_{th}(\beta)\propto \beta^{1-d}$ and thus
we find
\ba
S(\mathcal{T}^{\psi_1|\psi_2}_A)\leq \frac{1}{2}(S_{th}(\beta_1)+S_{th}(\beta_2)).
\ea
Thus the pseudo entropy is reduced when compared with the average of the original entanglement 
entropies.

\section{Two Harmonic Oscillators}\label{thppe}
Consider a system of two interacting harmonic oscillators whose Hamiltonian is given by 
\ba
H=a^\dagger a+b^\dagger b+\lambda(a^\dagger b^\dagger +ab),
\ea
where $\lambda$ is a parameter of the interaction.

The ground state can be found by a Bogoliubov transformation as usual and is found as 
\ba
|\Psi_\theta\lb=\frac{1}{\cosh\theta}\sum_{n=0}^\infty (-\tanh\theta)^n |n\lb_A |n\lb_B,
\ea
where we set
\ba
\lambda=\frac{2\sinh\theta\cosh\theta}{1+2\sinh^2\theta}.
\ea
Note also the standard definition $|n\lb_A=\frac{1}{\s{n!}}(a^\dagger)^n|0\lb_A$
and $|n\lb_B=\frac{1}{\s{n!}}(b^\dagger)^n|0\lb_B$.

The entanglement entropy when we trace out one of the harmonic oscillators becomes
\ba
S_A^\theta=\cosh^2\theta \log\cosh^2\theta-\sinh^2\theta \log\sinh^2\theta\equiv s(\theta).
\ea

Now we turn to the pseudo entropy for $|\Psi_{\theta_1}\lb$ and $|\Psi_{\theta_2}\lb$. First note 
\ba
\la \Psi_{\theta_2}|\Psi_{\theta_1}\lb=\frac{1}{\cosh(\theta_1-\theta_2)},
\ea
and
\ba
\mathcal{T}^{\Psi_{\theta_1}|\Psi_{\theta_2}}_A=\frac{\cosh(\theta_1-\theta_2)}{\cosh\theta_1\cosh\theta_2}
\cdot \sum_{n=0}^\infty (\tanh\theta_1\tanh\theta_2)^n|n\lb_A\la n|.
\ea
Thus the pseudo entropy reads
\ba
S(\mathcal{T}^{\Psi_{\theta_1}|\Psi_{\theta_2}}_A)=s(\ti{\theta}),
\ea
where $\ti{\theta}$ is defined such that
\ba
\tanh\theta_1\tanh \theta_2=\tanh^2\ti{\theta}.
\ea
It is again easy to see the inequality
\ba
S(\mathcal{T}^{\Psi_{\theta_1}|\Psi_{\theta_2}}_A)\leq \frac{1}{2}(S_A^{\theta_1}+S_A^{\theta_2}).
\ea
This is because the function  $s(\theta)$ is convex  as a function of $\log\tanh\theta$.


\section{Pseudo entropy for Haar random states} \label{App:Haar}

We here show that 
\begin{equation}
\mathbb{E}_{\ket{\varphi}, \ket{\psi} \sim {\sf H}} \bigl[ \Tr \bigl[ (\mathcal{T}^{\psi|\vv}_A)^q \bigr] \bigr] = \mathbb{E}_{\ket{\varphi} \sim {\sf H}} \bigl[ \Tr \bigl[ ( \varphi_A)^q \bigr] \bigr], \label{Eq:purity}
\end{equation}
for $q \in \mathbb{N}^+$, which readily implies \eqref{Eq:AvT}. As $q=1$ is trivial, we consider $q= 2,3,\dots$.

We start with a relation that, for any permutation $\sigma$ of degree $m$, and for any operators $O_1, \dots, O_m$ on a Hilbert space $\CH$,
\begin{equation}
\Tr \bigl[O_1 O_2 \dots O_m \bigr] = \Tr \bigl[ (O_{\sigma^{-1}(1)} \otimes O_{\sigma^{-1}(2)} \otimes \dots \otimes O_{\sigma^{-1}(m)} ) \mathbb{S}_{\rm \sigma} \bigr], \label{Eq:swaptrick}
\end{equation} 
where $\mathbb{S}_{\rm \sigma}$ is a unitary representation, defined on $\CH^{\otimes m}$, of the permutation $\sigma$. The action of $\mathbb{S}_{\rm \sigma}$ is given simply by $\mathbb{S}_{\sigma} \ket{i_1 i_2 \dots i_m} = \ket{i_{\sigma(1)} i_{\sigma(2)} \dots i_{\sigma(m)}}$. In particular, when $\sigma$ is a cyclic permutation, it follows that
\begin{equation}
\Tr \bigl[O_1 O_2 \dots O_m \bigr] = \Tr \bigl[ (O_1 \otimes O_2 \otimes \dots \otimes O_m ) \mathbb{S}_{\rm cyc} \bigr].
\end{equation} 
Note that this is equivalent to the replica trick. Using this, we have
\begin{align}
\Tr \bigl[ (\mathcal{T}^{\psi|\vv}_A)^q \bigr] 
&= \Tr \bigl[ (\mathcal{T}^{\psi|\vv}_A)^{\otimes q} \ \mathbb{S}_{\rm cyc} \bigr]\\
&= \Tr \biggl[ \biggl(\frac{|{\psi}\rangle\!\langle{\vv}|}{\braket{\vv|\psi}} \biggr)^{\otimes q} \bigl( \mathbb{S}_{\rm cyc} \otimes I_{B}^{\otimes q} \bigr) \biggr],
\end{align}
where $\mathbb{S}_{\rm cyc}$ is the cyclic operator on the $q$ copies of the system $A$, and $I_B$ is the identity operator on $B$. Due to this relation, it suffices to consider 
$W :=\mathbb{E}_{\ket{\varphi}, \ket{\psi} \sim {\sf H}} \bigl[\bigl(\frac{|{\psi}\rangle\!\langle{\vv}|}{\braket{\vv|\psi}} \bigr)^{\otimes q}  \bigr]$ to obtain the left-hand side of Eq.~\eqref{Eq:purity}.

To explicitly write down $W$, we use the fact that $W$ commutes with $U^{\otimes q}$ for any $U$ on $AB$ due to the unitary invariance of the Haar measure. Furthermore, since $\frac{|{\psi}\rangle\!\langle{\vv}|}{\braket{\vv|\psi}}$ is a rank-one operator, the support of $W$ is the symmetric subspace $\CH_{\rm sym} \subset \CH^{\otimes q}$, where $\CH = \CH_{A}\otimes \CH_B$. Recalling that the symmetric subspace is an irreducible representation of a unitary $U \in \mathcal{U}(\dim \CH)$ acting on $\CH^{\otimes q}$ as $U^{\otimes q}$, it follows from the Shur's lemma that 
\begin{equation}
\mathbb{E}_{\ket{\varphi}, \ket{\psi} \sim {\sf H}} \biggl[\biggl(\frac{|{\psi}\rangle\!\langle{\vv}|}{\braket{\vv|\psi}} \biggr)^{\otimes q}  \biggr] = \frac{\Pi_{\rm sym}}{\Tr [ \Pi_{\rm sym} ]},
\end{equation}
where $\Pi_{\rm sym}$ is the projection onto the symmetric subspace $\CH_{\rm sym}$, and the coefficient is determined from the fact that the trace of the left-hand side is $1$. Thus, we obtain
\begin{equation}
\mathbb{E}_{\ket{\varphi}, \ket{\psi} \sim {\sf H}} \bigl[ \Tr \bigl[ (\mathcal{T}^{\psi|\vv}_A)^q \bigr] \bigr] 
=
\Tr \biggl[\frac{\Pi_{\rm sym}}{\Tr [ \Pi_{\rm sym} ]} \bigl( \mathbb{S}_{\rm cyc} \otimes I_{B}^{\otimes q} \bigr) \biggr].
\end{equation}

We can use the same technique to compute $\mathbb{E}_{\ket{\varphi} \sim {\sf H}}  \Tr [ ( \varphi_A)^q ]$, resulting in a well-known relation that
\begin{equation}
\mathbb{E}_{\ket{\varphi} \sim {\sf H}} \bigl[ \Tr \bigl[ ( \varphi_A)^q \bigr] \bigr]
=
\Tr \biggl[\frac{\Pi_{\rm sym}}{\Tr [ \Pi_{\rm sym} ]} \bigl( \mathbb{S}_{\rm cyc} \otimes I_{B}^{\otimes q} \bigr) \biggr].
\end{equation}
Thus, we obtain \eqref{Eq:purity}.
As a concrete example, when $q=2$, the value can be directly computed to be
\begin{equation}
\mathbb{E}_{\ket{\varphi}, \ket{\psi} \sim {\sf H}} \bigl[ \Tr \bigl[ (\mathcal{T}^{\psi|\vv}_A)^q \bigr] \bigr] 
=
\mathbb{E}_{\ket{\varphi} \sim {\sf H}} \bigl[ \Tr \bigl[ ( \varphi_A)^q \bigr] \bigr]
=
\frac{\dim \CH_A + \dim \CH_B}{\dim \CH + 1}.
\end{equation}

\section{Pseudo Entropy in Perturbative External Field}\label{pertep}

Consider a quantum state $|\psi_0\lb$, which for example we can choose to be the ground state of a given quantum many-body system. We write the Schmidt decomposition of this state as 
\ba
|\psi_0\lb=\sum_i \s{\lambda_i}|i\lb_A|i\lb_B, \label{vacsc}
\ea
where $|i\lb_A$ and $|i\lb_B$ are orthonormal basis included in ${\mathcal H}_A$ and ${\mathcal H}_B$. 
Its reduced density matrix for $A$ is given by
\ba
\rho^{(0)}_A=\mbox{Tr}_B[|\psi_0\lb\la\psi_0|]=\sum_i \lambda_i|i\lb_A\la i|.  
\ea

We consider two states $|\psi_1\lb$ and  $|\psi_2\lb$ via infinitesimally small perturbations from  $|\psi_0\lb$. In terms of infinitesimally small parameters $\gamma_1$ and $\gamma_2$, we can write them 
\ba
|\psi_i\lb=\s{1-|\gamma_i|^2}|\psi_0\lb+\gamma_i|\chi_i\lb,\ \ \ (i=1,2),  \label{chife}
\ea
where $|\chi_i\lb$ are orthogonal to $|\psi_0\lb$ i.e. $\la \chi_i|\psi_0\lb=0$.
In this case the transition matrix looks like
\ba
&&\mathcal{T}^{\psi_1|\psi_2}_A\no
&&=\!\frac{\mbox{Tr}_B\!\left[\!\s{1-|\gamma_1|^2}\!\s{1-|\gamma_2|^2}|\psi_0\lb\!\la\psi_0|\!+\!\gamma_1\!
\s{1-|\gamma_2|^2}|\chi_1\lb\!\la\psi_0|\!+\!\gamma^*_2\!\s{1-|\gamma_1|^2}|\psi_0\lb\!\la\chi_2|\!+\!\gamma_1\gamma^*_2|\chi_1\lb\!\la\chi_2|\right]}
{\s{1-|\gamma_1|^2}\s{1-|\gamma_2|^2}+\gamma_1\gamma^*_2\la\chi_2|\chi_1\lb}\no
&&\simeq \frac{\mbox{Tr}_B\left[(1-|\gamma_1|^2/2-|\gamma_2|^2/2)|\psi_0\lb\la\psi_0|+\gamma_1|\chi_1\lb\la\psi_0|+\gamma^*_2|\psi_0\lb\la\chi_2|+\gamma_1\gamma^*_2|\chi_1\lb\la\chi_2|\right]}
{(1-|\gamma_1|^2/2-|\gamma_2|^2/2)+\gamma_1\gamma^*_2\la\chi_2|\chi_1\lb}\no
&&\simeq \rho^{(0)}_A-\gamma_1\gamma^*_2\la \chi_2|\chi_1\lb \rho^{(0)}_A
+\gamma_1\mbox{Tr}_B[|\chi_1\lb\la\psi_0|]+\gamma^*_2\mbox{Tr}_B[|\psi_0\lb\la\chi_2|]
+\gamma_1\gamma^*_2\mbox{Tr}_B[|\chi_1\lb\la\chi_2|],\no
\ea
where in the final expression we only keep terms up to quadratic terms of $\gamma$.

\subsection{First Order Perturbation}

If we write $\delta \mathcal{T}_{12,A}=\mathcal{T}^{\psi_1|\psi_2}_A-\rho^{(0)}_A$, the first order perturbation of pseudo entropy is estimated as the first law like relation:
\ba
S(\mathcal{T}^{\psi_1|\psi_2}_A)-S(\rho^{(0)}_A)\simeq \mbox{Tr}_A[(-\log\rho^{(0)}_A)\cdot \delta \mathcal{T}_{12,A}].  \label{firstlaw}
\ea
Thus we can find the $O(\gamma)$ contribution as follows:
\ba
S(\mathcal{T}^{\psi_1|\psi_2}_A)-S(\rho^{(0)}_A)\simeq \mbox{Tr}_A
\left[(-\log\rho^{(0)}_A)\cdot\left(\gamma_1\mbox{Tr}_B[|\chi_1\lb\la\psi_0|]+\gamma^*_2\mbox{Tr}_B[|\psi_0\lb\la\chi_2|]\right)
\right],
\ea
where $S(\rho^{(0)}_A)=\sum_i (-\log\lambda_i)\cdot \lambda_i$.

\subsection{External Field Perturbation}

Consider that $|\psi_1\lb$ and $|\psi_2\lb$ are created from $|\psi_0\lb$ by an infinitesimally small external field $J$ which couples to an operator $O$ which acts on $A$:
\ba
|\psi_i\lb={\mathcal{N}_i}e^{J_iO_i}|\psi_0\lb\simeq {\mathcal{N}_1}\left(1+J_iO_i+\frac{1}{2}J_i^2O_i^2\right)|\psi_0\lb,\ \ \ (i=1,2),
\ea
where ${\mathcal{N}_i}$ are the normalization factors such that $\la \psi_i|\psi_i\lb=1$. We regard $J_i$ as the perturbation parameter $\gamma_i$ up to a constant: $\gamma_i= J_i+O(J_i^2)$.

In this case, by comparing with (\ref{chife}), we can write
\ba
&& |\chi_1\lb=\sum_i \s{\lambda_i}|a_i\lb_A|i\lb_B,\no
&& |\chi_2\lb=\sum_i \s{\lambda_i}|b_i\lb_A|i\lb_B.
\ea
Note that  neither $|a_i\lb_A$ nor $|b_i\lb_A$ forms any orthogonal basis in general.

The conditions $\la\psi_0|\psi_0\lb=\la \chi_i|\chi_i\lb=1$ and $\la \psi_0|\chi_i\lb=0$ leads to 
\ba
&& \sum_i \lambda_i=1,\no
&& \sum_i \lambda_i \la a_i|a_i\lb=\sum_i \lambda_i \la b_i|b_i\lb=1, \no
&& \sum_i \lambda_i \la a_i|i\lb=\sum_i \lambda_i \la b_i|i\lb=0.
\ea

In this expression, the transition matrix looks like
\ba
&&\mathcal{T}^{\psi_1|\psi_2}_A \no
&&=(1-\gamma_1\gamma^*_2)\sum_i \lambda_i |i\lb\la i|+\gamma_1\sum_i \lambda_i|a_i\lb\la i |
+\gamma^*_2\sum_i \lambda_i|i\lb\la b_i |+\gamma_1\gamma^*_2 \sum_i \lambda_i|a_i\lb\la b_i |.\no
\ea

If we impose the $\mathbb{Z}_2$ symmetry of the operator $O_i$ such that $O_i\to -O_i$, then we find 
\ba
\la i|O_{1,2}|i\lb=0,
\ea
though in general we have $\la i|O_{1,2}|i\lb\neq 0$. Thus we can estimate the states $|a_i\lb$ and 
 $|b_i\lb$:
\ba
&& |a_i\lb=\eta_1\left[ O_1|i\lb+\frac{J_1}{2}O_1O_1|i\lb
-\frac{J_1}{2}\left(\sum_j \la j|O_1O_1|j\lb \lambda_j\right)|i\lb\right], \no
&&  |b_i\lb=\eta_2 \left[O_2|i\lb+\frac{J_2}{2}O_2O_2|i\lb
-\frac{J_2}{2}\left(\sum_j \la j|O_2O_2|j\lb \lambda_j\right)|i\lb\right], \no
\ea
where $\eta_{1,2}$ are $O(1)$ positive coefficients.
We have also similarly the inner products:
\ba
&& \la i|a_i\lb=\eta_1\left[ \frac{J_1}{2}\la i|O_1O_1|i\lb
-\frac{J_1}{2}\sum_{j}\lambda_j \la j|O_1O_1|j\lb\right]=O(\gamma_1),\no
&& \la i|b_i\lb=\eta_2\left[ \frac{J_2}{2}\la i|O_2O_2|i\lb
-\frac{J_2}{2}\sum_{j}\lambda_j \la j|O_2O_2|j\lb\right]=O(\gamma_2).\no
\ea
When $i\neq j$, we have 
\ba
\la i|a_j \lb=\gamma_1 \la i|O_1|j\lb=O(1),\ \ \ \la i|b_j \lb=\gamma_2 \la i|O_2|j\lb=O(1).
\ea

\subsection{Pseudo R\'{e}nyi Entropy in a Simple Example}

Let us consider the simplest setup of the previous external field perturbation.
Namely, we choose $\gamma_2=0$ and set $\gamma_1=\gamma$.  We also assume $|\chi_1\lb=|\chi_2\lb$ 
(i.e. $\gamma_1=\gamma_2(\equiv \gamma)$, 
which lead to $|a_i\lb_A=|b_i\lb_A$. In this case we have 
\ba
\mathcal{T}^{\psi_1|\psi_2}_A=\sum_i \lambda_i |i\lb\la i|+\gamma\sum_i \lambda_i|a_i\lb\la i |.
\ea
Since $\la i|a_i\lb=O(\gamma)$, the first order perturbation $O(\gamma)$ to 
the pseudo entanglement entropy is vanishing, we need to go beyond the first law  (\ref{firstlaw}).
Instead let us consider the Pseudo 2nd R\'{e}nyi entropy. This is estimated as 
\ba
S^{(2)}(\mathcal{T}^{\psi_1|\psi_2}_A)&=&-\log\left[\sum_i \lambda_i^2+2\gamma\sum_i \lambda_i^2\la i|a_i\lb
+\gamma^2\sum_{i,j}\lambda_i\lambda_j \la i|a_j\lb \la j|a_i\lb\right] \no
&\simeq& S^{(2)}(\rho^{(0)}_A)-\frac{2\gamma\sum_i \lambda_i^2\la i|a_i\lb}{\sum_i \lambda_i^2}
-\gamma^2\frac{\sum_{i,j} \lambda_i\lambda_j\la i|a_j\lb \la j|a_i\lb}{\sum_i \lambda_i^2}.
\ea
Thus we find that the difference $S^{(2)}(\mathcal{T}^{\psi_1|\psi_2}_A)-S^{(2)}(\rho^{(0)}_A)$ is 
$O(\gamma^2)$ and real valued. This is explicitly estimated by ($\eta$ is an $O(1)$ positive constant)
\ba
S^{(2)}(\mathcal{T}^{\psi_1|\psi_2}_A)-S^{(2)}(\rho^{(0)}_A)=-\eta\cdot C\cdot \gamma^2, 
\ea
where the coefficient $C$ is 
\ba
C=-\sum_{i}\lambda_i\la i|OO|i\lb+\frac{\sum_i \lambda_i^2\la i|OO|i\lb}{\sum_i \lambda_i^2}
+\frac{\sum_{i,j}\lambda_i\lambda_j|\la i|O|j\lb|^2}{\sum_i \lambda_i^2}.\label{coefre}
\ea

In AdS/CFT, we expect
\ba
\sum_i \lambda_i^2\la i|OO|i\lb-(\sum_{i}\lambda^2_i)\cdot (\sum_{j}\lambda_j\la j|OO|j\lb)> 0,
\ea
which leads to $C>0$. This is because in the holographic description of geodesic length approximation of the two point function we find $\la i|OO|i\lb$ gets monotonically decreasing as $|i\lb_A$ has a larger (modular) energy, and because the saddle point of $\sum_i \lambda_i^2\la i|OO|i\lb$ has larger $\lambda_i$ than that for the saddle point of  $\sum_i \lambda_i\la i|OO|i\lb$.

For the von Neumann pseudo entropy in the same setup, by using the following formula (note 
$\mbox{Tr}\delta\rho=0$)
\ba
&& \mbox{Tr}[-\log(\rho+\delta\rho)\cdot (\rho+\delta\rho)] \no
&& \simeq 
\mbox{Tr}[-\log \rho\cdot \rho]+\mbox{Tr}[\delta\rho(-\log\rho)]
\!-\!\mbox{Tr}\left[\int^\infty_0 dt\frac{t}{(t+\rho)^2} {\delta\rho} \frac{1}{t+\rho}\delta \rho\right] 
+O(\delta\rho^3),
\ea
we obtain
\ba
&& S(\mathcal{T}^{\psi_1|\psi_2}_A)-S(\rho^{(0)}_A)\no
&& \simeq  \gamma\sum_i(-\log\lambda_i)\lambda_i \la i|a_i\lb
+\gamma^2\sum_{i,j}\lambda_i\lambda_j\frac{\lambda_j-\lambda_i+\lambda_j\log(\lambda_i/\lambda_j)}{(\lambda_i-\lambda_j)^2} \la j|a_i\lb \la i|a_j\lb.  \label{pertwee}
\ea
This is also $O(\gamma^2)$ and is expected to be negative in the AdS/CFT setup.


\section{Pseudo Entropy for Operator Excited States}\label{opepea}

Consider a setup where a quantum state $|\psi_0\lb$, which is typically a ground state, is excited by acting an operator $O(x)$ at two different points $x_1$ and $x_2$, which are defined to be $|\psi_1\lb$ and  $|\psi_2\lb$
\ba
&& |\psi_1\lb={\mathcal N}_1O(x_1)|\psi_0\lb,\no
&& |\psi_2\lb={\mathcal N}_2O(x_2)|\psi_0\lb. \no
\ea
We decompose the total Hilbert space into  ${\mathcal H}_A$ and ${\mathcal H}_B$ by dividing the space in quantum field theory into two subregions $A$ and $B$ such that operator 
$O(x_1)$ and $O(x_2)$ both acts on ${\mathcal H}_A$.  In this case we can rewrite 
\ba
&& |\psi_1\lb=\sum_i \s{\lambda_i}|a_i\lb_A |i\lb_B,\no
&& |\psi_2\lb=\sum_i \s{\lambda_i}|b_i\lb_A |i\lb_B,\no
\ea
where we introduced 
\ba
&& |a_i\lb_A={\mathcal N}_1O(x_1)|i\lb_A,\no
&& |b_i\lb_A={\mathcal N}_2O(x_2)|i\lb_A.\no
\ea

By taking the inner products we obtain
\ba
&& \sum_{i}\lambda_i=\sum_i \lambda_i \la a_i|a_i\lb=\sum_i \lambda_i \la b_i|b_i\lb=1,\no
&& \sum_i \lambda_i \la b_i|a_i\lb=\la \psi_2|\psi_1\lb.
\ea
Note that $\la a_i|a_i\lb$ and  $\la b_i|b_i\lb$ are real valued and positive as is true for two point functions.

The reduce density matrices for $|\psi_1\lb$ and  $|\psi_2\lb$ read
\ba
&& \rho^{(1)}_A=\sum_i \lambda_i|a_i\lb_A\la a_i|,\ \ \ \rho^{(2)}_A=\sum_i \lambda_i|b_i\lb_A\la b_i|, \no
&& \rho^{(1)}_B=\sum_{i,j} \s{\lambda_i \lambda_j}\la a_j|a_i\lb |i\lb_B\la j|,
\ \ \ \rho^{(2)}_B=\sum_i \s{\lambda_i \lambda_j} \la b_j | b_i \lb |i\lb_B\la j|.
\ea 

In general we can write
\ba
&& \la a_i|a_j\lb=\delta_{ij}f^{(11)}_i+r^{(11)}_{ij},\ \ \  \la b_i|b_j\lb=\delta_{ij}f^{(22)}_i+r^{22}_{ij},\ \ \   \la a_i|b_j\lb=\delta_{ij}f^{(12)}_i+r^{12}_{ij},  \label{abin}
\ea
where the diagonal part $f^{(11,22,12)}_i$ are all real valued  and positive. Note $f^{(11)}_i\sim \la i|OO|i\lb$
and is thus $i$-dependent. 

We would like to argue that when the operator $O$ is light $\Delta_O\ll c$ in holographic CFTs, the diagonal term is dominant $f_i\gg r_{ij}$ as can be seen from the HHLL conformal block calculations. Below we assume this light operator approximation. In this approximation we have
\ba
&& \sum_i \lambda_i f^{(11)}_i=\sum_i \lambda_i f^{(22)}_i=1,\no
&& \sum_i \lambda_i f^{(12)}_i=\la \psi_2|\psi_1\lb.
\ea

\subsection{Pseudo R\'{e}nyi Entropy}

Let us evaluate the Pseudo 2nd R\'{e}nyi entropy for the transition 
between  $|\psi_1\lb$ and  $|\psi_2\lb$. 
This is essentially the trace of the square of the transition matrix, estimated as 
\ba
\mbox{Tr}\left(\mathcal{T}^{\psi_1|\psi_2}_A\right)^2
&= &\frac{\sum_{i,j}\lambda_i\lambda_j\la b_j|a_i\lb \la b_i|a_j\lb}
{\left(\sum_i \lambda_i \la b_i|a_j\lb\right)^2}\no
&\simeq & \frac{\sum_{i}(f^{(12)}_i)^2\lambda_i^2}{\left(\sum_i\lambda_if^{(12)}_i\right)^2}.
\ea

On the other hand, another trace is estimated as 
\ba
\mbox{Tr}\left[\mathcal{T}^{\psi_1|\psi_2}_A \left(\mathcal{T}^{\psi_1|\psi_2}_A\right)^{\dagger}\right]
&= &\frac{\sum_{i,j}\lambda_i\lambda_j\la a_j|a_i\lb \la b_i|b_j\lb}
{\left(\sum_i \lambda_i \la b_i|a_j\lb\right)^2}\no
&\simeq & \frac{\sum_{i}f^{(11)}_i f^{(22)}_i\lambda_i^2}{\left(\sum_i\lambda_i f^{(12)}_i\right)^2}.
\ea

These results show $\mbox{Tr}\left(\mathcal{T}^{\psi_1|\psi_2}_A\right)^2\neq \mbox{Tr}\left[\mathcal{T}^{\psi_1|\psi_2}_A\left(\mathcal{T}^{\psi_1|\psi_2}_A\right)^\dagger\right]$ in general and thus we find $\mathcal{T}^{\psi_1|\psi_2}_A$ is not hermitian.

\subsection{R\'{e}nyi Analogue of Fidelity}

To study the distinguishability between $|\psi_1\lb$ and  $|\psi_2\lb$ when we trace out a subsystem, we calculate a R\'{e}nyi analogue of fidelity \cite{Cardy:2014rqa, Suzuki:2019xdq}. When we trace out $B$ we obtain
\ba
I(\rho^{(1)}_A,\rho^{(2)}_A)&&=\frac{\mbox{Tr}[\rho^{(1)}_A \rho^{(2)}_A]}{\s{\mbox{Tr}[(\rho^{(1)}_A)^2]\mbox{Tr}[(\rho^{(2)}_A)^2]}}\no
&&=\frac{\sum_{i,j}\lambda_i\lambda_j|\la a_i|b_j\lb|^2}{\s{(\sum_{i,j}\lambda_i\lambda_j|\la a_i|a_j\lb|^2)\cdot (\sum_{i,j}\lambda_i\lambda_j|\la b_i|b_j\lb|^2)}}\no
&&\simeq \frac{\sum_i \lambda^2_i |f^{(12)}_i|^2} {\s{(\sum_i\lambda_i^2 |f^{(11)}_i|^2)(\sum_i\lambda_i^2 |f^{(22)}_i|^2)}}.
\ea
This is a non-trivial function which has a peak $I=1$ at $x_1=x_2$, which means we can distinguish two different points, as expected.

When we trace out $A$, on the other hand, we get
\ba
I(\rho^{(1)}_B,\rho^{(2)}_B)&&=\frac{\mbox{Tr}[\rho^{(1)}_B \rho^{(2)}_B]}{\s{\mbox{Tr}[(\rho^{(1)}_B)^2]\mbox{Tr}[(\rho^{(2)}_B)^2]}}\no
&&\simeq \frac{\sum_i \lambda^2_i f^{(11)}_i f^{(22)}_i} {\s{(\sum_i\lambda_i^2 |f^{(11)}_i|^2)(\sum_i\lambda_i^2 |f^{(22)}_i|^2)}}.
\ea
If we employ a saddle point approximation on the summation over $i$ with the assumption that 
the function $f_i$ does not depend on $i$ radically, which is equivalent to a probe approximation, then we simply find $I\simeq 1$. Therefore we cannot distinguish the two points. This is expected as we insert the operator $O$ in the subsystem $A$ and we cannot detect this from $\rho_B$. These results are consistent with the CFT derivation of entanglement wedge structure in holographic CFTs found in 
\cite{Suzuki:2019xdq}.

When the operator $O$ is heavy, the off diagonal term $r_{ij}$ in (\ref{abin}) gets important.
In this case we need a more complicated analysis, which we will not pursue in this paper.

\subsection{Fidelity}

Next we examine the genuine fidelity in the same way. When we trace out $B$ we can evaluate the fidelity as follows. First we note $\s{\sum_i\lambda_i |a_i\lb\la a_i|}\simeq \sum_i\s{\frac{\lambda_i}{f^{(11)}_i}}|a_i\lb\la a_i|$) and this leads to
\ba
&&\s{\s{\rho^{(1)}_A}\rho^{(2)}_A\s{\rho^{(1)}_A}}\no
&& \simeq \sum_i \frac{\lambda_i}{f_i}\cdot |\la a_i|b_i\lb|\cdot |a_i\lb\la a_i|.
\ea
Thus the fidelity is found as 
\ba
&&\mbox{Tr}\left[\s{\s{\rho^{(1)}_A}\rho^{(2)}_A\s{\rho^{(1)}_A}}\right] \no
&&\simeq \sum_{i}\lambda_i |f^{(12)}_i|\simeq \la \psi_2|\psi_1\lb.
\ea
In this way, even though we trace out $B$, the fidelity remains the same as the value for the original pure state.

On the other hand, if we trace out $B$, we get
\ba
&&\mbox{Tr}\left[\s{\s{\rho^{(1)}_B}\rho^{(2)}_B\s{\rho^{(1)}_B}}\right] \no
&&\simeq \sum_{i}\lambda_i \s{f^{(11)}_if^{(22)}_i}\simeq 1,
\ea
where in the last we again employed the saddle point approximation. This means that when we trace out $B$ we cannot distinguish the two points.

All these are consistent with the CFT derivation of entanglement wedge given in \cite{Suzuki:2019xdq}.

\section{Linearity of Pseudo Entropy from Heavy States}\label{app:linearity}
In this appendix, we give a sketch of proof for the linearity of pseudo entropy discussed in section \ref{subsec:linearity}. 
First, let us introduce the following pure states,
\begin{align}
\ket{\Psi_i}&=\sum^M_{j=1}\alpha_{ij}\ket{\mathcal{O}_{H_j}},\;\;\;(i=1,2),
\end{align}
and consider a transition matrix,
\begin{equation}
\mathcal{T}^{\Psi_1|\Psi_2}=\dfrac{\ket{\Psi_1}\hspace{-1mm}\bra{\Psi_2}}{\braket{\Psi_2|\Psi_1}}.\\
\end{equation}
In what follows, we will focus on holographic CFTs in two-dimension. 
Here each $\ket{\mathcal{O}_{H_j}}$ is a so-called heavy state supposed to be dual to a microstate of the BTZ black hole \cite{Hat}. We will mention local operators located on more general points {\it e.g.} $\mathcal{O}_{H_j}(x_i,\tau_i)\ket{0}$ later. 
Based on the replica trick, we first compute
\begin{equation}
\mathrm{Tr}_A(\mathcal{T}^{\Psi_1|\Psi_2})^n=\dfrac{\braket{\Psi_2|\sigma_n(0)\bar{\sigma}_n(\ell)|\Psi_1}}{\braket{\Psi_2|\Psi_1}^n}, \label{eq:replica}
\end{equation}
where we defined the reduced transition matrix $\mathcal{T}^{\Psi_1|\Psi_2}_A$ as
\begin{equation}
\mathcal{T}^{\Psi_1|\Psi_2}_{A}=\mathrm{Tr}_{\bar{A}}\mathcal{T}^{\Psi_1|\Psi_2}. 
\end{equation}
For a while, we focus on the numerator of \eqref{eq:replica}.
Let us unpack the summand, 
\begin{align}
&\braket{\Psi_2|\sigma_n(0)\bar{\sigma}_n(\ell)|\Psi_1}\nn\\
&=\sum_{a_1,\cdots a_M=0}^{n}\sum_{b_1,\cdots b_M=0}^{n} \alpha^{\ast a_{1}}_{21}\cdots \alpha^{\ast a_M}_{2M} \alpha^{b_1}_{11}\cdots \alpha^{b_M}_{1M}
\braket{0|\mathcal{O}^\dagger_{h(\{a_{i}\})}(\infty)\sigma_n(0)\bar{\sigma}_n(\ell)\mathcal{O}_{h(\{b_{i}\})}(-\infty)|0},
\end{align}
where
\begin{equation}
\mathcal{O}_{h(\{a_{i}\})}\equiv (\mathcal{O}_{H_1})^{a_{1}}\cdots (\mathcal{O}_{H_M})^{a_{M}}+\left(\dfrac{n!}{a_{1}!\cdots a_{M}!}-1\right)\textrm{permutations}.
\end{equation}
Here $\sum_ia_{i}=\sum_ib_{i}=n$. Due to the single conformal block approximation under the large $c$ limit\cite{Hartman:2013mia}, we can only maintain pieces with $a_i=b_i$. This approximation should be valid as long as we have $M\ll e^{\mathcal{O}(c)}$ \cite{Almheiri:2016blp}. 
The only difference from the usual entanglement entropy is that we now allow $\alpha_{1j}\neq \alpha_{2j}$. Therefore, the remaining calculation is completely parallel to one for the entanglement entropy in \cite{Almheiri:2016blp}. Thus, we can finally obtain linearity of pseudo-entropy,
\begin{equation}
S(\mathcal{T}^{\Psi_1|\Psi_2}_{A})=\sum^M_{i=1}\dfrac{\alpha^{\ast}_{2i}\alpha_{1i}}{\braket{\Psi_2|\Psi_1}}S(\mathcal{T}^{\mathcal{O}_{H_i}|\mathcal{O}_{H_i}}_{A})=\sum^M_{i=1}\dfrac{\alpha^{\ast}_{2i}\alpha_{1i}}{\braket{\Psi_2|\Psi_1}}S(\rho^{\mathcal{O}_{H_i}}_{A}),
\end{equation}
where
\begin{equation}
\rho^{\mathcal{O}_{H_i}}_{A}=\mathrm{Tr}_{\bar{A}}\ket{\mathcal{O}_{H_i}}\hspace{-1mm}\bra{\mathcal{O}_{H_i}},
\end{equation}
and each $S(\rho^{\mathcal{O}_{H_i}}_{A})$ is the standard entanglement entropy for the physical state $\rho^{\mathcal{O}_{H_i}}_{A}$. 

In the main text, we considered the heavy state located in a more general position, $\mathcal{O}_{H_j}(x_i,\tau_i)\ket{0}$. In particular, such a state is not yet normalized. If we started from a linear combination of the unnormalized states, 
\be
\ket{\Psi_i}=\sum_k\alpha_{ik}\ket{\tilde{\psi}_{ik}}\equiv\sum_k\alpha_{ik}\mathcal{O}_{H_k}(x_{ik},\tau_{ik})\ket{0},
\ee
the original linearity of entanglement entropy should become,
\be
S(\rho^{\Psi_i}_{A})\simeq \dfrac{\bra{\Psi_i}\frac{\hat{\mathcal{A}}}{4G_N}\ket{\Psi_i}}{\braket{\Psi_i|\Psi_i}}=\dfrac{\sum^M_{k=1}|\alpha_{ik}|^2\braket{\tilde{\psi}_{ik}|\tilde{\psi}_{ik}}\frac{\mathrm{A}(\Gamma^{\tilde{\psi}_{ik}}_A)}{4G_N}}{\sum^M_{k=1}|\alpha_{ik}|^2\braket{\tilde{\psi}_{ik}|\tilde{\psi}_{ik}}}
\ee
Of course, after redefinition of the each coefficient, it reduces to the original one. 
It is also the case for the pseudo-entropy,
\begin{equation}
S(\mathcal{T}^{\Psi_i|\Psi_j}_{A})\simeq \dfrac{\bra{\Psi_j}\frac{\hat{\mathcal{A}}}{4G_N}\ket{\Psi_i}}{\braket{\Psi_j|\Psi_i}}=\dfrac{\sum^M_{k=1}\alpha_{jk}^\ast\alpha_{ik}\braket{\tilde{\psi}_{jk}|\tilde{\psi}_{ik}}\frac{\mathrm{A}(\Gamma^{\tilde{\psi}_{jk}|\tilde{\psi}_{ik}}_A)}{4G_N}}{\sum^M_{k=1}\alpha_{jk}^\ast\alpha_{ik}\braket{\tilde{\psi}_{jk}|\tilde{\psi}_{ik}}}
\end{equation}
where $\Gamma^{\tilde{\psi}_{jk}|\tilde{\psi}_{ik}}_A$ corresponds to the minimal surface (homologous to $A$) in a geometry associated with $\braket{\tilde{\psi}_{jk}|\tilde{\psi}_{ik}}$. In our present example, it is equivalent to the two-point function of the heavy operators,
\begin{equation}
\braket{\tilde{\psi}_{jk}|\tilde{\psi}_{ik}}=\braket{0|\mathcal{O}^\dagger_{H_k}(x_j,\tau_j)\mathcal{O}_{H_k}(x_i,\tau_i)|0}\neq1.
\end{equation}
Note that in the numerator, this contribution universally appears from each HHLL conformal block.

\newpage

\end{document}